\documentclass[12pt]{article}
\usepackage[utf8]{inputenc}
\usepackage{amsmath}
\usepackage{amssymb}
\usepackage{enumitem}
\usepackage{amsfonts}
\usepackage{eurosym}
\usepackage{makeidx}
\usepackage{times}
\usepackage{anysize}
\usepackage{hyperref}
\usepackage{tikz}

\makeatletter
\marginsize{2cm}{2cm}{1cm}{1cm}
\newtheorem{theorem}{Theorem}[section]

\newtheorem{assumption}{Assumption}

\newtheorem{corollary}[theorem]{Corollary}

\newtheorem{definition}[theorem]{Definition}

\newtheorem{lemma}[theorem]{Lemma}

\newtheorem{proposition}[theorem]{Proposition}
\newtheorem{remark}[theorem]{Remark}

\newenvironment{proof}[1][Proof]{\noindent\textbf{#1.} }{\ \rule{0.5em}{0.5em}}
\makeatother

\begin{document}

\title{Quadratic Hamiltonians in Fermionic Fock Spaces}
\author{J.-B. Bru \and N. Metraud}
\date{\today }
\maketitle

\begin{abstract}
Quadratic Hamiltonians are important in quantum field theory and quantum
statistical mechanics. Their general studies, which go back to the sixties,
are relatively incomplete for the fermionic case studied here. Following
Berezin, they are quadratic in the fermionic field and in this way
well-defined self-adjoint operators acting on the fermionic Fock space. We
analyze their diagonalization by applying a novel elliptic operator-valued
differential equations on the one-particle Hilbert space studied in a
companion paper. This allows for their ($\mathrm{N}$--) diagonalization
under much weaker assumptions than before. Last but not least, in 1994 Bach,
Lieb and Solovej defined them to be generators of strongly continuous
unitary groups of Bogoliubov transformations. This is shown to be an
equivalent definition, as soon as the vacuum state belongs to the domain of
definition of these Hamiltonians. This second outcome is demonstrated to be
reminiscent to the celebrated Shale-Stinespring condition on Bogoliubov
transformations. \bigskip {}

\noindent \textbf{Keywords:} elliptic flow, Brocket-Wegner flow, quadratic
operators, Bogoliubov transformation, Shale-Stinespring condition.\bigskip {}

\noindent \textbf{2020 AMS Subject Classification: }81Q10, 34A34, 47D06
\end{abstract}

\tableofcontents%

\section{Introduction}

Quadratic Hamiltonians have been used for decades in theoretical physics,
starting with Bogoliubov's theory of superfluidity \cite{Bogoliubov1} and
the BCS theory of superconductivity \cite{BCS1,BCS2,BCS3,BogoforBCS} for
bosonic and fermionic systems, respectively. They can formally be defined
(up to a certain constant) as formal series: 
\begin{equation}
\mathrm{H}_{0}\doteq \sum_{k,l\in \mathbb{N}}\{\Upsilon
_{0}\}_{k,l}a_{k}^{\ast }a_{l}+\{D_{0}\}_{k,l}a_{k}^{\ast }a_{l}^{\ast
}+\epsilon \overline{\{D_{0}\}}_{k,l}a_{k}a_{l}\mathbf{\ ,}
\label{DefH0intro}
\end{equation}%
$\epsilon $ being $-1$ for fermions and $1$ for bosons, where $\{\Upsilon
_{0}\}_{k,l}$ and $\{D_{0}\}_{k,l}$ are possibly infinite matrix entries of
some operators $\Upsilon _{0}$ and $D_{0}$ acting on a separable Hilbert
space $\mathfrak{h}$ while for$\ k\in \mathbb{N}$, $a_{k}$ is the usual
annihilation operator of a fermion or boson with wave function $\varphi
_{k}\in \mathfrak{h}$, $\{\varphi _{k}\}_{k\in \mathbb{N}}$ being an
orthonormal basis of $\mathfrak{h}$.

In this paper, we only study the fermionic case. Indeed, after 1967, to our
knowledge, there is surprisingly no new general mathematical result on
fermionic quadratic Hamiltonians defined by (\ref{DefH0intro}). By contrast,
driven by the mathematical justification of Bogoliubov's theory or the
Brockett-Wegner flow, new studies on the bosonic case have been developed in
recent years, see \cite%
{bach-bru-memo,Solovej-Nam,Derezinski2,Bruneau-derezinski2007}, even if
bosonic quadratic Hamiltonians were neglected also for decades, from the
sixties \cite{Friedrichs,Berezin,Kato-Mugibayashi} until the year 2007 with
the paper \cite{Bruneau-derezinski2007}.

Making sense of the formal definition (\ref{DefH0intro}) as a self-adjoint
operator acting on the fermionic Fock space $\mathcal{F}$ over an infinite
dimensional Hilbert space $\mathfrak{h}$ can be a non-trivial task, in
general. This issue is already studied in \cite[Theorem 6.1]{Berezin}, but
the corresponding proof is not completely rigorous. Another proof is given
by \cite[Proposition 2.1]{Carey}, but under the assumption of bounded
operators $\Upsilon _{0}$, which is a too strong condition for applications.
In Proposition \ref{Hamilselfadjoint}, we show that $\mathrm{H}_{0}$ can be
well-defined as a self-adjoint operator on $\mathcal{F}$ even for unbounded
self-adjoint operators $\Upsilon _{0}$, provided $D_{0} $ is always a
Hilbert-Schmidt operator. This necessary outcome uses (a priori) nontrivial
arguments, which are, however, relatively direct to obtain by translating
the results of \cite[Theorem 5.3]{Bruneau-derezinski2007} (itself inspired
by \cite[Theorem 6.1]{Berezin}, following \cite{Carleman}) on bosonic
quadratic operators to the fermionic case.

The definition of fermionic quadratic operators via formal series like (\ref%
{DefH0intro}) refers to Berezin's approach \cite{Berezin}. An alternative
method to define fermionic quadratic operators has been used by Bach, Lieb
and Solovej in \cite{BLS}. Instead of (\ref{DefH0intro}), they are defined
to be the generators of strongly continuous unitary groups of Bogoliubov
transformations. In this definition, quadratic operators are automatically
self-adjoint, thanks to Stone's theorem. By contrast, an explicit form of
such quadratic Hamiltonians, as given by (\ref{DefH0intro}), becomes an open
problem. In fact, Bach, Lieb and Solovej's viewpoint is very general and it
is even not clear whether the (possibly unbounded) generator $h=h^{\ast }$
of the unitary group defining the Bogoliubov transformation in their
definition can be written as a block operator matrix in infinite dimension,
allowing to speak about $\Upsilon _{0}$ and $D_{0}$. This is a necessary
assumption one has to use. Up to this condition, we show in Theorem \ref%
{prop bach2 copy(1)} as well as in Corollaries \ref{Corollaire sympa
copy(1)bis} and \ref{Corollaire sympa copy(2)} that Berezin's approach and
Bach, Lieb and Solovej's viewpoint are equivalent to each other, as soon as
the vacuum state belongs to the domain of definition of quadratic
Hamiltonians. It corresponds to take a Hilbert-Schmidt operator $D_{0}$ in (%
\ref{DefH0intro}), which is a condition already used to prove the
self-adjointness of operators defined by (\ref{DefH0intro}). This is one of
the two main results of our paper.

The Shale-Stinespring condition on Bogoliubov transformations does not
necessarily imply a Hilbert-Schmidt condition on the off-diagonal elements
of its generator, but, by Theorems \ref{prop bach2 copy(1)} and \ref{prop
bach2} as well as Corollary \ref{Corollaire sympa copy(2)}, the
Hilbert-Schmidt condition on $D_{0}$ can be seen as Shale-Stinespring-like
condition for quadratic operators. Recall that the celebrated
Shale-Stinespring condition is a sufficient and necessary condition to
implement Bogoliubov transformations on the underlying indexing Hilbert
space of an abstract CAR $C^{\ast }$--algebras via unitary transformations
in the Fock space representation, see \cite[Theorems 3.1 and 6.2]%
{Ruijsenaars} or\ \cite[Theorem 7]{A70}.

Bogoliubov transformations are not only useful to define quadratic
Hamiltonians within Bach, Lieb and Solovej's approach. By definition, they
transform creation and annihilation operators in other creation and
annihilation operators, leading in theoretical physics to quasi-particle
interpretations. Implementable Bogoliubov transformations on Fock spaces are
well-known in relatively simple cases to ($\mathrm{N}$--) diagonalize
quadratic Hamiltonians $\mathrm{H}_{0}$ of the form (\ref{DefH0intro}). The $%
\mathrm{N}$--diagonalization of $\mathrm{H}_{0}$ refers to the construction
of a unitary operator $\mathrm{U}$ acting on the Fock space such that $%
\mathrm{UH}_{0}\mathrm{U}^{\ast }$ commutes with the particle number
operator $\mathrm{N}$. It means in this case that $\mathrm{UH}_{0}\mathrm{U}%
^{\ast }$ is a quadratic Hamiltonian of the form (\ref{DefH0intro}), up to a
certain constant, with $D_{0}=0$, i.e., the second quantization of a
self-adjoint operator acting on the one-particle Hilbert space (see (\ref%
{second quanti})).

Mathematical results on the $\mathrm{N}$--diagonalization of general
quadratic operators like (\ref{DefH0intro}) have been obtained in \cite%
{Friedrichs,Berezin,Kato-Mugibayashi}, but since the sixties no new
mathematical result has been performed. Note that Araki presented in 1968 
\cite{Araki} a very general method for the \textquotedblleft $\mathrm{N}$%
--diagonalization\textquotedblright\ of bilinear Hamiltonians, but it does
not solve the issue addressed here. For more details, see Section \ref%
{Araiki section}.

The second main result of this paper concerns the $\mathrm{N}$%
--diagonalization of fermionic quadratic Hamiltonians under much more
general conditions than before. Compare indeed the previously known result
given below by Theorem \ref{thm 1}, proven in \cite{Berezin,Kato-Mugibayashi}%
, with Theorem \ref{thm3}. For instance, observe that the \textquotedblleft
gap condition\textquotedblright, in the sense that $\Upsilon _{0}\geq \alpha 
\mathbf{1}$ with $\alpha \in \mathbb{R}^{+}$ and the Hilbert-Schmidt
condition on the commutator $[\Upsilon _{0},D_{0}]$ in Theorem \ref{thm 1}
are clearly very strong restrictions as compared to the hypotheses of
Theorem \ref{thm3}, which can be applied to lower semibounded $\Upsilon _{0}$%
. In some cases, when $\Upsilon _{0}D_{0}=D_{0}\Upsilon _{0}^{\top }$, we
are even able to fully characterize the kinetic one-particle operator which
determines the spectrum of the fermionic quadratic Hamiltonians. Note that
all these results are done by means of Bogoliubov transformations,
implemented in the Fock space, as expected.

While aware of the method used in \cite{Solovej-Nam} for the bosonic case,
we do no try to use it to handle the $\mathrm{N}$--diagonalization of
fermionic quadratic Hamiltonians formally defined by (\ref{DefH0intro}).
Here, we follow another path by using the Brockett-Wegner flow \cite%
{Brockett1,Wegner1,bach-bru,Opti} like in \cite{bach-bru-memo}, because \cite%
{Solovej-Nam} does not necessarily give explicit expressions for the $%
\mathrm{N}$-diagonal form of the Hamiltonian and, in any case, the
Brockett-Wegner method of proof is interesting in its own right, while
leading excellent results like explicit expressions in special cases.
Indeed, such differential equations on spaces of operators are very little
developed in Mathematics, being in general very challenging. For instance,
although the Brockett-Wegner flow has been often used in theoretical physics 
\cite{Kehrein}, there are only two rigorous results on the well-posedness of
the Brockett-Wegner flow and only one study of its asymptotics for unbounded
operators \cite{bach-bru-memo,bach-bru}. By using fermionic quadratic
operators we give another example of its well-posedness and asymptotics, by
Theorem \ref{theorem Ht=00003Dunitary orbite}, Proposition \ref{lemma H t
infinity copy(1)} and Corollary \ref{lemma uv-Bogoliubov transformation
copy(2)}. In fact, the Brockett-Wegner flow \cite%
{Brockett1,Wegner1,bach-bru,Opti} is used as a \emph{guideline} in order to
connect the diagonalization of quadratic Hamiltonians to a new, elegant flow
on operators acting on the one-particle Hilbert space. In \cite{EllipticFlow}
we demonstrate that this non-linear flow presents remarkable ellipticity
properties that turn out to be crucial for the study of the infinite-time
limit of its solution, which is proven under relatively weak hypotheses on
the initial data. It is the elliptic counterpart of an hyperbolic flow used
to ($\mathrm{N}$--) diagonalize bosonic quadratic Hamiltonians in \cite%
{bach-bru-memo}. Essential results of \cite{EllipticFlow} for the
diagonalization of quadratic Hamiltonians are shortly explained in Section %
\ref{Eliptic flow}, while the present paper gives a salient application of
this elliptic flow to quantum field theory and quantum statistical mechanics.

To conclude, our main results are the following: (a) Theorem \ref{thm3}; (b)
Theorem \ref{prop bach2 copy(1)}, and Corollaries \ref{Corollaire sympa
copy(1)bis} and \ref{Corollaire sympa copy(2)}. (a) focuses on the ($\mathrm{%
N}$--) diagonalization of (unbounded) quadratic operators, which results
from the infinite-time limit of an evolution equation on (unbounded)
quadratic operators, similar to \cite{bach-bru-memo} in the bosonic case.
Comparatively to \cite{Solovej-Nam} done also in the bosonic case, we obtain
here some expressions for the diagonal form as integrals of elements of the
elliptic flow, and in some cases even very explicit expressions (see 
\eqref{constant of
motion eq2bis} and \eqref{dfgdgdfgdfg}). (b) and related results belong to
the mathematical foundations of fermionic quadratic Hamiltonians, similar to
Bruneau and Derezinski's study \cite{Bruneau-derezinski2007,Derezinski2} on
bosonic quadratic Hamiltonians.

Finally, following some advice, we divided the article into two parts and
thus adopted a slightly more linear approach. This consists of two
sections, which are roughly the same size and each include their own
technical subsection:

\begin{itemize}
\item Section \ref{Diagonalization of Quadratic} explains the
diagonalization of fermionic quadratic Hamiltonians, by including our main
results on this topic (Theorem \ref{thm3}),\ while discussing the method of
proof via a non-trivial evolution equation named the Brockett-Wegner flow.
Note that most of the technical results are gathered in Section \ref{Section
tech7} while Section \ref{Apprendix} is an appendix. In particular, we
briefly expose in Section \ref{Eliptic flow} the pivotal results used on the
elliptical flow studied in \cite{EllipticFlow}, while in Section \ref%
{exemplesuperconductivity} we give a paradigmatic (and historical) example
of fermionic quadratic Hamiltonian.

\item Section \ref{Diagonalization of Quadraticbis} is devoted to the
mathematical foundations of quadratic Hamiltonians, presenting various
approaches. It presents our main results on this topic (Theorem \ref{prop
bach2 copy(1)}, Corollaries \ref{Corollaire sympa copy(1)bis}, \ref%
{Corollaire sympa copy(2)}). Observe also that most of the technical results
are gathered in Section \ref{Section tech7 bis}.
\end{itemize}

\section{Diagonalizing Flow on Quadratic Hamiltonians\label{Diagonalization
of Quadratic}}

\subsection{Preliminary Definitions}

To fix notation, let $\mathfrak{h}\doteq L^{2}(\mathcal{M})$ be a separable
complex Hilbert space which we assume to be realized as a space of
square-integrable (complex-valued) functions on a measure space $(\mathcal{M}%
,\mathfrak{m})$. The scalar product and norm on $\mathfrak{h}$ are given by 
\begin{equation}
\left\langle f,g\right\rangle _{\mathfrak{h}}\doteq \int_{\mathcal{M}}%
\overline{f\left( x\right) }g\left( x\right) \mathrm{d}\mathfrak{m}\left(
x\right) \qquad \text{and}\qquad \left\Vert f\right\Vert _{\mathfrak{h}%
}\doteq \langle f,f\rangle _{\mathfrak{h}}^{1/2}  \label{scalar product}
\end{equation}%
for any $f,g\in \mathfrak{h}$. For $f\in \mathfrak{h}$, we define its
complex conjugate $\bar{f}\in \mathfrak{h}$ by $\bar{f}\left( x\right)
\doteq \overline{f\left( x\right) }$, for all $x\in \mathcal{M}$.

For any bounded (linear) operator $X$ on $\mathfrak{h}$, using the Riesz
lemma \cite[Theorem II.4]{ReedSimon} we define its transpose $X^{\top }$ and
its complex conjugate $\bar{X}$ by their sesquilinear form $\langle
f,X^{\top }g\rangle _{\mathfrak{h}}\doteq \langle \bar{g},X\bar{f}\rangle _{%
\mathfrak{h}}$ and $\langle f,\bar{X}g\rangle _{\mathfrak{h}}\doteq 
\overline{\langle \bar{f},X\bar{g}\rangle _{\mathfrak{h}}}$ for $f,g\in 
\mathfrak{h}$. These definitions can be extended to unbounded operators $X$
with domain $\mathcal{D}(X)\subseteq \mathfrak{h}$. Note that $X\geq 0$ iff $%
X^{\top }\geq 0$ and the adjoint of an operator $X$ equals $X^{\ast }=%
\overline{X^{\top }}=\overline{X}^{\top }$, where it exists (taking into
account domain issues).

\begin{remark}
\label{remark idiote copy(1)}\mbox{}\newline
The hypothesis $\mathfrak{h}\doteq L^{2}(\mathcal{M})$ is used for the sake
of simplicity and, to be specific, is in fact only used in Equation (\ref%
{ddddd2}). So, one can simply take any separable complex Hilbert space $%
\mathfrak{h}$ endowed with a complex conjugation $\mathcal{C}$, which is an
antiunitary involution on $\mathfrak{h}$, i.e., an antilinear mapping $%
\mathfrak{h}\rightarrow \mathfrak{h}$ such that $\mathcal{C}^{2}=\mathbf{1}$
and 
\begin{equation*}
\left\langle \mathcal{C}f,\mathcal{C}g\right\rangle _{\mathfrak{h}%
}=\left\langle g,f\right\rangle _{\mathfrak{h}}\ ,\qquad f,g\in \mathfrak{h}%
\ .
\end{equation*}%
See, e.g., \cite[Chapter 7]{Bru-Pedra-livre}. Then, for any operator $X$ on $%
\mathfrak{h}$, $X^{\top }\doteq \mathcal{C}X^{\ast }\mathcal{C}$ and $%
\overline{X}\doteq \mathcal{C}X\mathcal{C}$.
\end{remark}

The Banach space of bounded operators acting on a Banach space $\left( 
\mathcal{X},\left\Vert \cdot \right\Vert _{\mathcal{X}}\right) $ is denoted
by $\mathcal{B}(\mathcal{X})$, with operator norm 
\begin{equation*}
\left\Vert X\right\Vert _{\mathrm{op}}\doteq \sup_{f\in \mathcal{X}%
:\left\Vert f\right\Vert _{\mathcal{X}}=1}\left\Vert Xf\right\Vert _{%
\mathcal{X}}\ ,\qquad X\in \mathcal{B}\left( \mathcal{X}\right) \ .
\end{equation*}%
The unit $\mathbf{1}\in \mathcal{B}(\mathcal{X})$ is the identity operator.
We also use the notation $\mathcal{L}^{2}(\mathfrak{h})$ for the Hilbert
space of Hilbert-Schmidt operators acting on the separable Hilbert space $%
\mathfrak{h}$. The norm in $\mathcal{L}^{2}(\mathfrak{h})$ is denoted by 
\begin{equation*}
\left\Vert X\right\Vert _{2}\doteq \sqrt{\mathrm{tr}\left( X^{\ast }X\right) 
}\ ,\qquad X\in \mathcal{L}^{2}(\mathfrak{h})\ ,
\end{equation*}%
where $\mathrm{tr}(\cdot )$ denotes the usual trace for operators.

\subsection{Fermionic Quadratic Hamiltonians\label{Setup of the Problem2}}

Let $D_{0}\in \mathcal{L}^{2}(\mathfrak{h})$ be any Hilbert-Schmidt operator
satisfying $D_{0}=-D_{0}^{\top }$. Take $E_{0}\in \mathbb{R}$ and a
self-adjoint operator $\Upsilon _{0}=\Upsilon _{0}^{\ast }$ that is bounded
from below on the (separable) Hilbert space $\mathfrak{h}$. Pick some real 
\footnote{%
Note here that $\mathcal{D}\left( \Upsilon _{0}\right) $ is implicitly
preserved by $\mathcal{C}$ to ensure the existence of a real basis in this
set. In the general setting, a real orthonormal basis $\left\{ \varphi
_{k}\right\} _{k=1}^{\infty }$ means that $\varphi _{k}=\mathcal{C}\varphi
_{k}$ \ for all $k$. If $\mathfrak{h}$ can be decomposed as $\mathfrak{h}=%
\mathfrak{X}\oplus \mathfrak{X}^{\ast }$ for some Hilbert space $\mathfrak{X}
$, then such a real basis always exists, thanks to Zorn's lemma. See \cite[%
Lemma 4.200]{Bru-Pedra-livre}. In any case, real orthonormal bases are not
required, but we assume this situation to slightly lighten the computations.}
orthonormal basis $\left\{ \varphi _{k}\right\} _{k=1}^{\infty }$ in the
dense domain $\mathcal{D}\left( \Upsilon _{0}\right) \subseteq \mathfrak{h}$
of $\Upsilon _{0}$ and, for any $k\in \mathbb{N}$, let $a_{k}\doteq a\left(
\varphi _{k}\right) $ be the corresponding annihilation operator acting on
the fermionic Fock space 
\begin{equation}
\mathcal{F}\doteq \bigoplus_{n=0}^{\infty }\wedge ^{n}\mathfrak{h}\ ,
\label{Fock}
\end{equation}%
the scalar product and the norm of which are respectively denoted by $%
\left\langle \cdot ,\cdot \right\rangle _{\mathcal{F}}$ and $\left\Vert
\cdot \right\Vert _{\mathcal{F}}$. Note that this scalar product is the sum
over $n\in \mathbb{N}$ of each canonical scalar product on the sector $%
\wedge ^{n}\mathfrak{h}$. Here, $\wedge ^{0}\mathfrak{h}\doteq \mathbb{C}$,
while, for $n\in \mathbb{N}$, $\wedge ^{n}\mathfrak{h}$ is the subspace of
totally antisymmetric $n$--particle wave functions in $\mathfrak{h}^{\otimes
n}$, the $n$--fold tensor product of $\mathfrak{h}$.

For any fixed $E_{0}\in \mathbb{R}$, the fermionic quadratic operator 
\footnote{%
One could simply take $E_{0}=0$, but since this constant will become
time-dependent and non-zero later on, we use it in the definition.} is
defined through the operators $\Upsilon _{0}=\Upsilon _{0}^{\ast }$ and $%
D_{0}=-D_{0}^{\top }$ by 
\begin{equation}
\mathrm{H}_{0}\doteq \sum_{k,l\in \mathbb{N}}\{\Upsilon
_{0}\}_{k,l}a_{k}^{\ast }a_{l}+\{D_{0}\}_{k,l}a_{k}^{\ast }a_{l}^{\ast }+\{%
\bar{D}_{0}\}_{k,l}a_{l}a_{k}+E_{0}\mathbf{1}  \label{DefH0}
\end{equation}%
with 
\begin{equation*}
\left\{ X\right\} _{k,l}\doteq \left\langle \varphi _{k},X\varphi
_{l}\right\rangle \ ,\qquad k,l\in \mathbb{N}\ ,
\end{equation*}%
for all operators $X$ acting on $\mathfrak{h}$. Since $\Upsilon _{0}$ is by
assumption self-adjoint and $D_{0}=-D_{0}^{\top }\in \mathcal{L}^{2}(%
\mathfrak{h})$ is Hilbert-Schmidt, we prove in Proposition \ref%
{Hamilselfadjoint} that the operator $\mathrm{H}_{0}$ is essentially
self-adjoint on the domain 
\begin{equation}
\mathcal{D}_{0}\doteq \bigcup_{N\in \mathbb{N}}\left(
\bigoplus_{n=0}^{N}\left( \wedge ^{n}\mathcal{D}\left( \Upsilon _{0}\right)
\right) \right)  \label{domain H0}
\end{equation}%
with $\wedge ^{0}\mathcal{D}\left( \Upsilon _{0}\right) \doteq \mathbb{C}$
while for $n\in \mathbb{N}$, $\wedge ^{n}\mathcal{D}\left( \Upsilon
_{0}\right) $ is the subspace of totally antisymmetric $n$--particle wave
functions in $\mathcal{D}\left( \Upsilon _{0}\right) ^{\otimes n}$, the $n$%
--fold tensor product of $\mathcal{D}\left( \Upsilon _{0}\right) ^{\otimes
n}\subseteq \mathfrak{h}^{\otimes n}$. We use again the notation $\mathrm{H}%
_{0}\equiv \mathrm{H}_{0}^{\ast \ast }$ for its (uniquely defined)
self-adjoint extension. Note additionally that $\mathrm{H}_{0}$ is
independent of the (possibily real) orthonormal basis $\left\{ \varphi
_{k}\right\} _{k=1}^{\infty }\subseteq \mathcal{D}\left( \Upsilon
_{0}\right) \subseteq \mathfrak{h}$ chosen, as explained after Proposition %
\ref{Hamilselfadjoint}. For a general form of the quadratic Hamiltonian $%
\mathrm{H}_{0}$ written within an arbitrary orthonormal basis in $\mathcal{D}%
\left( \Upsilon _{0}\right) $, see Equation (\ref{dddddd}).

\begin{remark}[Alternative definition]
\label{Defalternativequadra}\mbox{}\newline
In Definition \ref{def quadratic}, quadratic Hamiltonians are alternatively
defined as generators of strongly continuous (one-parameter) unitary groups
of Bogoliubov transformations. In this approach, they are automatically
self-adjoint, thanks to Stone's theorem \cite[Theorem 6.2]{Konrad}. The
relation to (\ref{DefH0}) is studied in Section \ref{Section Bach}.
\end{remark}

Observe that the condition $D_{0}=-D_{0}^{\top }$\ in (\ref{DefH0}) can be
taken without loss of generality because the family $\{a_{k}\}_{k\in \mathbb{%
N}}$ satisfies the Canonical Anticommutation Relations (CAR), i.e., 
\begin{equation}
a_{k}a_{l}+a_{l}a_{k}=0\ ,\quad a_{k}a_{l}^{\ast }+a_{l}^{\ast }a_{k}=\delta
_{k,l}\mathbf{1}\ ,  \label{CAR}
\end{equation}%
for all $k,l\in \mathbb{N}$. See \cite[p. 10]{BratteliRobinson}. Meanwhile,
the Hilbert-Schmidt condition $D_{0}\in \mathcal{L}^{2}(\mathfrak{h})$ is a
pivotal assumption to obtain a well-defined quadratic Hamiltonian. It is in
particular a necessary condition to be able to define $\mathrm{H}_{0}$ on $%
\mathcal{D}_{0}$ since, using the so-called vacuum state $\Psi \doteq \left(
1,0,\ldots \right) \in \mathcal{D}_{0}$, one gets 
\begin{equation}
\left\Vert \left( \mathrm{H}_{0}-E_{0}\mathbf{1}\right) \Psi \right\Vert _{%
\mathcal{F}}=\sqrt{2}\left\Vert D_{0}\right\Vert _{2}\ ,
\label{quadratic opereators vaccum}
\end{equation}%
thanks to Equation (\ref{eqJB20}).

Hamiltonians like (\ref{DefH0}) appear very often in theoretical physics and
can for instance encode in some sense interaction effects. In fact,
interparticle interactions usually play a pivotal role in physical
phenomena, but they make the\ physical system extremely difficult to study
in a mathematically rigorous way and effective models coming from different
approximations and ans\"{a}zte are usually used. It is exactly what happens
for the celebrated BCS theory proposed in the late 1950s to explain
conventional superconductivity, which leads to the explicit diagonalization
of a (very simple) quadratic operators like (\ref{DefH0}). See Section \ref%
{exemplesuperconductivity} for more details. Another example is given in the
context of the (generalized) Hartree-Fock theory \cite{BLS}, which has a
strong affinity with (fermionic) mean-field models and effective quadratic
Hamiltonians, as first discussed in \cite[Sections 5.7 and 6.10]%
{Bru-Pedra-livre} (with a research paper in preparation).

Fermionic quadratic Hamiltonians, as defined by (\ref{DefH0})--(\ref{domain
H0}), are generally unbounded in spite of the boundedness of fermionic
creation and annihilation operators \cite[Proposition 5.2.2]%
{BratteliRobinson}. For instance, if $\Upsilon _{0}=\mathbf{1}$ is the
identity mapping on $\mathfrak{h}$ and $D_{0}=0$, then one obtains from (\ref%
{DefH0}) the so-called \emph{particle number operator} 
\begin{equation}
\mathrm{N}=\sum_{k\in \mathbb{N}}a_{k}^{\ast }a_{k}\ ,
\label{particle number operator}
\end{equation}%
which is unbounded for infinite-dimensional Hilbert spaces $\mathfrak{h}$.
It is a very simple (well-known) quadratic Hamiltonian that can be
explicitly defined by its spectral properties: its spectrum is $\sigma (%
\mathrm{N})=\mathbb{N}_{0}$ with 
\begin{equation}
\mathrm{N}\varphi _{n}\doteq n\varphi _{n}\ ,\qquad n\in \mathbb{N}_{0},\
\varphi _{n}\in \wedge ^{n}\mathfrak{h}\subseteq \mathcal{F}\ ,
\label{domain N0}
\end{equation}%
while its domain is 
\begin{equation}
\mathcal{D}\left( \mathrm{N}\right) \doteq \left\{ \left( \varphi
_{n}\right) _{n\in \mathbb{N}_{0}}\in \mathcal{F}:\sum_{n\in \mathbb{N}%
_{0}}n^{2}\left\Vert \varphi _{n}\right\Vert ^{2}<\infty \right\} \supseteq 
\mathcal{D}_{0}  \label{domain N}
\end{equation}%
See, e.g., \cite[p. 7]{BratteliRobinson}. Equality (\ref{particle number
operator}) means in particular that $\mathcal{D}_{0}$ is a core of $\mathrm{N%
}$.

More generally, quadratic Hamiltonians of the form (\ref{DefH0}) for $%
D_{0}=0 $ are much easier Hamiltonians to study than general quadratic
Hamiltonians with possibly non-zero $D_{0}\neq 0$, because their spectral
features can be deduced from the spectral properties of the operator $%
\Upsilon _{0}$ acting on the one-particle Hilbert space $\mathfrak{h}$. As
is well-known, for $D_{0}=0$, the Hamiltonian $\mathrm{H}_{0}$ is the
so-called second-quantization of the one-particle Hamiltonian $\Upsilon _{0}$%
, see, e.g., \cite[Section 5.2.1]{BratteliRobinson}. Hamiltonians of this
form are named $\mathrm{N}$\emph{--diagonal} since they are the only ones
that commute with the particle number operator $\mathrm{N}$.

In the more general cases where $D_{0}\neq 0$, one can usually find a
unitary transformation $\mathrm{U}$ such that $\mathrm{U\mathrm{H}}_{0}%
\mathrm{U}^{\ast }$ is $\mathrm{N}$--diagonal. This was first shown for
fermions in the particular case of the BCS theory (Section \ref%
{exemplesuperconductivity}) by implementing\footnote{%
Bogoliubov transformations on CAR $C^{\ast }$-algebras are not necessarily
implementable on the Fock space, i.e., they do not necessarily well-define a
unitary transformation on the Fock space. This is discussed in detail in
Section \ref{Araiki section}, see in particular the Shale-Stinespring
theorem (Theorem \ref{Shale-Stinespring THM}).} on the Fock space an
algebraic transformation, named the Bogoliubov (unitary)\ transformation,
which defines a $\ast $-isomorphism of CAR $C^{\ast }$-algebras.

\subsection{Diagonalization of Quadratic Hamiltonians}

General studies of fermionic quadratic Hamiltonians go back to the sixties
with Berezin's book \cite{Berezin} published in 1966, following Friedrichs's
work \cite{Friedrichs} written in 1953. The main result on the $\mathrm{N}$%
--diagonalization of quadratic operators like (\ref{DefH0}) refers to \cite[%
Theorem 8.2]{Berezin}, which we reproduce for the reader's convenience:

\begin{theorem}[Berezin]
\label{thm 1}\mbox{}\newline
Let $E_{0}\in \mathbb{R}$. Take $D_{0}=\bar{D}_{0}=-D_{0}^{\top }\in 
\mathcal{L}^{2}(\mathfrak{h})$ and $\Upsilon _{0}=\Upsilon _{0}^{\top
}=\Upsilon _{0}^{\ast }\geq \alpha \mathbf{1}$ with strictly positive $%
\alpha \in \mathbb{R}^{+}$. Assume additionally that the commutator $%
[D_{0},\Upsilon _{0}]\in \mathcal{L}^{2}(\mathfrak{h})$ and 
\begin{equation}
\Upsilon _{0}^{2}+4D_{0}D_{0}^{\ast }\pm 2[D_{0},\Upsilon _{0}]>\alpha 
\mathbf{1}\ .  \label{berezin}
\end{equation}%
Then, there is a unitary transformation $\mathrm{U}$ such that $\mathrm{U%
\mathrm{H}}_{0}\mathrm{U}^{\ast }$ is $\mathrm{N}$--diagonal.
\end{theorem}

\noindent Note that this theorem is originally stated with the condition 
\begin{equation}
\Upsilon _{0}^{2}+4D_{0}D_{0}^{\ast }+2[D_{0},\Upsilon _{0}]>\alpha \mathbf{1%
}  \label{berezin2}
\end{equation}%
instead of (\ref{berezin}). But it seems to be a mistake, see \cite[Theorem 3%
]{Kato-Mugibayashi}. The unitary transformation of Theorem \ref{thm 1} is
given via a Bogoliubov transformation.

Note that Araki presented in 1968 \cite{Araki} a very general method for the
\textquotedblleft $\mathrm{N}$--diagonalization\textquotedblright\ of
bilinear Hamiltonians. The results are done in great generality by using an
algebraic approach, but they do not solve the issue addressed here. For more
details, we recommend Section \ref{Araiki section}, where Araki's work is
described in detail.

In fact, after 1967, to our knowledge, there is surprisingly no new general
results on the $\mathrm{N}$--diagonalization of fermionic quadratic
Hamiltonians described here. The same situation appears for the bosonic case
for which no new results on such Hamiltonians were performed since the
1960ies \cite{Friedrichs,Berezin,Kato-Mugibayashi}, until the year 2007 with
the paper \cite{Bruneau-derezinski2007}. Then, driven by the mathematical
justification of Bogoliubov's theory or the Brockett-Wegner flow, new
results on such models have been developed in recent years, see \cite%
{bach-bru-memo,Solovej-Nam,Derezinski2}, as on the $\mathrm{N}$%
--diagonalization of bosonic quadratic Hamiltonians \cite%
{bach-bru-memo,Solovej-Nam}.

By contrast, there is no recent activity on the (equally important)
fermionic case and we address the issue of the $\mathrm{N}$--diagonalization
of fermionic quadratic Hamiltonians by means of an elliptic operator-valued
flow (Section \ref{Eliptic flow}). It leads to a continuous family of
Bogoliubov transformations and gives new properties on the $\mathrm{N}$%
--diagonal form of $\mathrm{\mathrm{H}}_{0}$, allowing in some case to
compute it explicitly. This is explained in Section \ref{sectionBWflow}
while here we only sum up the final result on the $\mathrm{N}$%
--diagonalization of quadratic Hamiltonians:

\begin{theorem}[Diagonalization of quadratic Hamiltonians]
\label{thm3}\mbox{}\newline
Let $E_{0}\in \mathbb{R}$. Take $D_{0}=-D_{0}^{\top }\in \mathcal{L}^{2}(%
\mathfrak{h})$ and $\Upsilon _{0}=\Upsilon _{0}^{\ast }$, both acting on $%
\mathfrak{h}$. Assume that 
\begin{equation}
\Upsilon _{0}\geq -\left( \mu -\varepsilon \right) \mathbf{1}\qquad \text{and%
}\qquad \Upsilon _{0}+4D_{0}\left( \Upsilon _{0}^{\top }+\mu \mathbf{1}%
\right) ^{-1}D_{0}^{\ast }\geq \mu \mathbf{1}  \label{condition plus}
\end{equation}%
for some $\mu \in \mathbb{R}\backslash \{0\}$ and $\varepsilon \in \mathbb{R}%
^{+}$. Then, there is a unitary transformation $\mathrm{U}$ such that 
\begin{equation*}
\mathrm{U\mathrm{H}}_{0}\mathrm{U}^{\ast }=\sum_{k,l\in \mathbb{N}%
}\{\Upsilon _{\infty }\}_{k,l}a_{k}^{\ast }a_{l}+\left(
E_{0}-8\int_{0}^{\infty }\left\Vert D_{\tau }\right\Vert _{2}^{2}\mathrm{d}%
\tau \right) \mathbf{1}
\end{equation*}%
with the operator family $D$ and 
\begin{equation*}
\Upsilon _{\infty }\doteq \Upsilon _{0}+\Delta _{\infty }\geq \left\vert \mu
\right\vert \mathbf{1}
\end{equation*}%
being defined via Theorems \ref{lemma existence 2 copy(4)} and \ref{lemma
asymptotics1 copy(4)}. In particular, $\mathrm{U\mathrm{H}}_{0}\mathrm{U}%
^{\ast }$ is $\mathrm{N}$--diagonal.
\end{theorem}

\begin{proof}
Combine Corollary \ref{lemma uv-Bogoliubov transformation copy(2)} with
Theorem \ref{lemma asymptotics1 copy(4)}. In fact, Theorem \ref{thm3} gives
a nice application of the elliptic operator-valued flow studied in \cite%
{EllipticFlow} to quantum field theory and quantum statistical mechanics.
\end{proof}

\begin{remark}
\label{lemma asymptotics1 copy(5)}\mbox{}\newline
Let $\alpha \in \mathbb{R}^{+}$ and assume that $\Upsilon _{0}\geq \alpha 
\mathbf{1}$. Then, upon choosing $\varepsilon \doteq \alpha $ and $\mu
\doteq \alpha /2$, observe that (\ref{condition plus}) holds true. As a
result, Theorem \ref{lemma asymptotics1 copy(4)} can be applied to all
positive operators $\Upsilon _{0}$ with spectral gap with $0$ (meaning that $%
\Upsilon _{0}\geq \alpha \mathbf{1}$ with $\alpha >0$). In this case, note
that $\Upsilon _{\infty }\geq \alpha \mathbf{1}$.
\end{remark}

\begin{remark}
\label{lemma asymptotics1 copy(6)}\mbox{}\newline
In practice, one shall take the maximum value of $\mu \in \mathbb{R}$ such
that (\ref{condition plus}) holds true (provided such a $\mu $ exists) in
order to obtain a more accurate estimate of the spectral gap (with $0$) of $%
\Upsilon _{\infty }$.
\end{remark}

By Proposition \ref{lemma uv-Bogoliubov transformation} and Corollary \ref%
{lemma uv-Bogoliubov transformation copy(1)}, the unitary transformation $%
\mathrm{U}$ implements a Bogoliubov transformation. The $\mathrm{N}$%
--diagonal form of $\mathrm{\mathrm{H}}_{0}$ can be additionally computed in
some cases. For instance, using the constant of motion given by \cite[%
Equation (23)]{EllipticFlow} and Theorem \ref{lemma asymptotics1 copy(4)}
with $D_{0}^{\top }=\pm D_{0}$, we have 
\begin{equation}
\mathrm{tr}\left( \Upsilon _{\infty }^{2}-\Upsilon
_{0}^{2}-4D_{0}D_{0}^{\ast }\right) =0\ .  \label{constant of motion eq1bis}
\end{equation}%
at least if $\Upsilon _{0}$ is bounded. If $\Upsilon _{0}D_{0}=D_{0}\Upsilon
_{0}^{\top }$ on the domain $\mathcal{D}(\Upsilon _{0}^{\top })$, $\Upsilon
_{0}\geq -\left( \mu -\varepsilon \right) \mathbf{1}$ for some $\mu \in 
\mathbb{R}$ and $\varepsilon \in \mathbb{R}^{+}$ and 
\begin{equation}
\max \left\{ \Vert (\Upsilon _{0}+\mu \mathbf{1)}D_{0}(\Upsilon _{0}^{\top
}+\mu \mathbf{1)}^{-1}\Vert _{\mathrm{op}},\Vert (\Upsilon _{0}^{\top }+\mu 
\mathbf{1)}D_{0}^{\ast }(\Upsilon _{0}+\mu \mathbf{1)}^{-1}\Vert _{\mathrm{op%
}}\right\} \leq \mathrm{C}  \label{conditionplus}
\end{equation}%
for some strictly positive constant $\mathrm{C}\in \mathbb{R}^{+}$, then 
\cite[Equation (11)]{EllipticFlow} yield%
\begin{equation}
\Upsilon _{\infty }=\sqrt{\Upsilon _{0}^{2}+4D_{0}D_{0}^{\ast }}
\label{constant of motion eq2bis}
\end{equation}%
on the domain $\mathcal{D}(\Upsilon _{0})$ as well as 
\begin{equation}
8\int_{0}^{\infty }\left\Vert D_{\tau }\right\Vert _{2}^{2}\mathrm{d}\tau =%
\frac{1}{2}\mathrm{tr}\left( \sqrt{\Upsilon _{0}^{2}+4D_{0}D_{0}^{\ast }}%
-\Upsilon _{0}\right) \in \left[ 0,\infty \right) \ .  \label{dfgdgdfgdfg}
\end{equation}%
Note that the fact that the trace in the above equation is finite is a
non-trivial\footnote{%
For a direct proof under the condition $\Upsilon _{0}=\Upsilon _{0}^{\ast
}\geq \alpha \mathbf{1}$ with $\alpha \in \mathbb{R}^{+}$, see \cite[Lemma
8.1]{Berezin}.} consequence of the study. In other words, $\mathrm{U\mathrm{H%
}}_{0}\mathrm{U}^{\ast }$ can be fully characterized in this
\textquotedblleft commutative\textquotedblright\ case. This is for instance
used in Section \ref{exemplesuperconductivity} to recover the exact $\mathrm{%
N}$--diagonal form (\ref{BCSlimit}) of the BCS Hamiltonian.

Note that Theorem \ref{thm 1} does not make explicit neither the $\mathrm{N}$%
--diagonal form of fermionic quadratic Hamiltonians nor the Bogoliubov
transformation. In addition, observe that (\ref{condition plus}) looks
similar to Condition (\ref{berezin}) of Theorem \ref{thm 1}, but the
assumptions $\Upsilon _{0}\geq \alpha \mathbf{1}$ with $\alpha \in \mathbb{R}%
^{+}$ and $[\Upsilon _{0},D_{0}]\in \mathcal{L}^{2}(\mathfrak{h})$ in
Theorem \ref{thm 1} are clearly strong restrictions as compared to the
hypotheses of Theorem \ref{thm3}. See for instance Remark \ref{lemma
asymptotics1 copy(5)}. In fact, the asssumptions of Berezin's theorem
(Theorem \ref{thm 1}) are so strong that they may not even be satisfied by
the elementary quadratic Hamiltonians of the BCS theory, as discussed in
Section \ref{exemplesuperconductivity}. Indeed, for these Hamiltonians, the
condition $\Upsilon _{0}\geq \alpha \mathbf{1}$ may not necessarily hold for
a \emph{strictly positive} $\alpha \in \mathbb{R}^{+}$. Meanwhile, the
assumptions of Theorem \ref{thm3} are much more natural and general than
those of Theorem \ref{thm 1} and they include important cases with negative
spectra for $\Upsilon _{0}$.

To conclude, note that the conditions of Theorem \ref{thm3} are not
necessary (in principle). As already discussed in \cite{EllipticFlow} we
expect the limiting case to be when\footnote{$X>0$ means $X\geq 0$, with $0$
not being in the spectrum of $X$.} 
\begin{equation}
\Upsilon _{0}>0\text{\qquad and\qquad }\Upsilon _{0}+4D_{0}\left( \Upsilon
_{0}^{\top }\right) ^{-1}D_{0}^{\ast }\geq 0  \label{ghfhhg}
\end{equation}%
with $D_{0}\left( \Upsilon _{0}^{\top }\right) ^{-1}D_{0}^{\ast }$ being a
well-defined operator acting on $\mathfrak{h}$. Up to singular cases, these
assumptions may be necessary for the convergence of the flow because the
existence of $\Upsilon _{\infty }$ seems to generally imply its positivity
(up to singular cases). See the spectral properties of the operator family
studied in \cite[Theorem 3]{EllipticFlow}. They correspond to take $%
\varepsilon ,\mu \rightarrow 0$ in the conditions of Theorem \ref{thm3} with
a possible limit operator $\Upsilon _{\infty }\doteq \Upsilon _{0}+\Delta
_{\infty }\geq 0$ having a priori no spectral gap with $0$. The conditions
expressed by (\ref{ghfhhg}) look similar to the situation $\Omega _{0}\geq 0$%
, without spectral gap (with $0$), studied in the hyperbolic case \cite%
{bach-bru-memo}. By analogy with the hyperbolic case, one may assume that $%
\left( \Upsilon _{0}\right) ^{-1/2}D_{0}$ is a Hilbert-Schmidt operator to
ensure the existence of $\Delta _{\infty }$ as a trace-class operator. The
arguments to handle the non-spectral gap case (with $0$) in \cite%
{bach-bru-memo} were technically involved and long (with domain issues). We
do not expect any fundamental obstruction for the analogous study in the
fermionic case under Condition (\ref{ghfhhg}), but we refrain from doing so,
limiting in particular the length of this document.

\subsection{Brockett-Wegner Flow and Elliptic Operator Flows\label%
{sectionBWflow}}

We propose to tackle the diagonalization of quadratic fermionic Hamiltonians
by using the Brockett-Wegner flow \cite{Brockett1,Wegner1,bach-bru,Opti} as
a guideline. This method of proof is interesting in its own right. Indeed,
the Brockett-Wegner flow is very little developed in Mathematics, although
it has been often used in theoretical physics \cite{Kehrein}. In our
setting, it allows one to connect the $\mathrm{N}$--diagonalization of
fermionic quadratic Hamiltonians to an elegant non-linear elliptic flow on
operators acting on the one-particle Hilbert space, which is studied in
detail in \cite{EllipticFlow} (see Section \ref{Eliptic flow}).

In this section, we thus formally explain this connection. This allows the
interested reader to understand the general strategy to prove Theorem \ref%
{thm3}, before going through the rigorous arguments performed in Section \ref%
{Section tech7}. In addition, this demonstrates how the formal
Brockett-Wegner flow can produce novel and elegant ordinary differential
equations (ODEs) on operator spaces and this section can thus be viewed as
an invitation to develop this research direction.

The Brockett-Wegner flow \cite{Brockett1,Wegner1,bach-bru,Opti} is a
(quadratically) non-linear first-order differential equation for operators.
It is used to diagonalize self-adjoint operators. On a formal level, it is
easy to describe, but the mathematically rigorous treatment can be rather
involved, in particular for unbounded operators.

Aiming at the $\mathrm{N}$--diagonalization of fermionic quadratic
Hamiltonians, one can apply the Brockett-Wegner flow to a fermionic
quadratic Hamiltonian $\mathrm{H}_{0}$ via the particle number operator $%
\mathrm{N}$ defined by (\ref{particle number operator}), both acting on the
Fock space $\mathcal{F}$ (\ref{Fock}). Formally, it corresponds to the
evolution equation 
\begin{equation}
\forall t\in \mathbb{R}_{0}^{+}:\qquad \partial _{t}\mathrm{H}_{t}=\left[ 
\mathrm{H}_{t},\left[ \mathrm{H}_{t},\mathrm{N}\right] \right] \ ,\qquad 
\mathrm{H}_{t=0}\doteq \mathrm{H}_{0}\ ,  \label{brocket flow}
\end{equation}%
with $[A,B]\doteq AB-BA$ being the commutator between two operators $A$ and $%
B$ (provided it exists on some domain). This operator flow is very useful
because it is awaited that the solution to (\ref{brocket flow}) is a family
of unitarily equivalent quadratic operators: For any $s\in \mathbb{R}%
_{0}^{+} $ and $t\in \left[ s,\infty \right) $, $\mathrm{H}_{t}=\mathrm{U}%
_{t,s}\mathrm{\mathrm{H}}_{s}\mathrm{U}_{t,s}^{\ast }$ with $\left( \mathrm{U%
}_{t,s}\right) _{t\geq s}$ being a (evolution) family of unitary operators
solution to the non-autonomous (hyperbolic in Kato's terminology \cite%
{Kato,Kato1973}, named Kato-hyperbolic here to avoid confusions) evolution
equation 
\begin{equation}
\forall s\in \mathbb{R}_{0}^{+},\ t\in \left[ s,\infty \right) :\qquad
\partial _{t}\mathrm{U}_{t,s}=-i\mathrm{G}_{t}\mathrm{U}_{t,s}\ ,\qquad 
\mathrm{U}_{s,s}\doteq \mathbf{1}\ ,  \label{flow equationflow equation}
\end{equation}%
with generator $\mathrm{G}_{t}\doteq i\left[ \mathrm{N},\mathrm{H}_{t}\right]
$. What is more, we expect the convergence in some sense of $\mathrm{H}_{t}$
and $\mathrm{U}_{t,0}$ in the limit $t\rightarrow \infty $, leading to a
unitarily equivalent Hamiltonian $\mathrm{H}_{\infty }=\mathrm{U}_{\infty ,0}%
\mathrm{\mathrm{H}}_{0}\mathrm{U}_{\infty ,0}^{\ast }$ satisfying $\left[ 
\mathrm{H}_{\infty },\mathrm{N}\right] =0$.

Nonetheless, establishing the well-posedness of the Brockett-Wegner flow (%
\ref{brocket flow}) is highly non-trivial, because quadratic Hamiltonians
are generally unbounded operators. The only known general result in the
unbounded case concerns the local existence of its solution under some
restricted conditions \cite{bach-bru}, which cannot be applied here.
Moreover, even if the solution was proven to exist for all times, its
asymptotics would even be harder to study for general Brockett-Wegner flows. 
\newline

In order to avoid technical issues related to unboundedness and cumbersome
environment inherent to the Fock space, it is of course natural to transfer
the problem in the one-particle space. To this end, as in Section \ref%
{Araiki section}, on can use the space $\mathcal{H}=\mathfrak{h}\oplus 
\mathfrak{h}$ endowed with the antiunitary involution $\mathfrak{A}$ (\ref%
{ghghghg}), instead of the Fock space. In this context, the particle number
operator refers to the bounded (self-dual) operator 
\begin{equation*}
\mathcal{N}\doteq \frac{1}{2}\left( 
\begin{array}{cc}
\mathbf{1} & 0 \\ 
0 & -\mathbf{1}%
\end{array}%
\right) \in \mathcal{B}\left( \mathfrak{h}\oplus \mathfrak{h}\right) \ .
\end{equation*}%
Compare with Lemma \ref{Lemma quasi-free} for $\Upsilon _{0}=\mathbf{1}$ and 
$D_{0}=0$. Formally, the self-dual version of the Brockett-Wegner flow (\ref%
{brocket flow}) is the following differential equation in the strong sense
on (some domain of) $\mathcal{H}=\mathfrak{h}\oplus \mathfrak{h}$: 
\begin{equation}
\forall t\in \mathbb{R}_{0}^{+}:\qquad \partial _{t}h_{t}=4\left[ h_{t},%
\left[ h_{t},\mathcal{N}\right] \right] \ ,\qquad h_{t=0}\doteq \frac{1}{2}%
\left( 
\begin{array}{cc}
\Upsilon _{0} & 2D_{0} \\ 
2D_{0}^{\ast } & -\Upsilon _{0}^{\top }%
\end{array}%
\right) \ ,  \label{simplified flow}
\end{equation}%
where $D_{0}=-D_{0}^{\top }\in \mathcal{B}\left( \mathfrak{h}\right) $ and $%
\Upsilon _{0}=\Upsilon _{0}^{\ast }$ is a self-adjoint operator that is
bounded from below. Having Lemma \ref{Lemma quasi-free} in mind, one guesses
that the solution to the last differential equation should be of the form 
\begin{equation}
h_{t}=\frac{1}{2}\left( 
\begin{array}{cc}
\Upsilon _{t} & 2D_{t} \\ 
-2\overline{D}_{t} & -\Upsilon _{t}^{\top }%
\end{array}%
\right) =\frac{1}{2}\left( 
\begin{array}{cc}
\Upsilon _{t} & 2D_{t} \\ 
2D_{t}^{\ast } & -\Upsilon _{t}^{\top }%
\end{array}%
\right) \ ,\qquad t\in \mathbb{R}_{0}^{+}\ ,  \label{ghfghfh}
\end{equation}%
for some strongly continuous families of operators $\left( \Upsilon
_{t}\right) _{t\geq 0}$ and $\left( D_{t}\right) _{t\geq 0}$ such that $%
D_{t}=-D_{t}^{\top }\in \mathcal{B}\left( \mathfrak{h}\right) $ and $%
\Upsilon _{t}=\Upsilon _{t}^{\ast }$ is a self-adjoint operator that is
bounded from below, with always the time-independent domain $\mathcal{D}%
(\Upsilon _{0})$. In this case, one formally computes the commutators 
\begin{equation*}
g_{t}\doteq 2i\left[ \mathcal{N},h_{t}\right] =\frac{i}{2}\left( 
\begin{array}{cc}
0 & 4D_{t} \\ 
-4D_{t}^{\ast } & 0%
\end{array}%
\right) \ .
\end{equation*}%
Compare this generator with Lemma \ref{Lemma quasi-free} and Equation (\ref%
{generator}). As a consequence, 
\begin{equation*}
\partial _{t}h_{t}=4\left[ h_{t},\left[ h_{t},\mathcal{N}\right] \right] =2i%
\left[ h_{t},g_{t}\right] =\frac{1}{2}\left( 
\begin{array}{cc}
16D_{t}D_{t}^{\ast } & -4\left( \Upsilon _{t}D_{t}+D_{t}\Upsilon _{t}^{\top
}\right) \\ 
-4\left( \Upsilon _{t}^{\top }D_{t}^{\ast }+D_{t}^{\ast }\Upsilon _{t}\right)
& -16D_{t}^{\ast }D_{t}%
\end{array}%
\right)
\end{equation*}%
for all times $t\in \mathbb{R}_{0}^{+}$. By identification this leads
precisely to the flow (\ref{flow equation-quadratic deltabis}) (with $%
\Upsilon _{t}=\Upsilon _{0}+\Delta _{t}$) studied in \cite{EllipticFlow}.

Analogously to Equation (\ref{flow equationflow equation}) in the Fock
space, one defines an evolution family of unitary operators solution to the
(Kato-hyperbolic) non-autonomous evolution equation 
\begin{equation}
\forall s\in \mathbb{R}_{0}^{+},\ t\in \left[ s,\infty \right) :\qquad
\partial _{t}\mathcal{U}_{t,s}=-ig_{t}\mathcal{U}_{t,s}\ ,\qquad \mathcal{U}%
_{s,s}\doteq \mathbf{1}\ .
\end{equation}%
By (\ref{ghghghg}) and $D_{t}=-D_{t}^{\top }$ together with Remark \ref%
{remark idiote copy(1)}, $g_{t}\mathfrak{A}=\mathfrak{A}g_{t}$ for all $t\in 
\mathbb{R}_{0}^{+}$ and, via the uniqueness of the solution to the above
evolution equation, one can check that $\left( \mathcal{U}_{t,s}\right)
_{t\geq s}$ is a family of Bogoliubov transformations. In particular, $%
\mathcal{U}_{t,s}\mathfrak{A=A}\mathcal{U}_{t,s}$. Moreover, assuming a
unique solution to the flow (\ref{simplified flow}), we should have $h_{t}=%
\mathcal{U}_{t,0}h_{0}\mathcal{U}_{t,0}^{\ast }$ for all $t\in \mathbb{R}%
_{0}^{+}$. For instance, the Hilbert-Schmidt norm of $h_{(\cdot )}$ (if it
exists) is constant, which can be rewritten as 
\begin{equation*}
\frac{1}{4}\mathrm{tr}\left( \Upsilon _{t}^{2}+4D_{t}D_{t}^{\ast }\right) =%
\frac{1}{4}\mathrm{tr}\left( \Upsilon _{0}^{2}+4D_{0}D_{0}^{\ast }\right)
,\qquad t\in \mathbb{R}_{0}^{+}\ ,
\end{equation*}%
see (\ref{ghfghfh}). This gives an intuition of the \textbf{elliptic} nature
of the flow on the (self-dual) one-particle space. (In the bosonic case, one
does not use an inner product, but a symplectic form, leading heuristically
to an hyperbolic flow through the same argument.) Rigorous treatment of the
flow with generalized elliptic properties is developed in \cite{EllipticFlow}%
.

The Brockett-Wegner flow (\ref{simplified flow}) takes place in the
one-particle space, but we have to lift it again to the Fock space. In
particular, the above family $\left( \mathcal{U}_{t,s}\right) _{t\geq s}$ of
Bogoliubov transformations has to be implemented, as explained in Remark \ref%
{remark idiote copy(2)}. Mutatis mutandis for $\left( h_{t}\right) _{t\geq
0} $ and the relation $h_{t}=\mathcal{U}_{t,0}h_{0}\mathcal{U}_{t,0}^{\ast }$%
. This is done in Sections \ref{Section proof theroem important 2bis} and %
\ref{section unitarly equi H} whereas we present results on general
implementation criteria of quadratic Hamiltonians in Section \ref{Section
Bach}.

In fact, using the information we obtained in the one-particle space, the
solution to the Brockett-Wegner flow (\ref{brocket flow}) should be the
following fermionic quadratic operator: 
\begin{equation}
\mathrm{H}_{t}\doteq \sum_{k,l\in \mathbb{N}}\left\{ \Upsilon _{t}\right\}
_{k,l}a_{k}^{\ast }a_{l}+\left\{ D_{t}\right\} _{k,l}a_{k}^{\ast
}a_{l}^{\ast }+\left\{ \bar{D}_{t}\right\} _{k,l}a_{l}a_{k}+\left(
E_{0}-8\int_{0}^{t}\left\Vert D_{\tau }\right\Vert _{2}^{2}\mathrm{d}\tau
\right) \mathbf{1}  \label{DefH0bis}
\end{equation}%
with $\Upsilon _{t}\doteq \Upsilon _{0}+\Delta _{t}$ for each $t\in \mathbb{R%
}_{0}^{+}$ and where $\Delta $ and $D$ are \textbf{solution to the elliptic
operator-valued flow (\ref{flow equation-quadratic deltabis}) }studied in
detail in \cite{EllipticFlow}. Observe in this case that $D_{t}$\ needs to
be a Hilbert-Schmidt operator for all $t\in \mathbb{R}_{0}^{+}$ in order to
invoke Proposition \ref{Hamilselfadjoint} for the self-adjointness of $%
\mathrm{H}_{t}$. In fact, using additionally the CAR (\ref{CAR}), we should
have%
\begin{equation}
D_{t}=-D_{t}^{\top }\in \mathcal{L}^{2}(\mathfrak{h})\ ,\qquad t\in \mathbb{R%
}_{0}^{+}\ ,  \label{sdklfjdsklfjsdfkl}
\end{equation}%
which is satisfied provided $D_{0}=D_{0}^{\top }\in \mathcal{L}^{2}(%
\mathfrak{h})$, see Theorem \ref{lemma existence 2 copy(4)}. Therefore, if $%
D_{0}=-D_{0}^{\top }\in \mathcal{L}^{2}(\mathfrak{h})$ and $\Upsilon
_{0}=\Upsilon _{0}^{\ast }$ then $\Upsilon _{t}$ is always self-adjoint (by
definition of the elliptic flow) and the operator $\mathrm{H}_{t}$ is
essentially self-adjoint on the domain $\mathcal{D}_{0}$ (\ref{domain H0}),
thanks to Proposition \ref{Hamilselfadjoint}. We use again the notation $%
\mathrm{H}_{t}\equiv \mathrm{H}_{t}^{\ast \ast }$ for its self-adjoint
extension.

In other words, the Brockett-Wegner flow (\ref{brocket flow}) on the
many-fermion system can be indirectly analyzed via the Brockett-Wegner flow (%
\ref{simplified flow}), i.e., the elliptic flow (\ref{flow
equation-quadratic deltabis}) on operators acting on the one-particle
Hilbert space $\mathfrak{h}$. In particular, the existence of $\mathrm{H}%
_{\infty }$ and $\mathrm{U}_{\infty ,0}$ can be studied rigorously via a
detailed analysis of asymptotics of the operator families $\left( \Upsilon
_{t}\right) _{t\geq 0}$ and $\left( D_{t}\right) _{t\geq 0}$. For instance,
the convergence of the commutator $\left[ \mathrm{H}_{t},\mathrm{N}\right] $%
\ towards zero as $t\rightarrow \infty $ is related to the elimination of
the operator $D_{t}$ in (\ref{DefH0bis}). The same strategy was used for the
bosonic case in \cite{bach-bru-memo} by using an hyperbolic version of the
elliptic flow (\ref{flow equation-quadratic deltabis}) studied in \cite%
{EllipticFlow}.

In Section \ref{Section tech7}, we give the rigorous arguments based on this
strategy, starting with the existence and uniqueness of a solution to the
non-autonomous (Kato-hyperbolic) evolution equation (\ref{flow equationflow
equation}) associated with the Brockett-Wegner flow (\ref{brocket flow}).
This approach eventually leads to Theorem \ref{thm3}.

\subsection{Rigorous Implementation of the Brockett-Wegner Flow\label%
{Section tech7}}

This subsection constitutes the first technical part of the paper: After proving some preliminary results, we study the implementation of the
Brockett-Wegner flow (Section \ref{sectionBWflow}) in order to ultimately
prove Theorem \ref{thm3}. Before starting the proofs, we add a few general
definitions that are used in both technical sections.

First, $s\wedge t$\ and $s\vee t$ stand respectively for the minimum and the
maximum of real numbers $s,t\in \mathbb{R}$. This notation has not to be
mixed with the exterior product defining the subspace $\wedge ^{n}\mathfrak{h%
}$ of totally antisymmetric $n$--particle wave functions in $\mathfrak{h}%
^{\otimes n}$. The Banach space of bounded operators from a Banach space $%
\left( \mathcal{X},\left\Vert \cdot \right\Vert _{\mathcal{X}}\right) $ to
another Banach space $\left( \mathcal{Y},\left\Vert \cdot \right\Vert _{%
\mathcal{Y}}\right) $ is denoted by $\mathcal{B}(\mathcal{X},\mathcal{Y})$,
with operator norm 
\begin{equation*}
\left\Vert X\right\Vert _{\mathrm{op}}\doteq \sup_{f\in \mathcal{X}%
:\left\Vert f\right\Vert _{\mathfrak{h}}=1}\left\Vert Xf\right\Vert _{%
\mathcal{Y}}\ ,\qquad X\in \mathcal{B}(\mathcal{X},\mathcal{Y})\ .
\end{equation*}%
If $\mathcal{X=Y}$ then observe that $\mathcal{B}(\mathcal{X})\doteq 
\mathcal{B}(\mathcal{X},\mathcal{X})$ and recall that $\mathbf{1}\in 
\mathcal{B}(\mathcal{X})$ is the identity operator. Finally, $C\left( I;%
\mathcal{X}\right) $ denotes the Banach space of continuous functions from a
closed set $I\subseteq \mathbb{R}$ to a Banach space $\left( \mathcal{X}%
,\left\Vert \cdot \right\Vert _{\mathcal{X}}\right) $, along with the norm 
\begin{equation*}
\left\Vert X\right\Vert _{\infty }\doteq \sup_{t\in I}\left\Vert
X_{t}\right\Vert _{\mathcal{X}}\ ,\qquad X\in C\left( I;\mathcal{X}\right) \
.
\end{equation*}

\subsubsection{Self-adjointness of Quadratic Operators\label{Section tech7
copy(1)}}

In Equation (\ref{DefH0}) we define operators that are quadratic in the
creation and annihilation operators on the fermionic Fock space $\mathcal{F}$
(\ref{Fock}): Given $E_{0}\in \mathbb{R}$ and two operators $\Upsilon
_{0}=\Upsilon _{0}^{\ast }$ and $D_{0}=-D_{0}^{\top }\in \mathcal{L}^{2}(%
\mathfrak{h})$ acting on the separable Hilbert space $\mathfrak{h}$, we
formally define the operator 
\begin{equation}
\mathrm{H}_{0}\doteq d\Gamma (\Upsilon _{0})+\mathbf{D}+\mathbf{D}^{\ast
}+E_{0}\mathbf{1}\ ,  \label{sssdsd}
\end{equation}%
where 
\begin{equation}
d\Gamma \left( \theta \right) \doteq \sum_{k,l\in \mathbb{N}}\{\theta
\}_{k,l}a_{k}^{\ast }a_{l}  \label{second quanti}
\end{equation}%
is the (well-known) second-quantization of any self-adjoint operator $\theta
=\theta ^{\ast }$ (here with domain $\mathcal{D}\left( \theta \right)
\supseteq \left\{ \varphi _{k}\right\} _{k=1}^{\infty }$), while 
\begin{equation}
\mathbf{D}\doteq \sum_{k,l\in \mathbb{N}}\{\bar{D}_{0}\}_{k,l}a_{l}a_{k}\ .
\label{sdsds2}
\end{equation}%
Recall that $\left\{ X\right\} _{k,l}\doteq \left\langle \varphi
_{k},X\varphi _{l}\right\rangle _{\mathfrak{h}}$ for all operators $X$
acting on $\mathfrak{h}$, where $\left\{ \varphi _{k}\right\} _{k=1}^{\infty
}$ is a real orthonormal basis in the dense domain $\mathcal{D}\left(
\Upsilon _{0}\right) \subseteq \mathfrak{h}$ of $\Upsilon _{0}$. Since $%
\mathfrak{h}$ can have infinite dimension, it is not clear that such
quadratic operators are well-defined as (possibly unbounded) self-adjoint
operators on $\mathcal{F}$. This issue is already studied in the literature,
see for instance \cite[Theorem 6.1]{Berezin}. However, the proof of \cite[%
Theorem 6.1]{Berezin} is not completely rigorous. Another proof is given by 
\cite[Proposition 2.1]{Carey}, but under the assumption of bounded operators 
$\Upsilon _{0}\in \mathcal{B}(\mathfrak{h})$. As a consequence, we
explicitly establish the essential self-adjointness of $\mathrm{H}_{0}$ by
translating the results of \cite[Theorem 5.3]{Bruneau-derezinski2007}
(itself inspired by \cite[Theorem 6.1]{Berezin}, following \cite{Carleman})
on bosonic quadratic operators to the fermionic case.

To this end, we first need a technical lemma, which is not only useful to
the self-adjointness of fermionic quadratic operators but also pivotal to
study the Brockett-Wegner flow (\ref{brocket flow}) in Sections \ref{Section
proof theroem important 2bis}--\ref{section proof thm important 3}. In fact,
we first show that the operators $\mathbf{D}$ and $d\Gamma (\Delta )$ with $%
\Delta =\Delta ^{\ast }\in \mathcal{L}^{2}(\mathfrak{h})$ are well-defined
on the domain $\mathcal{D}\left( \mathrm{N}\right) $ of the particle number
operator $\mathrm{N}$, which is a self-adjoint operator that is explicitly
defined by (\ref{domain N0})--(\ref{domain N}), see, e.g., \cite[p. 7]%
{BratteliRobinson}. This is a simple corollary of the next assertion, which
is similar to \cite[Lemma 65]{bach-bru-memo} done in the bosonic case:

\begin{lemma}[Relative boundedness of quadratic operators -- I]
\label{lemmfermion}\mbox{}\newline
Let $\Delta =\Delta ^{\ast }\in \mathcal{L}^{2}(\mathfrak{h})$ and $%
D_{0}=-D_{0}^{\top }\in \mathcal{L}^{2}(\mathfrak{h})$ be two
Hilbert-Schmidt operators. Given a positive, invertible, operator $\theta
=\theta ^{\ast }$ acting on $\mathfrak{h}$ with domain $\mathcal{D}\left(
\theta \right) \supseteq \left\{ \varphi _{k}\right\} _{k=1}^{\infty }$, the
following inequalities hold true: 
\begin{eqnarray*}
\Vert \mathbf{D}\left( d\Gamma \left( \theta \right) +\mathbf{1}\right)
^{-1}\Vert _{\mathrm{op}} &\leq &\left\Vert \theta ^{-1/2}D_{0}\theta
^{-1/2}\right\Vert _{2}\ , \\
\Vert \mathbf{D}^{\ast }\left( d\Gamma \left( \theta \right) +\mathbf{1}%
\right) ^{-1}\Vert _{\mathrm{op}} &\leq &\sqrt{\left\Vert \theta
^{-1/2}D_{0}\theta ^{-1/2}\right\Vert _{2}^{2}}\ , \\
\Vert d\Gamma (\Delta )\left( d\Gamma \left( \theta \right) +\mathbf{1}%
\right) ^{-1}\Vert _{\mathrm{op}} &\leq &\sqrt{\left\Vert \theta
^{-1/2}\Delta \theta ^{-1/2}\right\Vert _{2}^{2}+\left\Vert \theta
^{-1/2}\Delta ^{2}\theta ^{-1/2}\right\Vert _{2}}\ .
\end{eqnarray*}
\end{lemma}

\begin{proof}
The proof is rather standard and we give only one when $\theta \geq \alpha 
\mathbf{1}$ for some $\alpha \in \mathbb{R}^{+}$ and%
\begin{equation*}
d\Gamma \left( \theta \right) =\sum_{k\in \mathbb{N}}\left\{ \theta \right\}
_{k,k}a_{k}^{\ast }a_{k}\equiv \sum_{k\in \mathbb{N}}\theta _{k}a_{k}^{\ast
}a_{k}
\end{equation*}%
where we use the shorter notation $\theta _{k}\equiv \left\{ \theta \right\}
_{k,k}$. The general case is completly identical, thanks to the spectral
theorem. For more details, we recommend \cite[Lemma 65]{bach-bru-memo}. In
fact, in this paper we only need the case $\theta =\mathbf{1}$, i.e., $%
\theta _{k}=1$ for $k\in \mathbb{N}$. Take once and for all $\varphi \in 
\mathcal{F}$, $\Delta =\Delta ^{\ast }\in \mathcal{L}^{2}(\mathfrak{h})$ and 
$D_{0}=-D_{0}^{\top }\in \mathcal{L}^{2}(\mathfrak{h})$. Using the
Cauchy-Schwarz inequality we estimate that 
\begin{align*}
\left\Vert \mathbf{D}\varphi \right\Vert _{\mathcal{F}}& \leq \sum_{k,l\in 
\mathbb{N}}\frac{\left\vert \{\bar{D}_{0}\}_{k,l}\right\vert }{\theta
_{k}^{1/2}\theta _{l}^{1/2}}\left\Vert \theta _{k}^{1/2}\theta
_{l}^{1/2}a_{k}a_{l}\varphi \right\Vert _{\mathcal{F}} \\
& \leq \left( \sum_{k,l\in \mathbb{N}}\theta _{k}^{-1}\left\vert
\{D_{0}\}_{k,l}\right\vert ^{2}\theta _{l}^{-1}\right) ^{1/2}\left(
\sum_{k,l\in \mathbb{N}}\theta _{k}\theta _{l}\left\Vert a_{k}a_{l}\varphi
\right\Vert _{\mathcal{F}}^{2}\right) ^{1/2}\ .
\end{align*}%
Then, by using the CAR (\ref{CAR}), it follows that 
\begin{multline*}
\left\Vert \mathbf{D}\varphi \right\Vert _{\mathcal{F}}^{2}\leq \left\Vert
\theta ^{-1/2}D_{0}\theta ^{-1/2}\right\Vert _{2}^{2}\sum_{k,l\in \mathbb{N}%
}\theta _{k}\theta _{l}\left\langle a_{l}\varphi ,a_{k}^{\ast
}a_{k}a_{l}\varphi \right\rangle _{\mathcal{F}} \\
\leq \left\Vert \theta ^{-1/2}D_{0}\theta ^{-1/2}\right\Vert
_{2}^{2}\sum_{k,l\in \mathbb{N}}\left\langle \theta _{l}a_{l}^{\ast
}a_{l}\varphi ,\theta _{k}a_{k}^{\ast }a_{k}\varphi \right\rangle _{\mathcal{%
F}}\leq \left\Vert \theta ^{-1/2}D_{0}\theta ^{-1/2}\right\Vert
_{2}^{2}\left\Vert d\Gamma \left( \theta \right) \varphi \right\Vert _{%
\mathcal{F}}^{2}\ ,\ 
\end{multline*}%
which proves the first inequality. Recall now that $\mathrm{tr}(\cdot )$
denotes the usual trace for operators. For the inequality with $\mathbf{D}%
^{\ast }$ we compute again from the CAR (\ref{CAR}) that 
\begin{eqnarray}
\left\Vert \mathbf{D}^{\ast }\varphi \right\Vert _{\mathcal{F}}^{2}
&=&\sum_{k,l,p,q\in \mathbb{N}}\{\bar{D}_{0}\}_{k,l}\left\{ D_{0}\right\}
_{p,q}\left\langle \varphi ,a_{l}a_{k}a_{p}^{\ast }a_{q}^{\ast }\varphi
\right\rangle _{\mathcal{F}}  \notag \\
&=&\sum_{k,l,p,q\in \mathbb{N}}\{\bar{D}_{0}\}_{k,l}\left\{ D_{0}\right\}
_{p,q}\left\langle a_{q}a_{p}\varphi ,a_{l}a_{k}\varphi \right\rangle _{%
\mathcal{F}}  \notag \\
&&-\sum_{k,p\in \mathbb{N}}\left\{ D_{0}D_{0}^{\ast }\right\}
_{p,k}\left\langle a_{p}\varphi ,a_{k}\varphi \right\rangle _{\mathcal{F}} 
\notag \\
&&+\sum_{l,p\in \mathbb{N}}\left\{ D_{0}\bar{D}_{0}\right\}
_{p,l}\left\langle a_{p}\varphi ,a_{l}\varphi \right\rangle _{\mathcal{F}} 
\notag \\
&&+\sum_{k,q\in \mathbb{N}}\{\bar{D}_{0}D_{0}\}_{k,q}\left\langle
a_{q}\varphi ,a_{k}\varphi \right\rangle _{\mathcal{F}}  \notag \\
&&-\sum_{l,q\in \mathbb{N}}\{D_{0}^{\ast }D_{0}\}_{l,q}\left\langle
a_{q}\varphi ,a_{l}\varphi \right\rangle _{\mathcal{F}}  \notag \\
&\leq &\sum_{k,l,p,q\in \mathbb{N}}\{\bar{D}_{0}\}_{k,l}\left\{
D_{0}\right\} _{p,q}\left\langle a_{q}a_{p}\varphi ,a_{l}a_{k}\varphi
\right\rangle _{\mathcal{F}}  \label{eqJB20}
\end{eqnarray}%
using that $D_{0}=-D_{0}^{\top }$. It follows from the Cauchy-Schwarz
inequality and explicit computations that 
\begin{equation}
\left\Vert \mathbf{D}^{\ast }\varphi \right\Vert _{\mathcal{F}}^{2}\leq
\left\Vert \theta ^{-1/2}D_{0}\theta ^{-1/2}\right\Vert _{2}^{2}\left\Vert
d\Gamma \left( \theta \right) \varphi \right\Vert _{\mathcal{F}}^{2}
\label{eqJB2}
\end{equation}%
from which we deduce the upper norm estimate related to the adjoint $\mathbf{%
D}^{\ast }$. It remains to study the norms of $d\Gamma (\Delta )$ relative
to $d\Gamma \left( \theta \right) $: By the CAR (\ref{CAR}) and the
Cauchy-Schwarz inequality, using the same trick to insert the coefficients
of $\theta $, we find that 
\begin{multline*}
\Vert d\Gamma (\Delta )\varphi \Vert _{\mathcal{F}}^{2}=-\sum_{k,l,p,q\in 
\mathbb{N}}\{\Delta \}_{k,l}\{\bar{\Delta}\}_{p,q}\left\langle
a_{k}a_{q}\varphi ,a_{p}a_{l}\varphi \right\rangle _{\mathcal{F}%
}+\sum_{l,q\in \mathbb{N}}\{\Delta ^{2}\}_{q,l}\left\langle a_{q}\varphi
,a_{l}\varphi \right\rangle _{\mathcal{F}} \\
\leq \left\Vert \theta ^{-1/2}\Delta \theta ^{-1/2}\right\Vert
_{2}^{2}\left\Vert d\Gamma \left( \theta \right) \varphi \right\Vert _{%
\mathcal{F}}^{2}+\left\Vert \theta ^{-1/2}\Delta ^{2}\theta
^{-1/2}\right\Vert _{2}\left\Vert d\Gamma \left( \theta \right)
^{1/2}\varphi \right\Vert _{\mathcal{F}}^{2}\ ,
\end{multline*}%
which directly yields the last inequality.
\end{proof}

\begin{corollary}[Relative boundedness of quadratic operators -- II]
\label{corofermion}\mbox{}\newline
Let $\Delta =\Delta ^{\ast }\in \mathcal{L}^{2}(\mathfrak{h})$ and $%
D_{0}=-D_{0}^{\top }\in \mathcal{L}^{2}(\mathfrak{h})$ be two
Hilbert-Schmidt operators. Then, the following inequalities hold true: 
\begin{eqnarray*}
\Vert \mathbf{D}\left( \mathrm{N}+\mathbf{1}\right) ^{-1}\Vert _{\mathrm{op}%
} &\leq &\left\Vert D_{0}\right\Vert _{2}\ , \\
\Vert \mathbf{D}^{\ast }\left( \mathrm{N}+\mathbf{1}\right) ^{-1}\Vert _{%
\mathrm{op}} &\leq &\left\Vert D_{0}\right\Vert _{2}\ ,
\\
\Vert d\Gamma (\Delta )\left( \mathrm{N}+\mathbf{1}\right) ^{-1}\Vert _{%
\mathrm{op}} &\leq &\sqrt{2}\left\Vert \Delta \right\Vert _{2}\ .
\end{eqnarray*}
\end{corollary}

\begin{proof}
This is an immediate application of Lemma \ref{lemmfermion}, upon choosing $%
\theta \doteq \mathbf{1}$ because $d\Gamma \left( \mathbf{1}\right) =\mathrm{%
N}$.
\end{proof}

\begin{remark}
\label{remark idiote}\mbox{}\newline
The bounds of Lemma \ref{lemmfermion} and Corollary \ref{corofermion} are
not supposed to be optimal or innovative, but are only sufficient for our
uses. They are in fact standard estimates, the proofs being given only for
completeness and to be as pedagogical as possible.
\end{remark}

Having now Corollary \ref{corofermion} at our disposal, we are in a position
to ensure the self-adjointness of $\mathrm{H}_{0}$ on the dense domain (\ref%
{domain H0}), that is, 
\begin{equation*}
\mathcal{D}_{0}\doteq \bigcup_{N\in \mathbb{N}}\left(
\bigoplus_{n=0}^{N}\left( \wedge ^{n}\mathcal{D}\left( \Upsilon _{0}\right)
\right) \right) \subseteq \mathcal{F}\ ,
\end{equation*}%
where $\mathcal{D}\left( \Upsilon _{0}\right) \subseteq \mathfrak{h}$ is the
(dense) domain of $\Upsilon _{0}=\Upsilon _{0}^{\ast }$, $\wedge ^{0}%
\mathcal{D}\left( \Upsilon _{0}\right) \doteq \mathbb{C}$ while for $n\in 
\mathbb{N}$, $\wedge ^{n}\mathcal{D}\left( \Upsilon _{0}\right) $ is the
(dense) subspace of totally antisymmetric $n$--particle wave functions in $%
\mathcal{D}\left( \Upsilon _{0}\right) ^{\otimes n}$, the $n$--fold tensor
product of $\mathcal{D}\left( \Upsilon _{0}\right) ^{\otimes n}\subseteq 
\mathfrak{h}^{\otimes n}$.

\begin{proposition}[Self--adjointness of quadratic operators]
\label{Hamilselfadjoint}\mbox{}\newline
Fix $E_{0}\in \mathbb{R}$. Let $\Upsilon _{0}=\Upsilon _{0}^{\ast }$ and $%
D_{0}=-D_{0}^{\top }\in \mathcal{L}^{2}(\mathfrak{h})$ be two operators
acting on the Hilbert space $\mathfrak{h}$. Then, the operator $\mathrm{H}%
_{0}$ defined by (\ref{sssdsd})--(\ref{sdsds2}) on the domain (\ref{domain
H0}) is essentially self-adjoint.
\end{proposition}

\begin{proof}
The proof is the same as the one of \cite[Theorem 5.3]%
{Bruneau-derezinski2007} (itself inspired by \cite[Theorem 6.1]{Berezin}),
up to obvious modifications, and is thus omitted. In particular, one shows $%
\ker \left( \mathrm{H}_{0}^{\ast }\pm i\right) =\left\{ 0\right\} $, which
is a well-known criterion for essential self-adjointness (see, e.g., \cite[%
Theorem VIII.3 and its corollary]{ReedSimon} or \cite[Proposition 3.8]%
{Konrad}). See \cite{Nathan} for the detailed proof in this case.
\end{proof}

We conclude this section by showing that the essentially self-adjoint
operator $\mathrm{H}_{0}$ defined by (\ref{sssdsd})--(\ref{sdsds2}) on the
domain (\ref{domain H0}) is independent of the (real) orthonormal basis $%
\left\{ \varphi _{k}\right\} _{k=1}^{\infty }\subseteq \mathcal{D}\left(
\Upsilon _{0}\right) \subseteq \mathfrak{h}$ chosen. So, take another
orthonormal basis $\left\{ \psi _{k}\right\} _{k=1}^{\infty }\subseteq 
\mathcal{D}\left( \Upsilon _{0}\right) \subseteq \mathfrak{h}$ and observe
that, for any $k\in \mathbb{N}$, 
\begin{equation}
a_{k}\doteq a(\varphi _{k})=\sum_{n\in \mathbb{N}}\left\langle \varphi
_{k},\psi _{n}\right\rangle _{\mathfrak{h}}a\left( \psi _{n}\right)
\label{definition}
\end{equation}%
by anti-linearity of the mapping $\varphi \mapsto a(\varphi )$ from $%
\mathfrak{h}$ to $\mathcal{B}\left( \mathcal{F}\right) $. The series are not
necessarily absolutely convergent but they are always well-defined because
of the CAR: 
\begin{equation*}
a(\psi )a(\varphi )+a(\varphi )a(\psi )=0\ ,\qquad a(\psi )a(\varphi )^{\ast
}+a(\varphi )^{\ast }a(\psi )=\langle \psi ,\varphi \rangle _{\mathfrak{h}}%
\mathbf{1}
\end{equation*}%
for any $\varphi ,\psi \in \mathfrak{h}$. These CAR are a generalization of (%
\ref{CAR}). Therefore, the essentially self-adjoint operator $\mathrm{H}_{0}$
defined by (\ref{sssdsd})--(\ref{sdsds2}) for $\Upsilon _{0}=\Upsilon
_{0}^{\ast }$ and $D_{0}=-D_{0}^{\top }$ is also equal to%
\begin{equation}
\mathrm{H}_{0}=\sum_{k,l\in \mathbb{N}}\left\{ \Upsilon _{0}\right\}
_{k,l}a\left( \psi _{k}\right) ^{\ast }a\left( \psi _{l}\right) +\left\{
D_{0}\mathcal{C}\right\} _{k,l}a\left( \psi _{k}\right) ^{\ast }a\left( \psi
_{l}\right) ^{\ast }+\overline{\left\{ D_{0}\mathcal{C}\right\} }%
_{k,l}a\left( \psi _{l}\right) a\left( \psi _{k}\right) +E_{0}\mathbf{1}
\label{dddddd}
\end{equation}%
on the domain (\ref{domain H0}), where $\mathcal{C}$ is the complex
conjugation (Remark \ref{remark idiote copy(1)}) and $\left\{ X\right\}
_{k,l}\doteq \left\langle \psi _{k},X\psi _{l}\right\rangle _{\mathfrak{h}}$
for all operators $X$ acting on $\mathfrak{h}$. As a consequence, the
quadratic Hamiltonian is written exactly like in (\ref{sssdsd})--(\ref%
{sdsds2}) when $\left\{ \psi _{k}\right\} _{k=1}^{\infty }$ is a \emph{real}
orthonormal basis. In fact, $\mathrm{H}_{0}$ is basis independent and
Equation (\ref{dddddd}) is the general form of the quadratic Hamiltonian $%
\mathrm{H}_{0}$ written with an arbitrary orthonormal basis in $\mathcal{D}%
\left( \Upsilon _{0}\right) $.

\subsubsection{Continuous Flow of Bogoliubov Transformations\label{Section
proof theroem important 2bis}}

We need to implement on the Fock space a continuous family of Bogoliubov
transformations generated by the solution to a flow on the one-particle
Hilbert space. At first glance, this could be done directly via Theorem \ref%
{Shale-Stinespring THM} and the Shale-Stinespring condition. For our
applications, however, we need a \textbf{non-autonomous} flow of Bogoliubov
transformations. To implement them, we invoke the theory of non-autonomous
(Kato-hyperbolic) evolution equations. So, we use it on the Fock space
because of the time-dependency of the generator in the Fock space.

Indeed, the continuous flow that implements Bogoliubov transformations is
constructed from the unique solution to the non-autonomous evolution
equation (\ref{flow equationflow equation}) with infinitesimal generator 
\begin{equation}
\mathrm{G}_{t}\doteq 2i\sum_{k,l\in \mathbb{N}}\left\{ D_{t}\right\}
_{k,l}a_{k}^{\ast }a_{l}^{\ast }+\left\{ \bar{D}_{t}\right\}
_{k,l}a_{k}a_{l}\ ,\qquad t\in \mathbb{R}_{0}^{+}\ ,  \label{generator}
\end{equation}%
for some continuous Hilbert-Schmidt operator family $D\in C(\mathbb{R}%
_{0}^{+};\mathcal{L}^{2}(\mathfrak{h}))$ satisfying $D_{t}=-D_{t}^{\top }\in 
\mathcal{L}^{2}(\mathfrak{h})$ for all $t\in \mathbb{R}_{0}^{+}$. In our
applications, $D$ is solution to the elliptic operator-valued flow (\ref%
{flow equation-quadratic deltabis}) studied in detail in \cite{EllipticFlow}%
. Note that the initial condition $D_{0}=-D_{0}^{\top }\in \mathcal{L}^{2}(%
\mathfrak{h})$ leads to $D_{t}=-D_{t}^{\top }\in \mathcal{L}^{2}(\mathfrak{h}%
)$ for all $t\in \mathbb{R}_{0}^{+}$, thanks to Theorem \ref{Corollary
existence}. Additionally, in this case, observe that the generator is
formally equal to the commutator $\mathrm{G}_{t}=i\left[ \mathrm{N},\mathrm{H%
}_{t}\right] $ with $\mathrm{N}$ and $\mathrm{H}_{t}$ being the particle
number operator and the general quadratic Hamiltonian respectively defined
by (\ref{particle number operator}) and (\ref{DefH0bis}).

We need to prove that the corresponding non-autonomous evolution equation is
well-posed with time-dependent generators (\ref{generator}). Given $D\in C(%
\mathbb{R}_{0}^{+};\mathcal{L}^{2}(\mathfrak{h}))$ with $D_{t}=-D_{t}^{\top
}\in \mathcal{L}^{2}(\mathfrak{h})$ for all $t\in \mathbb{R}_{0}^{+}$,
Proposition \ref{Hamilselfadjoint} proves that quadratic operators like the
infinitesimal generators $\mathrm{G}_{t}$, $t\in \mathbb{R}_{0}^{+}$, all
defined on the same (dense) domain 
\begin{equation*}
\mathcal{D}_{1}\doteq \bigcup_{N\in \mathbb{N}}\left(
\bigoplus_{n=0}^{N}\wedge ^{n}\mathfrak{h}\right) \supseteq \mathcal{D}_{0}
\end{equation*}%
(see (\ref{domain H0})), are essentially self-adjoint and, as already
mentioned, we use the notation $\mathrm{G}_{t}\equiv \mathrm{G}_{t}^{\ast
\ast }$, $t\in \mathbb{R}_{0}^{+}$, for their self-adjoint extension. In
particular, the domain $\mathcal{D}_{1}$ is a core for all these
self-adjoint operators, but also for the particle number operator $\mathrm{N}
$ (see (\ref{particle number operator})) and we infer from Corollary \ref%
{corofermion} that the (now self-adjoint) generators $\mathrm{G}_{t}$, $t\in 
\mathbb{R}_{0}^{+}$, are relatively bounded with respect to the particle
number operator $\mathrm{N}$: $\mathcal{D}(\mathrm{G}_{t})\supseteq \mathcal{%
D}(\mathrm{N})$ (see (\ref{domain N})) and
\begin{equation}
\left\Vert \mathrm{G}_{t}\left( \mathrm{N}+\mathbf{1}\right)
^{-1}\right\Vert _{\mathrm{op}}\leq 4\left\Vert D_{t}\right\Vert
_{2}<+\infty \ ,\qquad t\in \mathbb{R}_{0}^{+}\ ,  \label{generatorbound}
\end{equation}%
thanks to $D_{t}=-D_{t}^{\top }\in \mathcal{L}^{2}(\mathfrak{h})$ and
Corollary \ref{corofermion}.

The precise definition of infinitesimal generators of the non-autonomous
evolution equation (\ref{flow equationflow equation}) being established, we
show below the existence of a unique solution to this Kato-hyperbolic
evolution equation, analogously to the line of arguments used in \cite[%
Section VI.1]{bach-bru-memo}. Recall that $s\wedge t$\ and $s\vee t$ stand
respectively for the minimum and the maximum of real numbers $s,t\in \mathbb{%
R}$.

\begin{theorem}[Continuous Bogoliubov transformations]
\label{theorem Ht=00003Dunitary orbite copy(1)}\mbox{}\newline
Assume $D\in C(\mathbb{R}_{0}^{+};\mathcal{L}^{2}(\mathfrak{h}))$ with $%
D_{t}=-D_{t}^{\top }\in \mathcal{L}^{2}(\mathfrak{h})$ for all $t\in \mathbb{%
R}_{0}^{+}$. There is a family $(\mathrm{U}_{t,s})_{s,t\in \mathbb{R}%
_{0}^{+}}\subseteq \mathcal{B}\left( \mathcal{F}\right) $ of bounded
operators satisfying the following properties:

\begin{enumerate}
\item[\emph{(i)}] For any $s,t\in \mathbb{R}_{0}^{+}$, $\mathrm{U}_{t,s}=%
\mathrm{U}_{s,t}^{\ast }$ is a unitary operator.

\item[\emph{(ii)}] It satisfies the cocycle property\ $\mathrm{U}_{t,x}%
\mathrm{U}_{x,s}=\mathrm{U}_{t,s}$ for any $s,x,t\in \mathbb{R}_{0}^{+}$.

\item[\emph{(iii)}] It is jointly strongly continuous in $s$ and $t$ on $(%
\mathbb{R}_{0}^{+})^{2}$.

\item[\emph{(iv)}] There is $C\in \mathbb{R}^{+}$ such that,\ for any $%
s,t\in \mathbb{R}_{0}^{+}$, 
\begin{equation*}
\Vert (\mathrm{N}+\mathbf{1})\mathrm{U}_{t,s}(\mathrm{N}+\mathbf{1}%
)^{-1}\Vert _{\mathrm{op}}\leq \mathrm{\exp }\left\{ C\int_{s\wedge
t}^{s\vee t}\Vert D_{\tau }\Vert _{2}\mathrm{d}\tau \right\}
\end{equation*}%
and the mapping $\left( s,t\right) \mapsto (\mathrm{N}+\mathbf{1})\mathrm{U}%
_{t,s}(\mathrm{N}+\mathbf{1})^{-1}$ is strongly continuous on $(\mathbb{R}%
_{0}^{+})^{2}$.

\item[\emph{(v)}] It is the unique\footnote{$\partial _{t}\mathrm{U}_{t,s}=-i%
\mathrm{G}_{t}\mathrm{U}_{t,s}$ with $\mathrm{U}_{s,s}\doteq \mathbf{1}$ for 
$s,t\in \mathbb{R}_{0}^{+}$ has already a unique solution.}, bounded
evolution operator satisfying on the domain $\mathcal{D}\left( \mathrm{N}%
\right) $ the non-autonomous evolution equations 
\begin{equation*}
\forall s,t\in \mathbb{R}_{0}^{+}:\qquad \left\{ 
\begin{array}{llll}
\partial _{t}\mathrm{U}_{t,s}=-i\mathrm{G}_{t}\mathrm{U}_{t,s} & , & \mathrm{%
U}_{s,s}\doteq \mathbf{1} & . \\ 
\partial _{s}\mathrm{U}_{t,s}=i\mathrm{U}_{t,s}\mathrm{G}_{s} & , & \mathrm{U%
}_{t,t}\doteq \mathbf{1} & .%
\end{array}%
\right.
\end{equation*}
\end{enumerate}
\end{theorem}

\begin{proof}
The proof is done in several steps: \medskip {}

\noindent \underline{Step 1:} We first ensure the existence of a unique,
bounded evolution operator $\left( \mathrm{U}_{t,s}\right) _{t\geq s\geq 0}$
satisfying on the domain $\mathcal{D}\left( \mathrm{N}\right) $ the
non-autonomous evolution equations\footnote{%
The derivatives $\partial _{t}$ and $\partial _{s}$ on the borderline $t=s$
or $s=0$ have to be understood as either right or left derivatives.} 
\begin{equation}
\forall t\geq s\geq 0:\qquad \left\{ 
\begin{array}{llll}
\partial _{t}\mathrm{U}_{t,s}=-i\mathrm{G}_{t}\mathrm{U}_{t,s} & , & \mathrm{%
U}_{s,s}\doteq \mathbf{1} & . \\ 
\partial _{s}\mathrm{U}_{t,s}=i\mathrm{U}_{t,s}\mathrm{G}_{s} & , & \mathrm{U%
}_{t,t}\doteq \mathbf{1} & .%
\end{array}%
\right.  \label{flow equationbis-new}
\end{equation}%
This is done by using the general results of \cite[Section VII.1]%
{bach-bru-memo} for the infinitesimal generator $G_{t}=\mathrm{G}_{t}$, the
closed auxiliary operator $\Theta \doteq \mathrm{N}+\mathbf{1}$ and the
space 
\begin{equation}
\mathcal{Y}\doteq \mathcal{D}\left( \mathrm{N}\right) =\left\{ \varphi \in 
\mathcal{F}:\left\Vert \varphi \right\Vert _{\mathcal{Y}}\doteq \left\Vert
\left( \mathrm{N}+\mathbf{1}\right) \varphi \right\Vert <\infty \right\} \ .
\label{banach Y}
\end{equation}%
Since $\mathrm{G}_{t}$ is self-adjoint for all times $t\in \mathbb{R}%
_{0}^{+} $, the assumption B1 of \cite[Section VII.1]{bach-bru-memo} (with $%
m=1$ and $\beta _{0}(t)=0$) is immediate and we only need to verify a priori
two conditions: \cite[B2--B3 in Section VII.1]{bach-bru-memo}. It amounts to
prove that the relative norms $\Vert \mathrm{G}_{t}(\mathrm{N}+\mathbf{1}%
)^{-1}\Vert _{\mathrm{op}}$ and $\Vert \lbrack \mathrm{N},\mathrm{G}_{t}](%
\mathrm{N}+\mathbf{1})^{-1}\Vert _{\mathrm{op}}$ are bounded for any time $%
t\in \mathbb{R}_{0}^{+}$, with the function $t\mapsto \mathrm{G}_{t}(\mathrm{%
N}+\mathbf{1})^{-1}$ being norm-continuous on $\mathbb{R}_{0}^{+}$. Since a
formal computation using the CAR\ shows that 
\begin{equation}
\left[ \mathrm{N},\mathrm{G}_{t}\right] =4i\sum_{k,l\in \mathbb{N}}\left\{
D_{t}\right\} _{k,l}a_{k}^{\ast }a_{l}^{\ast }+\left\{ \bar{D}_{t}\right\}
_{k,l}a_{l}a_{k}\ ,  \label{formal computation2}
\end{equation}%
\cite[B2--B3 in Section VII.1]{bach-bru-memo} are direct consequences of $%
D\in C(\mathbb{R}_{0}^{+};\mathcal{L}^{2}(\mathfrak{h}))$ together with
Corollary \ref{corofermion} and Equation (\ref{generator}), which yield 
\begin{equation}
\max \left\{ \Vert \mathrm{G}_{t}\left( \mathrm{N}+\mathbf{1}\right)
^{-1}\Vert _{\mathrm{op}},\Vert \left[ \mathrm{N},\mathrm{G}_{t}\right]
\left( \mathrm{N}+\mathbf{1}\right) ^{-1}\Vert _{\mathrm{op}}\right\} \leq
C\Vert D_{t}\Vert _{2}\ ,\qquad t\in \mathbb{R}_{0}^{+}\ ,
\label{inegality BB 3}
\end{equation}%
for some (sufficiently large) constant $C\in \mathbb{R}^{+}$. Note that $%
\Vert D_{t}\Vert _{2}$ is uniformly bounded on compacta. As a consequence,
we can apply \cite[Theorem 88 and Lemma 92]{bach-bru-memo} and infer the
existence of a unique, bounded evolution operator $(\mathrm{U}_{t,s})_{t\geq
s\geq 0}$ satisfying the non-autonomous evolution equations (\ref{flow
equationbis-new}) in the strong sense on the domain $\mathcal{D}\left( 
\mathrm{N}\right) $. Moreover, one obtains from (\ref{inegality BB 3}) that,
for any $t\geq s\geq 0$, 
\begin{equation*}
\left\Vert (\mathrm{N}+\mathbf{1})\mathrm{U}_{t,s}(\mathrm{N}+\mathbf{1}%
)^{-1}\right\Vert _{\mathrm{op}}\leq \mathrm{\exp }\left\{
C\int_{s}^{t}\left\Vert D_{\tau }\right\Vert _{2}\mathrm{d}\tau \right\} \ .
\end{equation*}%
In particular, $\mathrm{U}_{t,s}$ conserves the domain $\mathcal{D}\left( 
\mathrm{N}\right) $ for all $t\geq s\geq 0$. The strong continuity of the
bounded operator family 
\begin{equation}
\left\{ (\mathrm{N}+\mathbf{1})\mathrm{U}_{t,s}(\mathrm{N}+\mathbf{1}%
)^{-1}\right\} _{t\geq s\geq 0}  \label{bounded operator family}
\end{equation}%
also results from \cite[Lemma 92]{bach-bru-memo}. \medskip

\noindent \underline{Step 2:} Observe that Step 1 neither implies that the
adjoint $\mathrm{U}_{t,s}^{\ast }$ of the evolution operator $\mathrm{U}%
_{t,s}$ conserves the dense domain $\mathcal{D}\left( \mathrm{N}\right) $ of
the particle number operator $\mathrm{N}$, nor that it is jointly strongly
continuous in $s$ and $t$ for all $t\geq s\geq 0$. Therefore, in the same
way we prove Step 1, one checks the existence of a unique, bounded evolution
operator $(\mathrm{V}_{t,s})_{t\geq s\geq 0}$ satisfying on the domain $%
\mathcal{D}\left( \mathrm{N}\right) $ the non-autonomous evolution equations%
\footnote{%
The derivatives $\partial _{t}$ and $\partial _{s}$ on the borderline $t=s$
or $s=0$ have to be understood as either right or left derivatives.} 
\begin{equation}
\forall t\geq s\geq 0:\qquad \left\{ 
\begin{array}{llll}
\partial _{s}\mathrm{V}_{t,s}=-i\mathrm{G}_{s}\mathrm{V}_{t,s} & , & \mathrm{%
V}_{t,t}\doteq \mathbf{1} & . \\ 
\partial _{t}\mathrm{V}_{t,s}=i\mathrm{V}_{t,s}\mathrm{G}_{t} & , & \mathrm{V%
}_{s,s}\doteq \mathbf{1} & .%
\end{array}%
\right.  \label{flow equationbis-new-adjoint}
\end{equation}%
Moreover, $\mathrm{V}_{t,s}$ conserves the domain $\mathcal{D}\left( \mathrm{%
N}\right) $ for all $t\geq s\geq 0$ as 
\begin{equation}
\Vert (\mathrm{N}+\mathbf{1})\mathrm{V}_{t,s}(\mathrm{N}+\mathbf{1}%
)^{-1}\Vert _{\mathrm{op}}\leq \mathrm{\exp }\left\{ C\int_{s}^{t}\Vert
D_{\tau }\Vert _{2}\mathrm{d}\tau \right\}
\end{equation}%
for some sufficiently large constant $C\in \mathbb{R}^{+}$, thanks to (\ref%
{inegality BB 3}). \medskip

\noindent \underline{Step 3:} We prove here the unitarity of the operators $%
\mathrm{U}_{t,s}$ for $t\geq s\geq 0$. For $t\geq s\geq 0$, note that \cite[%
Theorem 88]{bach-bru-memo} gives both the operators $\mathrm{V}_{t,s}$ and $%
\mathrm{U}_{t,s}$ as limit $\lambda \rightarrow \infty $ of (evolution)
operators $\mathrm{V}_{t,s,\lambda }=\mathrm{U}_{t,s,\lambda }^{\ast }$ and $%
\mathrm{U}_{t,s,\lambda }$ for $\lambda \gg 1$ defined like in \cite[%
Equation VII.6]{bach-bru-memo}, both having a norm convergent representation
as a Dyson series. From this, one deduces that $\mathrm{V}_{t,s}=\mathrm{U}%
_{t,s}^{\ast }$ for all $t\geq s\geq 0$. For more details, see the arguments
proving \cite[Lemma 68]{bach-bru-memo}. Using now Steps 1 and 2, we deduce
that, for any $t\geq s\geq 0$, 
\begin{equation*}
(\mathbf{1}-\mathrm{U}_{t,s}\mathrm{U}_{t,s}^{\ast })(\mathrm{N}+\mathbf{1}%
)^{-1}=\int_{s}^{t}\partial _{\tau }\left\{ \mathrm{U}_{t,\tau }\mathrm{U}%
_{t,\tau }^{\ast }\right\} (\mathrm{N}+\mathbf{1})^{-1}\mathrm{d}\tau =0
\end{equation*}%
and 
\begin{equation*}
(\mathbf{1}-\mathrm{U}_{t,s}^{\ast }\mathrm{U}_{t,s})(\mathrm{N}+\mathbf{1}%
)^{-1}=-\int_{s}^{t}\partial _{\tau }\left\{ \mathrm{U}_{\tau ,s}^{\ast }%
\mathrm{U}_{\tau ,s}\right\} (\mathrm{N}+\mathbf{1})^{-1}\mathrm{d}\tau =0\ .
\end{equation*}%
Thus, for all $t\geq s\geq 0$, $\mathrm{U}_{t,s}^{\ast }\mathrm{U}_{t,s}=%
\mathrm{U}_{t,s}\mathrm{U}_{t,s}^{\ast }=\mathbf{1}$ on the dense domain $%
\mathcal{D}\left( \mathrm{N}\right) \subseteq \mathcal{F}$, which
immediately implies the unitarity of $\mathrm{U}_{t,s}\in \mathcal{B}\left( 
\mathcal{F}\right) $. \medskip {}

\noindent \underline{Step 4:} We finally use the definition $\mathrm{U}%
_{t,s}\doteq \mathrm{U}_{s,t}^{\ast }$ for $s\geq t\geq 0$. Then, by
combining Steps 1-3, one deduces all the assertions of the theorem.
\end{proof}

We analyze now the effect of the unitary operators $\mathrm{U}_{t,s}$, $%
t,s\in \mathbb{R}_{0}^{+}$, on annihilation/creation operators $\left\{
a_{k}\right\} _{k\in \mathbb{N}}$ and $\left\{ a_{k}^{\ast }\right\} _{k\in 
\mathbb{N}}$. The aim is to relate the method used here to the so-called
Bogoliubov transformation. To this end, we define the bounded operators 
\begin{eqnarray}
\mathbf{u}_{t,s} &\doteq &\mathbf{1}+\sum_{n\in \mathbb{N}}\left( -1\right)
^{n}4^{2n}\int_{s}^{t}\mathrm{d}\tau _{1}\cdots \int_{s}^{\tau _{2n-1}}%
\mathrm{d}\tau _{2n}D_{\tau _{1}}D_{\tau _{2}}^{\ast }\cdots D_{\tau
_{2n-1}}D_{\tau _{2n}}^{\ast }\ ,  \label{u} \\
\mathbf{v}_{t,s} &\doteq &\sum_{n\in \mathbb{N}_{0}}\left( -1\right)
^{n}4^{2n+1}\int_{s}^{t}\mathrm{d}\tau _{1}\cdots \int_{s}^{\tau _{2n}}%
\mathrm{d}\tau _{2n+1}D_{{\tau _{1}}}D_{\tau _{2}}^{\ast }\cdots D_{\tau
_{2n}}^{\ast }D_{\tau _{2n+1}}\ ,  \label{v}
\end{eqnarray}%
with $\tau _{0}\doteq t$, while $D\in C(\mathbb{R}_{0}^{+};\mathcal{L}^{2}(%
\mathfrak{h}))$ with $D_{t}=-D_{t}^{\top }\in \mathcal{L}^{2}(\mathfrak{h})$%
. In our applications, it is solution to the elliptic operator flow (\ref%
{flow equation-quadratic deltabis}). Note that the mappings $(s,t)\mapsto (%
\mathbf{u}_{t,s}-\mathbf{1})$ and $(s,t)\mapsto \mathbf{v}_{t,s}$ from $(%
\mathbb{R}_{0}^{+})^{2}$ to $\mathcal{L}^{2}(\mathfrak{h})$ are well-defined
and continuous in the Hilbert-Schmidt topology, thanks to the bounds 
\begin{eqnarray*}
\left\Vert \mathbf{u}_{t_{1},s_{1}}-\mathbf{u}_{t_{2},s_{2}}\right\Vert _{2}
&\leq &\cosh \left( 4\int_{t_{1}\wedge t_{2}}^{t_{1}\vee t_{2}}\left\Vert
D_{\tau }\right\Vert _{2}\mathrm{d}\tau \right) +\cosh \left(
4\int_{s_{1}\wedge s_{2}}^{s_{1}\vee s_{2}}\left\Vert D_{\tau }\right\Vert
_{2}\mathrm{d}\tau \right) -2 \\
\left\Vert \mathbf{v}_{t_{1},s_{1}}-\mathbf{v}_{t_{2},s_{2}}\right\Vert _{2}
&\leq &\sinh \left( 4\int_{t_{1}\wedge t_{2}}^{t_{1}\vee t_{2}}\left\Vert
D_{\tau }\right\Vert _{2}\mathrm{d}\tau \right) +\sinh \left(
4\int_{s_{1}\wedge s_{2}}^{s_{1}\vee s_{2}}\left\Vert D_{\tau }\right\Vert
_{2}\mathrm{d}\tau \right)
\end{eqnarray*}%
for all $s_{1},s_{2},t_{1},t_{2}\in \mathbb{R}_{0}^{+}$. (Note that the
above inequalities are straightforwardly deduced from the triangle
inequality.) In particular, the series in (\ref{u}) and (\ref{v}) are
absolutely convergent in $\mathcal{L}^{2}(\mathfrak{h})$. These coefficients
allow one to define a Bogoliubov transformation which is implemented (in the
sense of Theorem \ref{Shale-Stinespring THM}) by the unitary evolution
system $(\mathrm{U}_{t,s})_{s,t\in \mathbb{R}_{0}^{+}}$ of Theorem \ref%
{theorem Ht=00003Dunitary orbite copy(1)}:

\begin{proposition}[Implementation of non-autonomous Bogoliubov
transformations]
\label{lemma uv-Bogoliubov transformation}\mbox{}\newline
Assume $D\in C(\mathbb{R}_{0}^{+};\mathcal{L}^{2}(\mathfrak{h}))$ with $%
D_{t}=-D_{t}^{\top }\in \mathcal{L}^{2}(\mathfrak{h})$ for all $t\in \mathbb{%
R}_{0}^{+}$. For all $s,t\in \mathbb{R}_{0}^{+}$, 
\begin{equation}
\mathrm{U}_{t,s}a\left( \varphi \right) \mathrm{U}_{t,s}^{\ast }=a(\mathbf{u}%
_{t,s}\varphi )+a^{\ast }(\mathbf{v}_{t,s}\overline{\varphi })\ ,\qquad
\varphi \in \mathfrak{h}\ ,  \label{generalized Bog transf 4}
\end{equation}%
with the operators $\mathbf{u}_{t,s}$ and $\mathbf{v}_{t,s}$ satisfying the
following equalities: 
\begin{eqnarray}
\mathbf{u}_{t,s}^{\ast }\mathbf{u}_{t,s}+\mathbf{v}_{t,s}^{\top }\overline{%
\mathbf{v}}_{t,s} &=&\mathbf{1}\ ,\text{\qquad }\mathbf{u}_{t,s}^{\ast }%
\mathbf{v}_{t,s}+\mathbf{v}_{t,s}^{\top }\overline{\mathbf{u}}_{t,s}=0\ ,
\label{CCRu,v} \\
\mathbf{u}_{t,s}\mathbf{u}_{t,s}^{\ast }-\mathbf{v}_{t,s}\mathbf{v}%
_{t,s}^{\ast } &=&\mathbf{1}\ ,\qquad \mathbf{u}_{t,s}\mathbf{v}_{t,s}^{\top
}+\mathbf{v}_{t,s}\mathbf{u}_{t,s}^{\top }=0\ .  \label{bog2bis}
\end{eqnarray}
\end{proposition}

\begin{proof}
Since $(D_{t})_{t\in \mathbb{R}_{0}^{+}}\in C(\mathbb{R}_{0}^{+};\mathcal{L}%
^{2}(\mathfrak{h}))$ , the operator families $(\mathbf{u}_{t,s})_{s,t\in 
\mathbb{R}_{0}^{+}}$ and $(\mathbf{v}_{t,s})_{s,t\in \mathbb{R}_{0}^{+}}$
satisfy\footnote{%
The derivatives $\partial _{t}$ and $\partial _{s}$ on the borderline $t=0$
or $s=0$ have to be understood as right derivatives.} 
\begin{equation}
\partial _{t}\mathbf{u}_{t,s}=4D_{t}\overline{\mathbf{v}}_{t,s}\qquad \text{%
and}\qquad \partial _{t}\mathbf{v}_{t,s}=4D_{t}\overline{\mathbf{u}}_{t,s}
\label{derivation t uv}
\end{equation}%
for all $s,t\in \mathbb{R}_{0}^{+}$, in the Hilbert-Schmidt topology.
Therefore, using the continuity\footnote{%
This follows from the inequalities $\Vert A\Vert _{2}=\Vert A^{\ast }\Vert
_{2}=\Vert A^{\top }\Vert _{2}$ for any $A\in \mathcal{L}^{2}(\mathfrak{h})$.%
} of the mappings $A\mapsto A^{\ast }$ and $A\mapsto A^{\top }$ in $\mathcal{%
L}^{2}(\mathfrak{h})$, one computes that%
\begin{equation*}
\partial _{t}\left\{ \mathbf{u}_{t,s}^{\ast }\mathbf{u}_{t,s}+\mathbf{v}%
_{t,s}^{\top }\overline{\mathbf{v}}_{t,s}\right\} =0\qquad \text{and}\qquad
\partial _{t}\left\{ \mathbf{u}_{t,s}^{\ast }\mathbf{v}_{t,s}+\mathbf{v}%
_{t,s}^{\top }\overline{\mathbf{u}}_{t,s}\right\} =0\ ,
\end{equation*}%
which in turn imply that 
\begin{equation*}
\mathbf{u}_{t,s}^{\ast }\mathbf{u}_{t,s}+\mathbf{v}_{t,s}^{\top }\overline{%
\mathbf{v}}_{t,s}=\mathbf{u}_{s,s}^{\ast }\mathbf{u}_{s,s}+\mathbf{v}%
_{s,s}^{\top }\overline{\mathbf{v}}_{s,s}=\mathbf{1}
\end{equation*}%
as well as 
\begin{equation*}
\mathbf{u}_{t,s}^{\ast }\mathbf{v}_{t,s}+\mathbf{v}_{t,s}^{\top }\overline{%
\mathbf{u}}_{t,s}=\mathbf{u}_{s,s}^{\ast }\mathbf{v}_{s,s}+\mathbf{v}%
_{s,s}^{\top }\overline{\mathbf{u}}_{s,s}=0\ .
\end{equation*}%
I.e., one gets Equation (\ref{CCRu,v}). To prove Equation (\ref{bog2bis}),
one performs similar computations together with the observation that 
\begin{equation*}
\partial _{s}\mathbf{u}_{t,s}=4\mathbf{v}_{t,s}D_{s}^{\ast }\qquad \text{and}%
\qquad \partial _{s}\mathbf{v}_{t,s}=4\mathbf{u}_{t,s}D_{s}\ ,
\end{equation*}%
for all $s,t\in \mathbb{R}_{0}^{+}$. Note that these derivatives\footnote{%
The derivatives $\partial _{t}$ and $\partial _{s}$ on the borderline $t=0$
or $s=0$ have to be understood as right derivatives.} follow from $%
D_{t}=-D_{t}^{\top }$ and the equalities (\ref{u}) and (\ref{v}) rewritten
as 
\begin{eqnarray*}
\mathbf{u}_{t,s} &=&\mathbf{1}+\sum_{n\in \mathbb{N}}\left( -1\right)
^{n}4^{2n}\int_{s}^{t}\mathrm{d}\tau _{1}\cdots \int_{\tau _{2n-1}}^{t}%
\mathrm{d}\tau _{2n}D_{\tau _{2n}}D_{\tau _{2n-1}}^{\ast }\cdots D_{\tau
_{2}}D_{\tau _{1}}^{\ast }\ , \\
\mathbf{v}_{t,s} &=&\sum_{n\in \mathbb{N}_{0}}\left( -1\right)
^{n}4^{2n+1}\int_{s}^{t}\mathrm{d}\tau _{1}\cdots \int_{\tau _{2n}}^{t}%
\mathrm{d}\tau _{2n+1}D_{{\tau _{2n+1}}}D_{\tau _{2n}}^{\ast }\cdots D_{\tau
_{2}}^{\ast }D_{\tau _{1}}\ ,
\end{eqnarray*}%
with $\tau _{0}\doteq s$.

Now, recall that $\left\{ \varphi _{k}\right\} _{k=1}^{\infty }$ is some
real orthonormal basis in $\mathcal{D}\left( \Upsilon _{0}\right) \subseteq 
\mathfrak{h}$. So, to prove Equation (\ref{generalized Bog transf 4}) for
all $\varphi \in \mathfrak{h}$, it suffices to prove it for any basis
element $\varphi _{k}$, thanks to the antilinearity of the mapping $\varphi
\mapsto a(\varphi )$ from $\mathfrak{h}$ to $\mathcal{B}\left( \mathcal{F}%
\right) $. So, for each $k\in \mathbb{N}$ and $s,t\in \mathbb{R}_{0}^{+}$,
we define the bounded operator $\tilde{a}_{t,s,k}\in \mathcal{B}\left( 
\mathcal{F}\right) $ by 
\begin{equation}
\tilde{a}_{t,s,k}\doteq a(\mathbf{u}_{t,s}\varphi _{k})+a^{\ast }(\mathbf{v}%
_{t,s}\varphi _{k})=\sum_{\ell \in \mathbb{N}}\left\{ \mathbf{u}_{t,s}^{\ast
}\right\} _{k,\ell }a_{\ell }+\left\{ \mathbf{v}_{t,s}^{\top }\right\}
_{k,\ell }a_{\ell }^{\ast }\ ,  \label{generalized Bog transf 0}
\end{equation}%
where $a_{k}\doteq a\left( \varphi _{k}\right) $ is the annihilation
operator acting on the fermionic Fock space $\mathcal{F}$. Because of (\ref%
{CCRu,v})--(\ref{bog2bis}), for each $t\in \mathbb{R}_{0}^{+}$ the operator
family $\{\tilde{a}_{t,s,k},\tilde{a}_{t,s,k}^{\ast }\}_{k=1}^{\infty }$
satisfies the CAR and since 
\begin{equation}
\left\Vert a(\varphi )\right\Vert _{\mathrm{op}}=\left\Vert \varphi
\right\Vert _{\mathfrak{h}}\ ,\qquad \varphi \in \mathfrak{h}\ ,
\label{poipoi}
\end{equation}%
and $D_{t}=-D_{t}^{\top }$, we have the equalities 
\begin{equation}
\partial _{t}\tilde{a}_{t,s,k}=a(4D_{t}\overline{\mathbf{v}}_{t,s}\varphi
_{k})+a^{\ast }(4D_{t}\overline{\mathbf{u}}_{t,s}\varphi _{k})  \label{iuiu}
\end{equation}%
in the Banach space $\mathcal{B}(\mathcal{F})$ for each $k\in \mathbb{N}$
and $t\in \mathbb{R}_{0}^{+}$, thanks to Equation (\ref{derivation t uv}).
Meanwhile, observe from a formal computation using the CAR that, for all $%
k\in \mathbb{N}$ and $t\in \mathbb{R}_{0}^{+}$, 
\begin{equation*}
\left[ \mathrm{N}\mathbf{,}\tilde{a}_{t,s,k}\right] =\sum_{\ell \in \mathbb{N%
}}\left\{ -\mathbf{u}_{t,s}^{\ast }\right\} _{k,\ell }a_{\ell }+\left\{ 
\mathbf{v}_{t,s}^{\top }\right\} _{k,\ell }a_{\ell }^{\ast }=a(-\mathbf{u}%
_{t,s}\varphi _{k})+a^{\ast }(\mathbf{v}_{t,s}\varphi _{k})\ ,
\end{equation*}%
and from this observation and (\ref{generalized Bog transf 0})--(\ref{poipoi}%
) one in fact checks that 
\begin{equation}
\left\Vert (\mathrm{N}+\mathbf{1)}\tilde{a}_{t,s,k}(\mathrm{N}+\mathbf{1)}%
^{-1}\right\Vert _{\mathrm{op}}\leq 2\Vert \mathbf{u}_{t,s}\Vert _{\mathrm{op%
}}+2\Vert \mathbf{v}_{t,s}\Vert _{\mathrm{op}}<\infty \ .  \label{rtrtrtrt1}
\end{equation}%
Note that another formal computation using the CAR shows that 
\begin{eqnarray}
i\left[ \tilde{a}_{t,s,k},\mathrm{G}_{t}\right] &=&4\sum_{\ell \in \mathbb{N}%
}\left\{ \mathbf{v}_{t,s}^{\top }D_{t}^{\ast }\right\} _{k,\ell }a_{\ell
}-4\sum_{\ell \in \mathbb{N}}\left\{ \mathbf{u}_{t,s}^{\ast }D_{t}\right\}
_{k,\ell }a_{\ell }^{\ast }  \label{rtrtrtrt2} \\
&=&a(4D_{t}\overline{\mathbf{v}}_{t,s}\varphi _{k})+a^{\ast }(4D_{t}%
\overline{\mathbf{u}}_{t,s}\varphi _{k})\ ,  \notag
\end{eqnarray}%
which is in fact well-defined on $\mathcal{D}(\mathrm{N})$.Then, by
Equations (\ref{iuiu})--(\ref{rtrtrtrt2}) and Theorem \ref{theorem
Ht=00003Dunitary orbite copy(1)}, we arrive at the equality 
\begin{equation}
\partial _{t}\left\{ \mathrm{U}_{t,s}^{\ast }\tilde{a}_{t,s,k}\mathrm{U}%
_{t,s}\psi \right\} =\mathrm{U}_{t,s}^{\ast }\left( \partial _{t}\tilde{a}%
_{t,s,k}-i\left[ \tilde{a}_{t,s,k},\mathrm{G}_{t}\right] \right) \mathrm{U}%
_{t,s}\psi =0  \label{deri bog}
\end{equation}%
for all vectors $\psi \in \mathcal{D}(\mathrm{N})$ and $s,t\in \mathbb{R}%
_{0}^{+}$. Using again Theorem \ref{theorem Ht=00003Dunitary orbite copy(1)}%
, we deduce that 
\begin{equation}
\tilde{a}_{t,s,k}\psi =\mathrm{U}_{t,s}\tilde{a}_{k}\mathrm{U}_{t,s}^{\ast
}\psi  \label{extension equality}
\end{equation}%
for all $k\in \mathbb{N}$, $\psi \in \mathcal{D}(\mathrm{N})$ and $s,t\in 
\mathbb{R}_{0}^{+}$. The domain $\mathcal{D}(\mathrm{N})\subseteq \mathcal{F}
$ is dense and both $\tilde{a}_{t,s,k}$ and $\mathrm{U}_{t,s}\tilde{a}_{k}%
\mathrm{U}_{t,s}^{\ast }$ are bounded operators. As a consequence, using a
standard continuity argument one deduces that (\ref{extension equality})
holds true for all $\psi \in \mathcal{F}$.
\end{proof}

Proposition \ref{lemma uv-Bogoliubov transformation} expresses the fact that
the isospectral flow defined via the unitary evolution system $\left( 
\mathrm{U}_{t,s}\right) _{s,t\in \mathbb{R}_{0}^{+}}$ is a time-dependent
family of Bogoliubov $\ast $--automorphisms. In other words, this evolution
system implements on the Fock space the family of Bogoliubov transformations%
\begin{equation*}
\mathcal{U}_{t,s}=\left( 
\begin{array}{cc}
\mathbf{u}_{t,s} & \mathbf{v}_{t,s} \\ 
\overline{\mathbf{v}}_{t,s} & \overline{\mathbf{u}}_{t,s}%
\end{array}%
\right) \ ,\qquad s,t\in \mathbb{R}_{0}^{+}\ ,
\end{equation*}%
defined on $\mathcal{H}\doteq \mathfrak{h}\oplus \mathfrak{h}$. Compare
Proposition \ref{lemma uv-Bogoliubov transformation} with Theorem \ref%
{Shale-Stinespring THM}.

\begin{remark}
\mbox{}\newline
There is a small inconsistency or typo in the first assertion of \cite[Lemma
71]{bach-bru-memo} as one can see from its proof (which is correct). \cite[%
Lemma 71]{bach-bru-memo} done in the bosonic case has to be analogous to
Proposition \ref{lemma uv-Bogoliubov transformation}.
\end{remark}

\subsubsection{Solution to the Brockett-Wegner Flow on Quadratic
Hamiltonians \label{section unitarly equi H}}

The continuous flow of Bogoliubov transformations used here is based on an
elliptic flow (\ref{flow equation-quadratic deltabis}) on operators acting
on the one-particle Hilbert space. This flow is not studied here, but in 
\cite{EllipticFlow}, some important results of which are shortly explained
in Section \ref{Eliptic flow}. Therefore, \textbf{the kernel of our proofs
on the diagonalization of quadratic Hamiltonians lies on the one-particle
Hilbert space}.

This being said, we had to implement on the Fock space the corresponding
Bogoliubov transformations generating by the solution to the elliptic flow.
This was done in Section \ref{Section proof theroem important 2bis} and it
remains to shows that the Brockett-Wegner flow (\ref{brocket flow}), applied
to a fermionic quadratic Hamiltonian $\mathrm{H}_{0}$ and the particle
number operator $\mathrm{N}$, has a solution as a unitary equivalent
quadratic Hamiltonian. As a consequence, we have to deal here with domain
issues on the Fock space, which can be sometimes cumbersome. However, as
explained in Section \ref{sectionBWflow}, the heuristic idea is very easy to
understand and thus mathematically elegant from our viewpoint.

In fact, our aim here is \textbf{not only} to obtain a diagonalization of
quadratic Hamiltonians, but also to highlight the Brockett-Wegner flow \cite%
{Brockett1,Wegner1,bach-bru,Opti} by giving another rigorous solution (and
even its asymptotics) of this (usually) formal method for an \textbf{%
unbounded} case. In particular, trying to diagonalize quadratic Hamiltonians
by adapting the method developed in \cite{Solovej-Nam} for the bosonic case
is a priori a natural idea, but it would definitively suppress our second
aim here, namely the study of the Brockett-Wegner flow and ODEs on operator
spaces like the elliptic flow (\ref{flow equation-quadratic deltabis}).
Indeed, differential equations on non-commutative algebras of operators
(like the Brockett-Wegner flow \cite{Brockett1,Wegner1,bach-bru, Opti}) are
very little developed in Mathematics (in particular for infinite-dimensional
spaces). Hence, any mathematical result on relevant differential equations
for operators is useful and possibly liminal. This is one important
objective of \cite{EllipticFlow} and the present paper, the diagonalization
of quadratic Hamiltonians under much more general conditions being one
salient application achieved via asymptotics results of Section \ref{section
proof thm important 3}. Moreover, note that comparatively to \cite%
{Solovej-Nam} (done in the bosonic case), we obtain here some expressions
for the diagonal form as integrals of elements of the elliptic flow, and in
some cases a very explicit expressions (see 
\eqref{constant of
motion eq2bis} and \eqref{dfgdgdfgdfg}).

To this end, in this section we aim at proving that the Brockett-Wegner flow
(\ref{brocket flow}), applied to a fermionic quadratic Hamiltonian $\mathrm{H%
}_{0}$ and the particle number operator $\mathrm{N}$, has a solution as a
unitary equivalent quadratic Hamiltonian $\mathrm{H}_{t}=\mathrm{U}_{t,s}%
\mathrm{\mathrm{H}}_{s}\mathrm{U}_{t,s}^{\ast }$ for any $s,t\in \mathbb{R}%
_{0}^{+}$, with the evolution system $\left( \mathrm{U}_{t,s}\right)
_{s,t\in \mathbb{R}_{0}^{+}}$ being the family of unitary operators
described in Theorem \ref{theorem Ht=00003Dunitary orbite copy(1)}. This
refers to the following assertion:\ 

\begin{theorem}[Solution to the Brockett-Wegner flow on quadratic
Hamiltonians]
\label{theorem Ht=00003Dunitary orbite}\mbox{}\newline
The quadratic Hamiltonians defined by (\ref{DefH0bis}) satisfy the equality 
\begin{equation*}
\mathrm{H}_{t}=\mathrm{U}_{t,s}\mathrm{\mathrm{H}}_{s}\mathrm{U}_{t,s}^{\ast
}\ ,\qquad s,t\in \mathbb{R}_{0}^{+},
\end{equation*}%
with $\left( \mathrm{U}_{t,s}\right) _{s,t\in \mathbb{R}_{0}^{+}}$ being the
(evolution) family of unitary operators of Theorem \ref{theorem
Ht=00003Dunitary orbite copy(1)} for the solution $D$ to the elliptic
operator-valued flow (\ref{flow equation-quadratic deltabis}) with initial
condition $D_{0}=-D_{0}^{\top }\in \mathcal{L}^{2}(\mathfrak{h})$.
\end{theorem}

\begin{proof}
Note first that the solution $D$ to the elliptic operator-valued flow (\ref%
{flow equation-quadratic deltabis}) with initial condition $%
D_{0}=-D_{0}^{\top }\in \mathcal{L}^{2}(\mathfrak{h})$ satisfies $D\in C(%
\mathbb{R}_{0}^{+};\mathcal{L}^{2}(\mathfrak{h}))$ and $D_{t}=-D_{t}^{\top
}\in \mathcal{L}^{2}(\mathfrak{h})$ for all $t\in \mathbb{R}_{0}^{+}$,
thanks to Theorem \ref{Corollary existence}. We can in particular apply
Theorem \ref{theorem Ht=00003Dunitary orbite copy(1)} for this continuous
family of Hilbert-Schmidt operators. Observe next that the theorem is
equivalent to the assertion that 
\begin{equation}
\left( \mathrm{H}_{t}+i\lambda \mathbf{1}\right) ^{-1}=\mathrm{U}%
_{t,s}\left( \mathrm{H}_{s}+i\lambda \mathbf{1}\right) ^{-1}\mathrm{U}%
_{t,s}^{\ast }\ ,\qquad s,t\in \mathbb{R}_{0}^{+}\ ,
\end{equation}%
for some $\lambda \in \mathbb{R}\backslash \{0\}$. The proof is the same as
the one done in various lemmata of \cite[Section VI.2]{bach-bru-memo},
except one has to replace \cite[Lemma 65]{bach-bru-memo} with Corollary \ref%
{corofermion}. For the reader's convenience, we give below the important
steps, but without much details. \medskip

\noindent \underline{Step 1:} For any $t\in \mathbb{R}_{0}^{+}$, one writes
the operator 
\begin{equation}
(\mathrm{N}+\mathbf{1})(\mathrm{H}_{t}+i\lambda \mathbf{1})^{-1}(\mathrm{N}+%
\mathbf{1})^{-1}=%
\Big(%
\mathrm{H}_{t}-i\mathrm{G}_{t}(\mathrm{N}+\mathbf{1})^{-1}+i\lambda \mathbf{1%
}%
\Big)%
^{-1}  \label{condition lambda0}
\end{equation}%
as a norm convergent Neumann series (see \cite[Equation (VI.100)]{bach-bru})
for any $\lambda \in \mathbb{R}$ such that 
\begin{equation}
\left\vert \lambda \right\vert >C\Vert D_{t}\Vert _{2}\geq \Vert \mathrm{G}%
_{t}\left( \mathrm{N}+\mathbf{1}\right) ^{-1}\Vert _{\mathrm{op}}
\label{fffgfgfg}
\end{equation}%
$C\in \mathbb{R}^{+}$ being some fixed, sufficiently large, time-independent
constant (see (\ref{inegality BB 3})). It yields 
\begin{eqnarray}
\Vert (\mathrm{N}+\mathbf{1})(\mathrm{H}_{t}+i\lambda \mathbf{1})^{-1}(%
\mathrm{N}+\mathbf{1})^{-1}\Vert _{\mathrm{op}} &\leq &\frac{1}{|\lambda
|-C\Vert D_{t}\Vert _{2}}  \label{sssd} \\
\Vert (\mathrm{N}+\mathbf{1})\mathrm{H}_{t}(\mathrm{H}_{t}+i\lambda \mathbf{1%
})^{-1}(\mathrm{N}+\mathbf{1})^{-1}\Vert _{\mathrm{op}} &\leq &\frac{%
|\lambda |\left( 1+C\Vert D_{t}\Vert _{2}\right) }{|\lambda |-C\Vert
D_{t}\Vert _{2}}  \label{sssd2}
\end{eqnarray}%
for any $t\in \mathbb{R}_{0}^{+}$ and $\lambda \in \mathbb{R}$ satisfying (%
\ref{fffgfgfg}). \medskip {}

\noindent \underline{Step 2:} By applying Corollary \ref{corofermion}, one
directly arrives at 
\begin{equation}
\Vert (\mathrm{H}_{t}-\mathrm{H}_{s})(\mathrm{N}+\mathbf{1)}^{-1}\Vert _{%
\mathrm{op}}\leq C\left( \int_{s\wedge t}^{s\vee t}\Vert D_{\tau }\Vert
_{2}^{2}\mathrm{d}\tau +\Vert D_{t}-D_{s}\Vert _{2}\right)  \label{V.316a}
\end{equation}%
for some fixed (time-independent) constant $C\in \mathbb{R}^{+}$ and any $%
s,t\in \mathbb{R}_{0}^{+}$. Using carefully the resolvent identity together
with Step 1 and Equation (\ref{V.316a}), one proves that 
\begin{equation}
t\mapsto \left( \mathrm{H}_{t}+i\lambda \mathbf{1}\right) ^{-1}\in C\left( %
\left[ s,T\right] ;\mathcal{B}(\mathcal{Y},\mathcal{F}_{b})\right)
\label{sdklfjsdklfjsdfklj0}
\end{equation}%
for any $s\in \mathbb{R}_{0}^{+}$, $T\in \left( s,\infty \right) $ and $%
\lambda \in \mathbb{R}$ satisfying 
\begin{equation}
\left\vert \lambda \right\vert >C\sup_{t\in \left[ s,T\right] }\Vert
D_{t}\Vert _{2}\ .  \label{sdklfjsdklfjsdfklj}
\end{equation}%
In Equation (\ref{sdklfjsdklfjsdfklj0}), note that $\mathcal{Y}=\mathcal{D}%
\left( \mathrm{N}\right) $ is the dense domain of the particle number
operator $\mathrm{N}$, see (\ref{banach Y}). In particular, the resolvent $%
t\mapsto \left( \mathrm{H}_{t}+i\lambda \mathbf{1}\right) ^{-1}$ is strongly
continuous on $\left[ 0,T\right] $ for any fixed $T\in \mathbb{R}^{+}$ and $%
\lambda \in \mathbb{R}$ satisfying (\ref{sdklfjsdklfjsdfklj}) at $s=0$.
\medskip {}

\noindent \underline{Step 3:} Using Steps 1 and 2 with a cautious attention
to domain issues, one proves that, for any non-zero initial time $s\in 
\mathbb{R}^{+}$, $T\in \left( s,\infty \right) $ and $\lambda \in \mathbb{R}$
satisfying (\ref{sdklfjsdklfjsdfklj}), the resolvent $\left( \mathrm{H}%
_{t}+i\lambda \mathbf{1}\right) ^{-1}$ satisfies, for any $\varphi ,\psi \in 
\mathcal{D}\left( \mathrm{N}\right) $ and $t\in \left[ s,T\right] $, the
differential equation 
\begin{equation*}
\langle \psi ,(\partial _{t}Y_{t})\varphi \rangle =\langle \psi ,(i\left[
Y_{t},\mathrm{G}_{t}\right] )\varphi \rangle \ ,\quad Y_{s}\doteq \left( 
\mathrm{H}_{s}+i\lambda \mathbf{1}\right) ^{-1}\ .
\end{equation*}%
Note that Theorem \ref{lemma existence 2 copy(4)} and Corollary \ref%
{corofermion} are pivotal to arrive at this assertion. \medskip {}

\noindent \underline{Step 4:} Now, using Steps 1--3, one shows that, for any
non-zero initial time $s\in \mathbb{R}^{+}$, $T\in \left( s,\infty \right) $
and $\lambda \in \mathbb{R}$ satisfying (\ref{sdklfjsdklfjsdfklj}), 
\begin{equation}
\langle \left( \mathrm{N}+\mathbf{1}\right) ^{-1}\psi ,\partial _{t}\{%
\mathrm{U}_{t,s}^{\ast }\left( \mathrm{H}_{t}+i\lambda \mathbf{1}\right)
^{-1}\mathrm{U}_{t,s}\}\left( \mathrm{N}+\mathbf{1}\right) ^{-1}\varphi
\rangle =0  \label{eq sup encore new 1}
\end{equation}%
for any $\varphi ,\psi \in \mathcal{F}$ and all $t\in \left[ s,T\right] $,
i.e., 
\begin{equation}
\langle \left( \mathrm{N}+\mathbf{1}\right) ^{-1}\psi ,(\mathrm{U}%
_{t,s}^{\ast }\left( \mathrm{H}_{t}+i\lambda \mathbf{1}\right) ^{-1}\mathrm{U%
}_{t,s}-\left( \mathrm{H}_{s}+i\lambda \mathbf{1}\right) ^{-1})\left( 
\mathrm{N}+\mathbf{1}\right) ^{-1}\varphi \rangle =0\ .
\label{eq sup encore new 2}
\end{equation}%
Since the domain $\mathcal{D}\left( \mathrm{N}\right) $ is dense and the
resolvent $\left( \mathrm{H}_{t}+i\lambda \mathbf{1}\right) ^{-1}$ as well
as the unitary operator $\mathrm{U}_{t,s}$ are bounded, we infer from (\ref%
{eq sup encore new 2}) that 
\begin{equation}
\mathrm{U}_{t,s}^{\ast }\left( \mathrm{H}_{t}+i\lambda \mathbf{1}\right)
^{-1}\mathrm{U}_{t,s}=\left( \mathrm{H}_{s}+i\lambda \mathbf{1}\right) ^{-1}
\label{eq sup encore new 5}
\end{equation}%
for all $s\in \mathbb{R}^{+}$, $T\in \left( s,\infty \right) $, $t\in \left[
s,T\right] $ and $\lambda \in \mathbb{R}$ satisfying (\ref%
{sdklfjsdklfjsdfklj}). It remains to perform in the above equality the limit 
$s\rightarrow 0^{+}$, by using Theorems \ref{lemma existence 2 copy(4)} and %
\ref{theorem Ht=00003Dunitary orbite copy(1)} as well as the fact that $%
t\mapsto \left( \mathrm{H}_{t}+i\lambda \mathbf{1}\right) ^{-1}$ is strongly
continuous on $\left[ 0,T\right] $ for all $T\in \mathbb{R}^{+}$ and $%
\lambda \in \mathbb{R}$ satisfying (\ref{sdklfjsdklfjsdfklj}) with $s=0$.
This leads to the assertion.
\end{proof}

\subsubsection{Asymptotics of the Solution to the Brockett-Wegner Flow}

Keeping in mind Brockett-Wegner's strategy applied to the diagonalization of
quadratic Hamiltonians, we study the limit $t\rightarrow \infty $ of the
solution $\left( \mathrm{H}_{t}\right) _{t\in \mathbb{R}_{0}^{+}}$ to the
Brockett-Wegner flow (\ref{brocket flow}). See Equation (\ref{DefH0bis}) and
Theorem \ref{theorem Ht=00003Dunitary orbite}. It is directly related to the
square-integrability of the mapping $t\mapsto \Vert D_{t}\Vert _{2}$ on $%
\mathbb{R}_{0}^{+}$, where $D$ is solution to the elliptic operator flow (%
\ref{flow equation-quadratic deltabis}). Sufficient conditions to ensure
this feature are given by Theorem \ref{lemma asymptotics1 copy(4)}.

The square-integrability of $\Vert D_{t}\Vert _{2}$ allows us to define the
self-adjoint operator 
\begin{equation*}
\Upsilon _{\infty }\doteq \Upsilon _{0}+16\int_{0}^{\infty }D_{\tau }D_{\tau
}^{\ast }\mathrm{d}\tau
\end{equation*}%
with domain $\mathcal{D}\left( \Upsilon _{\infty }\right) =\mathcal{D}\left(
\Upsilon _{0}\right) $ (the integral being absolutely convergent in the
space $\mathcal{L}^{1}(\mathfrak{h})$ of trace-class operators)\ and
therefore, 
\begin{equation}
\mathrm{H}_{\infty }\doteq \sum_{k,l\in \mathbb{N}}\left\{ \Upsilon _{\infty
}\right\} _{k,l}a_{k}^{\ast }a_{l}+\left( E_{0}-8\int_{0}^{\infty
}\left\Vert D_{\tau }\right\Vert _{2}^{2}\mathrm{d}\tau \right) \mathbf{1}%
=d\Gamma \left( \Upsilon _{\infty }\right) +\left( E_{0}-8\int_{0}^{\infty
}\left\Vert D_{\tau }\right\Vert _{2}^{2}\mathrm{d}\tau \right) \mathbf{1}\ ,
\label{Hinfini}
\end{equation}%
on the domain $\mathcal{D}_{0}$ (\ref{domain H0}), keeping always in mind
that $D$ is solution to the elliptic operator flow (\ref{flow
equation-quadratic deltabis}). See (\ref{DefH0bis}) extended to $t=\infty $.
Note $\mathrm{H}_{\infty }\equiv \mathrm{H}_{\infty }^{\ast \ast }=\mathrm{H}%
_{\infty }^{\ast }$ is well-defined, for instance via Proposition \ref%
{Hamilselfadjoint}. Up to a constant, it is the well-known
second-quantization $d\Gamma (\Upsilon _{\infty })$ (\ref{second quanti}) of 
$\Upsilon _{\infty }$ and its spectral properties can be directly obtained
from the ones of the one-particle self-adjoint operator $\Upsilon _{\infty }$%
. In fact, $\mathrm{H}_{\infty }$ is $\mathrm{N}$--diagonal since it
commutes with the particle number operator $\mathrm{N}$. Last but not least,
it is the limit $t\rightarrow \infty $, in the strong resolvent sense, of
the family $\left( \mathrm{H}_{t}\right) _{t\geq 0}$ of unitary equivalent
quadratic Hamiltonians.

\begin{proposition}[Quasi $\mathrm{N}$--diagonalization of quadratic
Hamiltonians]
\label{lemma H t infinity copy(1)}\mbox{}\newline
Assume the square-integrability\footnote{%
This condition is clearly satisfied under the assumptions of Theorem \ref%
{lemma asymptotics1 copy(4)}.} of the mapping $t\mapsto \Vert D_{t}\Vert
_{2} $ on $\mathbb{R}_{0}^{+}$, $D$ being solution to (\ref{flow
equation-quadratic deltabis}) with $D_{0}=-D_{0}^{\top }\in \mathcal{L}^{2}(%
\mathfrak{h})$. Then, 
\begin{equation}
\underset{t\rightarrow \infty }{\lim }\left\Vert (\mathrm{H}_{\infty }-%
\mathrm{H}_{t})(\mathrm{N}+\mathbf{1)}^{-1}\right\Vert _{\mathrm{op}}=0
\label{llimit1}
\end{equation}%
and for any non-zero real $\lambda \in \mathbb{R}\backslash \{0\}$ and
vector $\varphi \in \mathcal{F}$, 
\begin{equation}
\underset{t\rightarrow \infty }{\lim }\left\Vert \{(\mathrm{H}_{\infty
}+i\lambda \mathbf{1})^{-1}-(\mathrm{H}_{t}+i\lambda \mathbf{1}%
)^{-1}\}\varphi \right\Vert _{\mathcal{F}}=0\ .  \label{llimit2}
\end{equation}
\end{proposition}

\begin{proof}
Because of the square-integrability of the mapping $t\mapsto \Vert
D_{t}\Vert _{2}$ on $\mathbb{R}_{0}^{+}$, we can extend (\ref{V.316a}) to
all $t,s\in \mathbb{R}_{0}^{+}\cup \{\infty \}$ with $D_{\infty }\doteq 0$
and obviously deduce (\ref{llimit1}). To prove next that $\left( \mathrm{H}%
_{t}\right) _{t\geq 0}$ converges to $\mathrm{H}_{\infty }$ in the strong
resolvent sense, observe first that, for any non-zero $\lambda \in \mathbb{R}%
\backslash \{0\}$ and $C\in \mathbb{R}^{+}$, there is $s_{\lambda }\in 
\mathbb{R}_{0}^{+}$ such that 
\begin{equation*}
\left\vert \lambda \right\vert >C\sup_{t\in \left[ s_{\lambda },\infty
\right) }\Vert D_{t}\Vert _{2}\ ,
\end{equation*}%
because, by assumption, the Hilbert-Schmidt norm $\Vert D_{t}\Vert _{2}$
vanishes when $t\rightarrow \infty $. If the mapping $t\mapsto \Vert
D_{t}\Vert _{2}$ is square-integrable on $\mathbb{R}_{0}^{+}$ then Equations
(\ref{sssd})--(\ref{sssd2}) are satisfied for $t=\infty $ and sufficiently
large $s\in \mathbb{R}_{0}^{+}$. As a consequence, using the resolvent
identity, for any $\lambda \in \mathbb{R}\backslash \{0\}$, sufficiently
large initial time $s\in \mathbb{R}_{0}^{+}$ and $t\in \left[ s,\infty
\right) $, 
\begin{eqnarray}
&&\left\Vert \{(\mathrm{H}_{\infty }+i\lambda \mathbf{1})^{-1}-(\mathrm{H}%
_{t}+i\lambda \mathbf{1})^{-1}\}(\mathrm{N}+\mathbf{1})^{-1}\right\Vert _{%
\mathrm{op}}  \notag \\
&\leq &\Vert \left( \mathrm{H}_{\infty }+i\lambda \mathbf{1}\right)
^{-1}\Vert _{\mathrm{op}}\Vert \left( \mathrm{H}_{\infty }-\mathrm{H}%
_{t}\right) \left( \mathrm{N}+\mathbf{1}\right) ^{-1}\Vert _{\mathrm{op}} 
\notag \\
&&\Vert \left( \mathrm{N}+\mathbf{1}\right) \left( \mathrm{H}_{t}+i\lambda 
\mathbf{1}\right) ^{-1}\left( \mathrm{N}+\mathbf{1}\right) ^{-1}\Vert _{%
\mathrm{op}}\ ,  \label{eq converge resolvent 1}
\end{eqnarray}%
which in turn yields the limit 
\begin{equation}
\underset{t\rightarrow \infty }{\lim }\left\Vert \{(\mathrm{H}_{\infty
}+i\lambda \mathbf{1})^{-1}-(\mathrm{H}_{t}+i\lambda \mathbf{1})^{-1}\}(%
\mathrm{N}+\mathbf{1})^{-1}\right\Vert _{\mathrm{op}}=0
\label{eq converge resolvent 4}
\end{equation}%
for any $\lambda \in \mathbb{R}\backslash \{0\}$, thanks to (\ref{sssd})--(%
\ref{sssd2}) for $t=\infty $ and (\ref{llimit1}). The domain $\mathcal{D}(%
\mathrm{N})$ is dense in $\mathcal{F}$ and both resolvents $(\mathrm{H}%
_{\infty }+i\lambda \mathbf{1})^{-1}$ and $(\mathrm{H}_{t}+i\lambda \mathbf{1%
})^{-1}$ in (\ref{eq converge resolvent 4}) are bounded operators for any $%
t\in \mathbb{R}_{0}^{+}$ and for any fixed $\lambda \in \mathbb{R}\backslash
\{0\}$. As a consequence, the strong resolvent convergence follows from (\ref%
{eq converge resolvent 4}) and a standard continuity argument.
\end{proof}

By Equations (\ref{generator})--(\ref{generatorbound}), for any $t\in 
\mathbb{R}_{0}^{+}$, the non-$\mathrm{N}$--diagonal part\footnote{%
Recall that Proposition \ref{Hamilselfadjoint} shows that $\mathrm{Z}_{t}$, $%
t\in \mathbb{R}_{0}^{+}$, as operators defined on the domain (\ref{domain H0}%
), are essentially self-adjoint and we use again the notation $\mathrm{Z}%
_{t}\equiv \mathrm{Z}_{t}^{\ast \ast }$, $t\in \mathbb{R}_{0}^{+}$, for
their self-adjoint extension.} 
\begin{equation}
\mathrm{Z}_{t}\doteq \sum_{k,l\in \mathbb{N}}\{D_{t}\}_{k,l}a_{k}^{\ast
}a_{l}^{\ast }+\{\bar{D}_{t}\}_{k,l}a_{l}a_{k}  \label{non-N--diagonal part}
\end{equation}%
of the quadratic operator $\mathrm{H}_{t}=\mathrm{U}_{t,0}\mathrm{H}_{0}%
\mathrm{U}_{t,0}^{\ast }$ originally defined by (\ref{DefH0bis}) satisfies $%
\mathcal{D}(\mathrm{Z}_{t})\supseteq \mathcal{D}(\mathrm{N})$ with 
\begin{equation*}
\Vert \mathrm{Z}_{t}\left( \mathrm{N}+\mathbf{1}\right) ^{-1}\Vert _{\mathrm{%
op}}\leq (1+\sqrt{3})\Vert D_{t}\Vert _{2}\ ,\qquad t\in \mathbb{R}_{0}^{+}\
.
\end{equation*}%
Hence, if the Hilbert-Schmidt norm $\left\Vert D_{t}\right\Vert _{2}$
vanishes in the limit $t\rightarrow \infty $ then the non-$\mathrm{N}$%
--diagonal part $\mathrm{Z}_{t}$ of the quadratic operator $\mathrm{H}_{t}$
vanishes on the domain $\mathcal{D}(\mathrm{N})$, as $t\rightarrow \infty $.
In this sense, Proposition \ref{lemma H t infinity copy(1)} asserts a \emph{%
quasi }$\mathrm{N}$--diagonalization of $\mathrm{H}_{0}$. In other words,
performing the limit $t\rightarrow +\infty $ removes the non-$\mathrm{N}$%
--diagonal part of the Hamiltonian, but at this stage we have no information
on the link between $\mathrm{H}_{\infty }$ and $\mathrm{H}_{0}$. This study
is the aim of the next section.

\subsubsection{Diagonalization of Quadratic Hamiltonians\label{section proof
thm important 3}}

Following Brockett-Wegner's strategy -- or Theorem \ref{theorem
Ht=00003Dunitary orbite} and Proposition \ref{lemma H t infinity copy(1)} in
our example -- the diagonalization of quadratic Hamiltonians $\mathrm{H}_{0}$
(\ref{DefH0}) via a Bogoliubov $\mathbf{u}$--$\mathbf{v}$ transformation
should be achieved by performing the limits $t\rightarrow \infty $ of
bounded operator families $\left( \mathrm{U}_{t,s}\right) _{t,s\in \mathbb{R}%
_{0}^{+}}$ (Theorems \ref{theorem Ht=00003Dunitary orbite copy(1)}) and $%
\left( \mathbf{u}_{t,s}\right) _{t,s\in \mathbb{R}^{+}}$, $\left( \mathbf{v}%
_{t,s}\right) _{t,s\in \mathbb{R}_{0}^{+}}$ (see (\ref{u})--(\ref{v}) and
Proposition \ref{lemma uv-Bogoliubov transformation}). Recall indeed that
the limit $t\rightarrow \infty $ of the solution $\left( \mathrm{H}%
_{t}\right) _{t\in \mathbb{R}_{0}^{+}}$ to the Brockett-Wegner flow (\ref%
{brocket flow}) leads to the $\mathrm{N}$--diagonal quadratic Hamiltonian $%
\mathrm{H}_{\infty }$ (\ref{Hinfini}), thanks to Proposition \ref{lemma H t
infinity copy(1)}. This refers to a quasi $\mathrm{N}$--diagonalization of
fermionic quadratic Hamiltonians. To get a \emph{true} $\mathrm{N}$%
--diagonalization we prove that $\mathrm{H}_{\infty }$ is unitarily
equivalent to any $\mathrm{H}_{t}$, $t\in \mathbb{R}_{0}^{+}$.

To this end, we first show that $\left( \mathrm{U}_{t,s}\right) _{t,s\in 
\mathbb{R}_{0}^{+}}$ can be extended to infinite times $s,t=\{\infty \}$ by
using now the integrability of the mapping $t\mapsto \left\Vert
D_{t}\right\Vert _{2}$ on $\mathbb{R}_{0}^{+}$, instead of its
square-integrability only, which is used in Proposition \ref{lemma H t
infinity copy(1)}.

\begin{theorem}[Extension of the unitary flow to infinite times]
\label{lemma H t infinity copy(2)}\mbox{}\newline
Take $D\in C(\mathbb{R}_{0}^{+};\mathcal{L}^{2}(\mathfrak{h}))$ with $%
D_{t}=-D_{t}^{\top }\in \mathcal{L}^{2}(\mathfrak{h})$ for all $t\in \mathbb{%
R}_{0}^{+}$ and assume the integrability of the mapping $t\mapsto \Vert
D_{t}\Vert _{2}$ on $\mathbb{R}_{0}^{+}$. Then the family of Theorem \ref%
{theorem Ht=00003Dunitary orbite copy(1)} can be extended to a family $(%
\mathrm{U}_{t,s})_{s,t\in \mathbb{R}_{0}^{+}\cup \left\{ \infty \right\} }$
of unitary operators satisfying, on $\mathbb{R}_{0}^{+}\cup \left\{ \infty
\right\} $, all properties (i)--(v) of Theorem \ref{theorem Ht=00003Dunitary
orbite copy(1)}.
\end{theorem}

\begin{proof}
From straightforward, albeit tedious, estimates using the triangle
inequality, Equations (\ref{generator}), (\ref{inegality BB 3}), Theorem \ref%
{theorem Ht=00003Dunitary orbite copy(1)} and Corollary \ref{corofermion},
we obtain that, for any $s_{1},t_{1},s_{2},t_{2}\in \mathbb{R}_{0}^{+}$, 
\begin{eqnarray}
&&\Vert (\mathrm{U}_{t_{2},s_{2}}-\mathrm{U}_{t_{1},s_{1}})(\mathrm{N}+%
\mathbf{1})^{-1}\Vert _{\mathrm{op}}  \notag \\
&\leq &C\int_{s_{1}\wedge s_{2}}^{s_{1}\vee s_{2}}\Vert D_{\tau }\Vert _{2}\ 
\mathrm{d}\tau +C\int_{t_{1}\wedge t_{2}}^{t_{1}\vee t_{2}}\Vert D_{\tau
}\Vert _{2}\mathrm{e}^{C\int_{s_{2}\wedge \tau }^{s_{2}\vee \tau }\Vert
D_{\alpha }\Vert _{2}\mathrm{d}\alpha }\ \mathrm{d}\tau
\label{sdsdssjhkljkljkl} \\
&=&C\int_{s_{1}\wedge s_{2}}^{s_{1}\vee s_{2}}\Vert D_{\tau }\Vert _{2}\ 
\mathrm{d}\tau +\mathrm{e}^{C\int_{s_{2}}^{\left( s_{2}\vee \left(
t_{1}\wedge t_{2}\right) \right) \wedge \left( t_{1}\vee t_{2}\right) }\Vert
D_{\tau }\Vert _{2}\mathrm{d}\tau }\left( \mathrm{e}^{C\int_{\left(
s_{2}\vee \left( t_{1}\wedge t_{2}\right) \right) \wedge \left( t_{1}\vee
t_{2}\right) }^{t_{1}\vee t_{2}}\Vert D_{\tau }\Vert _{2}\mathrm{d}\tau
}-1\right)  \notag \\
&&+\mathrm{e}^{-C\int_{s_{2}}^{\left( s_{2}\vee \left( t_{1}\wedge
t_{2}\right) \right) \wedge \left( t_{1}\vee t_{2}\right) }\Vert D_{\tau
}\Vert _{2}\mathrm{d}\tau }\left( \mathrm{e}^{-C\int_{\left( s_{2}\vee
\left( t_{1}\wedge t_{2}\right) \right) \wedge \left( t_{1}\vee t_{2}\right)
}^{t_{1}\wedge t_{2}}\Vert D_{\tau }\Vert _{2}\mathrm{d}\tau }-1\right)
\label{flow equation convergence unitary 1}
\end{eqnarray}%
for some time-independent constant $C\in \mathbb{R}^{+}$. In particular, if
the mapping $t\mapsto \Vert D_{t}\Vert _{2}$ is integrable on $\mathbb{R}%
_{0}^{+}$ then the operator 
\begin{equation*}
\{\mathrm{U}_{t_{2},s_{2}}-\mathrm{U}_{t_{1},s_{1}}\}(\mathrm{N}+\mathbf{1}%
)^{-1}
\end{equation*}%
converges to zero in the norm topology either when $t_{1}\wedge
t_{2}\rightarrow \infty $ and $s_{1}=s_{2}$ or when $s_{1}\wedge
s_{2}\rightarrow \infty $ and $t_{1}=t_{2}$. Hence, since the domain $%
\mathcal{D}\left( \mathrm{N}\right) $ is dense with 
\begin{equation}
(\mathrm{U}_{t,s})_{s,t\in \mathbb{R}_{0}^{+}}\subseteq \mathcal{B}\left( 
\mathcal{F}\right) \ ,  \label{operator family}
\end{equation}%
one infers from a continuity argument the existence of families $\left( 
\mathrm{U}_{\infty ,s}\right) _{s\in \mathbb{R}_{0}^{+}}$ and $\left( 
\mathrm{U}_{t,\infty }\right) _{t\in \mathbb{R}_{0}^{+}}$ of bounded
operators respectively defined, for any $\varphi \in \mathcal{F}$ and $%
s,t\in \mathbb{R}_{0}^{+}$, by the strong limits 
\begin{equation}
\mathrm{U}_{\infty ,s}\varphi \doteq \underset{t\rightarrow \infty }{\lim }%
\left\{ \mathrm{U}_{t,s}\varphi \right\} \ ,\quad \mathrm{U}_{t,\infty
}\varphi \doteq \underset{s\rightarrow \infty }{\lim }\left\{ \mathrm{U}%
_{t,s}\varphi \right\} \ .  \label{flow equation convergence unitary 1bisbis}
\end{equation}%
In particular, by the unitarity of the operator family (\ref{operator family}%
),%
\begin{equation}
\max \left\{ \left\Vert \mathrm{U}_{\infty ,s}\right\Vert _{\mathrm{op}%
},\left\Vert \mathrm{U}_{t,\infty }\right\Vert _{\mathrm{op}}\right\} \leq 1
\label{bound for limit unitarity}
\end{equation}%
for any $s,t\in \mathbb{R}_{0}^{+}$. From (\ref{flow equation convergence
unitary 1}) we also obtain that, for any $\varphi \in \mathcal{F}$, 
\begin{equation}
\mathrm{U}_{\infty ,\infty }\varphi \doteq \underset{t\rightarrow \infty }{%
\lim }\underset{s\rightarrow \infty }{\lim }\left\{ \mathrm{U}_{t,s}\varphi
\right\} =\underset{s\rightarrow \infty }{\lim }\underset{t\rightarrow
\infty }{\lim }\left\{ \mathrm{U}_{t,s}\varphi \right\} =\underset{%
t\rightarrow \infty }{\lim }\left\{ \mathrm{U}_{t,t}\varphi \right\}
=\varphi \ .  \label{flow equation convergence unitary 1bis}
\end{equation}%
In particular, $\mathrm{U}_{\infty ,\infty }=\mathbf{1}$. It is also
straightforward to verify from (\ref{flow equation convergence unitary 1})
and (\ref{flow equation convergence unitary 1bisbis}) that $\left( \mathrm{U}%
_{\infty ,s}\right) _{s\in \mathbb{R}_{0}^{+}}$ and $\left( \mathrm{U}%
_{t,\infty }\right) _{t\in \mathbb{R}_{0}^{+}}$ are strongly continuous in $%
s $ and $t$, respectively, while $\mathrm{U}_{t,s}=\mathrm{U}_{t,x}\mathrm{U}%
_{x,s}$ for any $x,t,s\in \mathbb{R}_{0}^{+}\cup \left\{ \infty \right\} $.
In the same way, one easily checks that they are family of unitary operators
with $\mathrm{U}_{\infty ,s}^{\ast }=\mathrm{U}_{s,\infty }$ for any $s\in 
\mathbb{R}_{0}^{+}$. Moreover, using the triangle inequality, one gets that,
for any $\epsilon \in \mathbb{R}\backslash \{0\}$ with sufficiently small $%
\left\vert \epsilon \right\vert >0$, all times $s,t\in \mathbb{R}^{+}$ and
any vector $\varphi \in \mathcal{D}\left( \mathrm{N}^{2}\right) \subseteq 
\mathcal{D}\left( \mathrm{N}\right) \subseteq \mathcal{D}\left( \mathrm{G}%
_{s}\right) $,%
\begin{eqnarray}
&&\Vert \{\epsilon ^{-1}(\mathrm{U}_{\infty ,s+\epsilon }-\mathrm{U}_{\infty
,s})-i\mathrm{U}_{\infty ,s}\mathrm{G}_{s}\}\varphi \Vert _{\mathcal{F}} 
\notag \\
&\leq &\Vert \{\mathrm{U}_{\infty ,s}-\mathrm{U}_{t,s}\}\mathrm{G}%
_{s}\varphi \Vert _{\mathcal{F}}+\epsilon ^{-1}\int_{s}^{s+\epsilon }\Vert \{%
\mathrm{G}_{\tau }-\mathrm{G}_{s}\}\varphi \Vert _{\mathcal{F}}\mathrm{d}\tau
\notag \\
&&+\epsilon ^{-1}\int_{s}^{s+\epsilon }\Vert \{\mathrm{U}_{t,\tau }-\mathrm{U%
}_{t,s}\}\left( \mathrm{N}+\mathbf{1}\right) ^{-1}\Vert _{\mathrm{op}}\Vert 
\mathrm{G}_{s}\left( \mathrm{N}+\mathbf{1}\right) ^{-1}\Vert _{\mathrm{op}%
}\left\Vert \left( \mathrm{N}+\mathbf{1}\right) ^{2}\varphi \right\Vert _{%
\mathcal{F}}\mathrm{d}\tau  \notag \\
&&+\epsilon ^{-1}\int_{s}^{s+\epsilon }\Vert \{\mathrm{U}_{t,\tau }-\mathrm{U%
}_{t,s}\}\left( \mathrm{N}+\mathbf{1}\right) ^{-1}\Vert _{\mathrm{op}}\Vert %
\left[ \mathrm{N},\mathrm{G}_{s}\right] \left( \mathrm{N}+\mathbf{1}\right)
^{-1}\Vert _{\mathrm{op}}\left\Vert \left( \mathrm{N}+\mathbf{1}\right)
^{2}\varphi \right\Vert _{\mathcal{F}}\mathrm{d}\tau  \notag \\
&&+\epsilon ^{-1}\Vert \{\mathrm{U}_{\infty ,s}-\mathrm{U}_{t,s}\}\varphi
\Vert _{\mathcal{F}}+\epsilon ^{-1}\Vert \{\mathrm{U}_{\infty ,s+\epsilon }-%
\mathrm{U}_{t,s+\epsilon }\}\varphi \Vert _{\mathcal{F}}\ .
\label{flow equation convergence unitary 2}
\end{eqnarray}%
Since $D\in C(\mathbb{R}_{0}^{+};\mathcal{L}^{2}(\mathfrak{h}))$, by
Corollary \ref{corofermion} and Equation (\ref{generator}), 
\begin{equation}
\mathrm{G}_{t}\in C(\mathbb{R}_{0}^{+};\mathcal{B}(\mathcal{Y},\mathcal{F}%
))\qquad \text{and}\qquad \left[ \mathrm{N},\mathrm{G}_{t}\right] \left( 
\mathrm{N}+\mathbf{1}\right) ^{-1}\in \mathcal{B}\left( \mathcal{F}\right)
\label{sdfsdf}
\end{equation}%
with $\mathcal{Y\doteq D}\left( \mathrm{N}\right) $, see, e.g., Equation (%
\ref{inegality BB 3}). Therefore, we take $t=t(\epsilon )\rightarrow \infty $
as $\epsilon \rightarrow 0$ in (\ref{flow equation convergence unitary 2})
and use (\ref{flow equation convergence unitary 1}) together with standard
arguments to get that $\partial _{s}\mathrm{U}_{\infty ,s}\varphi =i\mathrm{U%
}_{\infty ,s}\mathrm{G}_{s}\varphi $ for any $\varphi \in \mathcal{D}\left( 
\mathrm{N}^{2}\right) \subseteq \mathcal{D}\left( \mathrm{G}_{s}\right) $
and $s\in \mathbb{R}^{+}$. Since $\mathcal{D}\left( \mathrm{N}^{2}\right) $
is a core for the particle number operator $\mathrm{N}$, using a continuity
argument together with (\ref{inegality BB 3}), (\ref{flow equation
convergence unitary 1}) and (\ref{bound for limit unitarity}), we deduce
that $\partial _{s}\mathrm{U}_{\infty ,s}=i\mathrm{U}_{\infty ,s}\mathrm{G}%
_{s}$ holds true on the domain $\mathcal{D}\left( \mathrm{N}\right) $, for
any $s\in \mathbb{R}^{+}$. The case $s=0$ is done exactly in the same way by
using only $\epsilon >0$.

We need now to prove that 
\begin{equation}
s\mapsto (\mathrm{N}+\mathbf{1})\mathrm{U}_{\infty ,s}(\mathrm{N}+\mathbf{1}%
)^{-1}\qquad \text{and}\qquad t\mapsto (\mathrm{N}+\mathbf{1})\mathrm{U}%
_{t,\infty }(\mathrm{N}+\mathbf{1})^{-1}  \label{mapping}
\end{equation}%
are strongly continuous mappings from $\mathbb{R}_{0}^{+}$ to $\mathcal{B}%
\left( \mathcal{F}\right) $. First, in the same way one proves (\ref%
{inegality BB 3}), we obtain from the CAR (\ref{CAR}) and Corollary \ref%
{corofermion} that 
\begin{equation*}
\Vert \lbrack (\mathrm{N}+\mathbf{1)}^{2},\mathrm{G}_{t}](\mathrm{N}+\mathbf{%
1)}^{-2}\Vert _{\mathrm{op}}\leq C\Vert D_{t}\Vert _{2}
\end{equation*}%
for some (time-independent) constant $C\in \mathbb{R}^{+}$. Similar to the
proof of Theorem \ref{theorem Ht=00003Dunitary orbite copy(1)} (iv), one
then finds that, for any $s,t\in \mathbb{R}_{0}^{+}$, 
\begin{equation}
\Vert (\mathrm{N}+\mathbf{1})^{2}\mathrm{U}_{t,s}(\mathrm{N}+\mathbf{1}%
)^{-2}\Vert _{\mathrm{op}}\leq \mathrm{\exp }\left\{ C\int_{s\wedge
t}^{s\vee t}\Vert D_{\tau }\Vert _{2}\mathrm{d}\tau \right\} \ ,
\label{inegality BB 4}
\end{equation}%
keeping in mind that $C\in \mathbb{R}^{+}$ is some fixed (time-independent)
constant. Next, using again the triangle inequality, Equations (\ref%
{generator}), (\ref{inegality BB 3}), (\ref{inegality BB 4}), Theorem \ref%
{theorem Ht=00003Dunitary orbite copy(1)} and Corollary \ref{corofermion},
like in Inequality (\ref{flow equation convergence unitary 1}), we get that,
for any $s_{1},t_{1},s_{2},t_{2}\in \mathbb{R}_{0}^{+}$, 
\begin{eqnarray}
&&\left\Vert (\mathrm{N}+\mathbf{1})(\mathrm{U}_{t_{2},s_{2}}-\mathrm{U}%
_{t_{1},s_{1}})(\mathrm{N}+\mathbf{1})^{-2}\right\Vert _{\mathrm{op}}  \notag
\\
&\leq &C\int_{s_{1}\wedge s_{2}}^{s_{1}\vee s_{2}}\left\Vert D_{\tau
}\right\Vert _{2}\mathrm{e}^{C\int_{t_{1}\wedge \tau }^{t_{1}\vee \tau
}\left\Vert D_{\alpha }\right\Vert _{2}\mathrm{d}\alpha }\mathrm{d}\tau
+C\int_{t_{1}\wedge t_{2}}^{t_{1}\vee t_{2}}\left\Vert D_{\tau }\right\Vert
_{2}\mathrm{e}^{C\int_{s_{2}\wedge \tau }^{s_{2}\vee \tau }\left\Vert
D_{\alpha }\right\Vert _{2}\mathrm{d}\alpha }\mathrm{d}\tau
\label{flow equation convergence unitary 1flow equation convergence unitary 1}
\end{eqnarray}%
for some time-independent constant $C\in \mathbb{R}^{+}$. Compare (\ref{flow
equation convergence unitary 1flow equation convergence unitary 1}) with (%
\ref{sdsdssjhkljkljkl}). In particular, an explicit expression for the upper
bound of (\ref{flow equation convergence unitary 1flow equation convergence
unitary 1}) can be obtained, similar to (\ref{flow equation convergence
unitary 1}). The particle number operator $\mathrm{N}$ is self-adjoint and
in particular a closed operator. Using this property and the last inequality
together with a continuity argument and (\ref{flow equation convergence
unitary 1bisbis})--(\ref{flow equation convergence unitary 1bis}), it
follows that, for any $\varphi \in \mathcal{F}$ and $s,t\in \mathbb{R}%
_{0}^{+}$, 
\begin{eqnarray*}
(\mathrm{N}+\mathbf{1})\mathrm{U}_{\infty ,s}(\mathrm{N}+\mathbf{1}%
)^{-1}\varphi &=&\underset{t\rightarrow \infty }{\lim }\left\{ (\mathrm{N}+%
\mathbf{1})\mathrm{U}_{t,s}(\mathrm{N}+\mathbf{1})^{-1}\varphi \right\} \ ,
\\
(\mathrm{N}+\mathbf{1})\mathrm{U}_{t,\infty }(\mathrm{N}+\mathbf{1}%
)^{-1}\varphi &=&\underset{s\rightarrow \infty }{\lim }\left\{ (\mathrm{N}+%
\mathbf{1})\mathrm{U}_{t,s}(\mathrm{N}+\mathbf{1})^{-1}\varphi \right\} \ ,
\end{eqnarray*}%
as well as 
\begin{multline*}
\underset{t\rightarrow \infty }{\lim }\underset{s\rightarrow \infty }{\lim }%
\left\{ (\mathrm{N}+\mathbf{1})\mathrm{U}_{t,s}(\mathrm{N}+\mathbf{1}%
)^{-1}\varphi \right\} =\underset{s\rightarrow \infty }{\lim }\underset{%
t\rightarrow \infty }{\lim }\left\{ (\mathrm{N}+\mathbf{1})\mathrm{U}_{t,s}(%
\mathrm{N}+\mathbf{1})^{-1}\varphi \right\} \\
=\varphi =(\mathrm{N}+\mathbf{1})\mathrm{U}_{\infty ,\infty }(\mathrm{N}+%
\mathbf{1})^{-1}\ .
\end{multline*}%
It is then straightforward to verify that the two mappings of Equations (\ref%
{mapping}) are strongly continuous in $s$ and $t$, respectively. Moreover,
using again the triangle inequality and Theorem \ref{theorem
Ht=00003Dunitary orbite copy(1)}, one finds that, for any $\epsilon \in 
\mathbb{R}$ with sufficiently small $\left\vert \epsilon \right\vert >0$,
all times $t,s\in \mathbb{R}^{+}$ and any vector $\varphi \in \mathcal{D}%
\left( \mathrm{N}^{2}\right) $, 
\begin{eqnarray}
&&\Vert \{\epsilon ^{-1}(\mathrm{U}_{t+\epsilon ,\infty }-\mathrm{U}%
_{t,\infty })+i\mathrm{G}_{t}\mathrm{U}_{t,\infty }\}\varphi \Vert _{%
\mathcal{F}}  \notag \\
&\leq &\Vert \mathrm{G}_{t}\left( \mathrm{N}+\mathbf{1}\right) ^{-1}\Vert _{%
\mathrm{op}}\Vert \left( \mathrm{N}+\mathbf{1}\right) \{\mathrm{U}_{t,\infty
}-\mathrm{U}_{t,s}\}\varphi \Vert _{\mathcal{F}}  \notag \\
&&+\epsilon ^{-1}\int_{t}^{t+\epsilon }\Vert \{\mathrm{G}_{\tau }-\mathrm{G}%
_{t}\}\left( \mathrm{N}+\mathbf{1}\right) ^{-1}\Vert _{\mathrm{op}}\Vert
\left( \mathrm{N}+\mathbf{1}\right) \mathrm{U}_{\tau ,s}\left( \mathrm{N}+%
\mathbf{1}\right) ^{-1}\varphi \Vert _{\mathcal{F}}\mathrm{d}\tau  \notag \\
&&+\epsilon ^{-1}\int_{t}^{t+\epsilon }\Vert \mathrm{G}_{t}\left( \mathrm{N}+%
\mathbf{1}\right) ^{-1}\Vert _{\mathrm{op}}\Vert \left( \mathrm{N}+\mathbf{1}%
\right) \left( \mathrm{U}_{\tau ,s}-\mathrm{U}_{t,s}\right) \left( \mathrm{N}%
+\mathbf{1}\right) ^{-1}\varphi \Vert _{\mathrm{op}}\mathrm{d}\tau  \notag \\
&&+\epsilon ^{-1}\Vert \{\mathrm{U}_{t+\epsilon ,\infty }-\mathrm{U}%
_{t+\epsilon ,s}\}\varphi \Vert _{\mathcal{F}}+\epsilon ^{-1}\Vert \{\mathrm{%
U}_{t,\infty }-\mathrm{U}_{t,s}\}\varphi \Vert _{\mathcal{F}}\ .
\label{flow equation convergence unitary 2bis}
\end{eqnarray}%
Therefore, we take $s=s(\epsilon )\rightarrow \infty $ as $\epsilon
\rightarrow 0$ in (\ref{flow equation convergence unitary 2bis}) and use (%
\ref{flow equation convergence unitary 1flow equation convergence unitary 1}%
) together with (\ref{inegality BB 3}), Theorem \ref{Corollary existence}
and standard arguments to get that $\partial _{t}\mathrm{U}_{t,\infty
}\varphi =-i\mathrm{G}_{t}\mathrm{U}_{t,\infty }\varphi $ for any $\varphi
\in \mathcal{D}\left( \mathrm{N}^{2}\right) $. Using again a continuity
argument together with (\ref{inegality BB 3}), (\ref{flow equation
convergence unitary 1}) and (\ref{bound for limit unitarity}), we deduce
that $\partial _{t}\mathrm{U}_{t,\infty }=-i\mathrm{G}_{t}\mathrm{U}%
_{t,\infty }$ holds true on the domain $\mathcal{D}\left( \mathrm{N}\right) $%
, for any $t\in \mathbb{R}^{+}$. The case $t=0$ is done exactly in the same
way by using $\epsilon >0$.
\end{proof}

Note that Theorem \ref{lemma H t infinity copy(2)} for the \emph{bosonic}
case is only partially proven in \cite{bach-bru-memo}, even though it can
also be shown in this situation. Remark additionally that $n\in \mathbb{N}$
iterations of the formal computation (\ref{formal computation2}) to compute
the $n$-folds muticommutators $[\mathrm{N},[\mathrm{N},\cdots ,[\mathrm{N},%
\mathrm{G}_{t}]\cdots ]$ together with Corollary \ref{corofermion} yield 
\begin{equation*}
\Vert \lbrack (\mathrm{N}+\mathbf{1)}^{n},\mathrm{G}_{t}](\mathrm{N}+\mathbf{%
1)}^{-n}\Vert _{\mathrm{op}}\leq C_{n}\Vert D_{t}\Vert _{2}\ ,\qquad t\in 
\mathbb{R}_{0}^{+}\ ,
\end{equation*}%
for some (time-independent) constant $C_{n}\in \mathbb{R}^{+}$. By using the
same arguments of the proof of Theorem \ref{lemma H t infinity copy(2)} one
thus checks that, for any $n\in \mathbb{N}$ and $s,t\in \mathbb{R}%
_{0}^{+}\cup \left\{ \infty \right\} $, 
\begin{equation*}
\Vert (\mathrm{N}+\mathbf{1})^{n}\mathrm{U}_{t,s}(\mathrm{N}+\mathbf{1}%
)^{-n}\Vert _{\mathrm{op}}\leq \mathrm{\exp }\left\{ C_{n}\int_{s\wedge
t}^{s\vee t}\Vert D_{\tau }\Vert _{2}\mathrm{d}\tau \right\} \ ,
\end{equation*}%
and both 
\begin{equation*}
s\mapsto (\mathrm{N}+\mathbf{1})^{n}\mathrm{U}_{t,s}(\mathrm{N}+\mathbf{1}%
)^{-n},\quad t\in \mathbb{R}_{0}^{+}\cup \left\{ \infty \right\} ,\quad 
\text{and}\quad t\mapsto (\mathrm{N}+\mathbf{1})^{n}\mathrm{U}_{t,s}(\mathrm{%
N}+\mathbf{1})^{-n},\quad s\in \mathbb{R}_{0}^{+}\cup \left\{ \infty
\right\} ,
\end{equation*}%
are strongly continuous mapping from $\mathbb{R}_{0}^{+}$ to $\mathcal{B}%
\left( \mathcal{F}\right) $. These properties can be used to compute higher
order derivatives of the evolution system $(\mathrm{U}_{t,s})_{s,t\in 
\mathbb{R}_{0}^{+}\cup \left\{ \infty \right\} }$. Mutatis mutandis for the
bosonic case studied in \cite{bach-bru-memo}.

Theorem \ref{lemma H t infinity copy(2)} means in particular that, as $%
t\rightarrow \infty $ (resp. $s\rightarrow \infty $), the unitary operator
family $(\mathrm{U}_{t,s})_{s,t\in \mathbb{R}_{0}^{+}}$ strongly (pointwise)
converges to a unitary operator family $\left( \mathrm{U}_{\infty ,s}\right)
_{s\in \mathbb{R}_{0}^{+}}$ (resp. $\left( \mathrm{U}_{t,\infty }\right)
_{t\in \mathbb{R}_{0}^{+}}=\left( \mathrm{U}_{\infty ,t}^{\ast }\right)
_{t\in \mathbb{R}_{0}^{+}}$). The limit families are strongly continuous in $%
s\in \mathbb{R}_{0}^{+}\cup \left\{ \infty \right\} $ or $t\in \mathbb{R}%
_{0}^{+}\cup \left\{ \infty \right\} $ and they satisfy the non-autonomous
evolution equations 
\begin{equation*}
\forall s,t\in \mathbb{R}_{0}^{+}:\qquad \left\{ 
\begin{array}{llll}
\partial _{t}\mathrm{U}_{t,\infty }=-i\mathrm{G}_{t}\mathrm{U}_{t,\infty } & 
, & \mathrm{U}_{\infty ,\infty }\doteq \mathbf{1} & , \\ 
\partial _{s}\mathrm{U}_{\infty ,s}=i\mathrm{U}_{\infty ,s}\mathrm{G}_{s} & ,
& \mathrm{U}_{\infty ,\infty }\doteq \mathbf{1} & ,%
\end{array}%
\right.
\end{equation*}%
on the domain $\mathcal{D}\left( \mathrm{N}\right) $ and the cocycle
property $\mathrm{U}_{t,s}=\mathrm{U}_{t,x}\mathrm{U}_{x,s}$ for $x,t,s\in 
\mathbb{R}_{0}^{+}\cup \left\{ \infty \right\} $. What is more, the extended
family of Theorem \ref{lemma H t infinity copy(2)} still realizes a
Bogoliubov $\mathbf{u}$--$\mathbf{v}$ transformation and the solution $%
\left( \mathrm{H}_{t}\right) _{t\in \mathbb{R}_{0}^{+}}$, defined by (\ref%
{DefH0bis}), to the Brockett-Wegner flow (\ref{brocket flow}) and its
asymptotics $\mathrm{H}_{\infty }$, defined by (\ref{Hinfini}) (see
Proposition \ref{lemma H t infinity copy(1)}), are unitarily equivalent:

\begin{corollary}[Implementation of non-autonomous Bogoliubov transformations%
]
\label{lemma uv-Bogoliubov transformation copy(1)}\mbox{}\newline
Take $D\in C(\mathbb{R}_{0}^{+};\mathcal{L}^{2}(\mathfrak{h}))$ with $%
D_{t}=-D_{t}^{\top }\in \mathcal{L}^{2}(\mathfrak{h})$ for all $t\in \mathbb{%
R}_{0}^{+}$ and assume the integrability of the mapping $t\mapsto \Vert
D_{t}\Vert _{2}$ on $\mathbb{R}_{0}^{+}$. Then, Proposition \ref{lemma
uv-Bogoliubov transformation} can be extended to all $s,t\in \mathbb{R}%
_{0}^{+}\cup \left\{ \infty \right\} $, where $\mathbf{u}_{\infty ,\infty
}\doteq \mathbf{1}$ and $\mathbf{v}_{\infty ,\infty }\doteq 0$.
\end{corollary}

\begin{proof}
If the mapping $t\mapsto \Vert D_{t}\Vert _{2}$ is integrable on $\mathbb{R}%
_{0}^{+}$ then the operator families $(\mathbf{u}_{t,s}-\mathbf{1})_{s,t\in 
\mathbb{R}_{0}^{+}\cup \left\{ \infty \right\} }$ and $(\mathbf{v}%
_{t,s})_{s,t\in \mathbb{R}_{0}^{+}\cup \left\{ \infty \right\} }$, with the
definition $\mathbf{u}_{\infty ,\infty }\doteq \mathbf{1}$ and $\mathbf{v}%
_{\infty ,\infty }\doteq 0$, are well-defined in $\mathcal{L}^{2}(\mathfrak{h%
})$ by (\ref{u})-(\ref{v}) and continuous in the Hilbert-Schmidt topology.\
Using the same arguments of Proposition \ref{lemma uv-Bogoliubov
transformation} one verifies (\ref{CCRu,v})--(\ref{bog2bis}) for all $s,t\in 
\mathbb{R}_{0}^{+}\cup \left\{ \infty \right\} $. Additionally, the bounded
operators (\ref{generalized Bog transf 0}) can also be extended to $s,t\in 
\mathbb{R}_{0}^{+}\cup \left\{ \infty \right\} $. In particular, Proposition %
\ref{lemma uv-Bogoliubov transformation} and Theorem \ref{lemma H t infinity
copy(2)} together with the integrability of the mapping $t\mapsto \Vert
D_{t}\Vert _{2}$ on $\mathbb{R}_{0}^{+}$ yield 
\begin{equation*}
\mathrm{U}_{\infty ,s}a\left( \varphi \right) \mathrm{U}_{\infty ,s}^{\ast
}=\lim_{t\rightarrow \infty }\mathrm{U}_{t,s}a\left( \varphi \right) \mathrm{%
U}_{t,s}^{\ast }=\lim_{t\rightarrow \infty }\left\{ a(\mathbf{u}%
_{t,s}\varphi )+a^{\ast }(\mathbf{v}_{t,s}\overline{\varphi })\right\} =a(%
\mathbf{u}_{\infty ,s}\varphi )+a^{\ast }(\mathbf{v}_{\infty ,s}\overline{%
\varphi })
\end{equation*}%
for all $\varphi \in \mathcal{F}$ and $s,t\in \mathbb{R}_{0}^{+}\cup \left\{
\infty \right\} $, in the strong operator topology.
\end{proof}

\begin{corollary}[$\mathrm{N}$--diagonalization of quadratic Hamiltonians]
\label{lemma uv-Bogoliubov transformation copy(2)}\mbox{}\newline
Let $D$ be the solution to (\ref{flow equation-quadratic deltabis}) with
initial condition $D_{0}=-D_{0}^{\top }\in \mathcal{L}^{2}(\mathfrak{h})$.
Assume the integrability\footnote{%
Again, this condition is clearly satisfied under the assumptions of Theorem %
\ref{lemma asymptotics1 copy(4)}.} of the mapping $t\mapsto \Vert D_{t}\Vert
_{2}$ on $\mathbb{R}_{0}^{+}$. Then, quadratic Hamiltonians defined by (\ref%
{DefH0bis}) satisfy 
\begin{equation*}
\mathrm{H}_{t}=\mathrm{U}_{t,s}\mathrm{\mathrm{H}}_{s}\mathrm{U}_{t,s}^{\ast
}\ ,\qquad s,t\in \mathbb{R}_{0}^{+}\cup \left\{ \infty \right\} \ ,
\end{equation*}%
with $\left( \mathrm{U}_{t,s}\right) _{s,t\in \mathbb{R}_{0}^{+}\cup \left\{
\infty \right\} }$ being the (evolution) family of unitary operators of
Theorem \ref{lemma H t infinity copy(2)}.
\end{corollary}

\begin{proof}
Combine Theorem \ref{theorem Ht=00003Dunitary orbite}, Proposition \ref%
{lemma H t infinity copy(1)} and Theorem \ref{lemma H t infinity copy(2)}
with elementary estimates using the triangle inequality. We omit the
details. Note only that the solution $D$ to the elliptic operator-valued
flow (\ref{flow equation-quadratic deltabis}) with initial condition $%
D_{0}=-D_{0}^{\top }\in \mathcal{L}^{2}(\mathfrak{h})$ satisfies $D\in C(%
\mathbb{R}_{0}^{+};\mathcal{L}^{2}(\mathfrak{h}))$ and $D_{t}=-D_{t}^{\top
}\in \mathcal{L}^{2}(\mathfrak{h})$ for all $t\in \mathbb{R}_{0}^{+}$,
thanks to Theorem \ref{Corollary existence}. We can in particular apply
Theorems \ref{theorem Ht=00003Dunitary orbite copy(1)} and \ref{lemma H t
infinity copy(2)} for this continuous family of Hilbert-Schmidt operators.
\end{proof}

\noindent Corollaries \ref{lemma uv-Bogoliubov transformation copy(1)} and %
\ref{lemma uv-Bogoliubov transformation copy(2)} mean in particular that 
\begin{equation*}
\mathrm{H}_{\infty }=\mathrm{U}_{\infty ,0}\mathrm{H}_{0}\mathrm{U}_{\infty
,0}^{\ast }
\end{equation*}%
is the result of the $\mathrm{N}$--diagonalization of fermionic quadratic
Hamiltonians via unitary Bogoliubov $\mathbf{u}$--$\mathbf{v}$
transformations. In contrast with Proposition \ref{lemma H t infinity
copy(1)} (which needs the \emph{square} integrability of the mapping $%
t\mapsto \Vert D_{t}\Vert _{2}$), the integrability of the mapping $t\mapsto
\Vert D_{t}\Vert _{2}$ is crucial here to get always the existence of the
limit operators $\mathrm{U}_{\infty ,0}$ and $\mathrm{U}_{\infty ,0}^{\ast }$%
.

\subsection{Appendix\label{Apprendix}}

\subsubsection{The Elliptic Operator-Valued Flow\label{Eliptic flow}}

In this appendix, we gather only the important results of \cite{EllipticFlow}
associated with its application to the $\mathrm{N}$--diagonalization of
fermionic quadratic Hamiltonians.

Recall first that the Banach space of bounded operators acting on $\mathfrak{%
h}$ is denoted by $\mathcal{B}(\mathfrak{h})$, while $\mathcal{L}^{1}(%
\mathfrak{h})$\ and $\mathcal{L}^{2}(\mathfrak{h})$ are the spaces of
trace-class and Hilbert-Schmidt operators, respectively. In the paper \cite%
{EllipticFlow} we study for lower semibounded self-adjoint operators $%
\Upsilon _{0}$ and (non-zero) bounded operators $D_{0}\in \mathcal{B}(%
\mathfrak{h})$ the non-linear system of differential equations 
\begin{equation}
\left\{ 
\begin{array}{llll}
\partial _{t}\Delta _{t}\varphi =16D_{t}D_{t}^{\ast }\varphi \ , & \Delta
_{t=0}\doteq 0\ , & t\in \mathbb{R}_{0}^{+}\ , & \varphi \in \mathfrak{h}\ ,
\\ 
\partial _{t}D_{t}\varphi =-2\left( \left( \Delta _{t}+\Upsilon _{0}\right)
D_{t}+D_{t}\left( \Delta _{t}+\Upsilon _{0}\right) ^{\top }\right) \varphi \
, & D_{t=0}\doteq D_{0}\ , & t\in \mathbb{R}^{+}\ , & \varphi \in \mathcal{D}%
(\Upsilon _{0}^{\top })\ ,%
\end{array}%
\right.  \label{flow equation-quadratic deltabis}
\end{equation}%
either on the Banach space $\mathcal{B}(\mathfrak{h})^{2}$ or in $\mathcal{L}%
^{1}(\mathfrak{h})\times \mathcal{L}^{2}(\mathfrak{h})$. Here, $\mathcal{D}%
(\Upsilon _{0}^{\top })\subseteq \mathfrak{h}$ is the (dense) domain of the
self-adjoint operator $\Upsilon _{0}^{\top }$. For the application to the $%
\mathrm{N}$--diagonalization of quadratic Hamiltonians, we only need the
results of \cite{EllipticFlow} in $\mathcal{L}^{2}(\mathfrak{h})$. In this
case, the existence of a global solution to the non-linear system of
differential equations (\ref{flow equation-quadratic deltabis}) is given by 
\cite[Theorems 1--2]{EllipticFlow}, which is reproduced below: \ 

\begin{theorem}[Well-posedness of the flow -- Hilbert-Schmidt topology]
\label{Corollary existence}\label{lemma existence 2 copy(4)}\mbox{}\newline
Assume $\Upsilon _{0}=\Upsilon _{0}^{\ast }\geq -\mu \mathbf{1}$ with $\mu
\in \mathbb{R}$ and $D_{0}\in \mathcal{L}^{2}(\mathfrak{h})$ ($D_{0}\neq 0$%
). Then, there exists a unique solution $(\Delta ,D)\in C(\mathbb{R}_{0}^{+};%
\mathcal{L}^{1}(\mathfrak{h})\times \mathcal{L}^{2}(\mathfrak{h}))$ to 
\begin{equation*}
\left\{ 
\begin{array}{llllll}
\partial _{t}\Delta _{t}=16D_{t}D_{t}^{\ast } & , & \quad \Delta
_{t=0}\doteq 0 & , & \quad t\in \mathbb{R}_{0}^{+} & , \\ 
\partial _{t}D_{t}=-2\left( \left( \Delta _{t}+\Upsilon _{0}\right)
D_{t}+D_{t}\left( \Delta _{t}+\Upsilon _{0}\right) ^{\top }\right) & , & 
\quad D_{t=0}\doteq D_{0} & , & \quad t\in \mathbb{R}^{+} & ,%
\end{array}%
\right.
\end{equation*}%
in $\mathcal{L}^{1}(\mathfrak{h})\times \mathcal{L}^{2}(\mathfrak{h})$,
i.e., in the trace class topology for $\Delta $, and in the Hilbert-Schmidt
topology for $D$. Additionally, if $D_{0}^{\top }=\pm D_{0}$ then $%
D_{t}^{\top }=\pm D_{t}$ for any $t\in \mathbb{R}_{0}^{+}$.
\end{theorem}

The assumption $D_{0}\in \mathcal{L}^{2}(\mathfrak{h})$ can be replaced with 
$D_{0}\in \mathcal{L}^{2p}(\mathfrak{h})$ for some $p\in \lbrack 1,\infty ]$%
, where $\mathcal{L}^{q}(\mathfrak{h})$ is the $L^{q}$--spaces constructed
from Schatten norms. In this case, the operator-valued flow is a well-posed
system of differential equations on $\mathcal{L}^{p}\left( \mathfrak{h}%
\right) \times \mathcal{L}^{2p}\left( \mathfrak{h}\right) $, at least when $%
\Upsilon _{0}\in \mathcal{B}(\mathfrak{h})$. See \cite[Theorem 14]%
{EllipticFlow}. However, the case $p=1$ is the only relevant one in view of
the application of the flow given here.

The convergence of the flow is a difficult task also solved in \cite%
{EllipticFlow} under sufficient, albeit still general, conditions. This
refers to \cite[Theorem 4]{EllipticFlow}, which corresponds to the following
theorem in the Hilbert-Schmidt case: \ 

\begin{theorem}[Asymptotics of the operator-valued flow]
\label{lemma asymptotics1 copy(4)}\mbox{}\newline
Take $D_{0}=\pm D_{0}^{\top }\in \mathcal{L}^{2}(\mathfrak{h})$ and $%
\Upsilon _{0}=\Upsilon _{0}^{\ast }$, both acting on $\mathfrak{h}$. Assume
that 
\begin{equation*}
\Upsilon _{0}\geq -\left( \mu -\varepsilon \right) \mathbf{1}\qquad \text{and%
}\qquad \mathfrak{D}_{0}\doteq \Upsilon _{0}+4D_{0}\left( \Upsilon
_{0}^{\top }+\mu \mathbf{1}\right) ^{-1}D_{0}^{\ast }\geq \mu \mathbf{1}
\end{equation*}%
for some $\mu \in \mathbb{R}\backslash \{0\}$ and $\varepsilon \in \mathbb{R}%
^{+}$. Then, as $t\rightarrow \infty $, $D$ \emph{exponentially} converges
in the Hilbert-Schmidt topology (i.e., in $\mathcal{L}^{2}(\mathfrak{h})$)
to zero, while $\Delta $ \emph{exponentially} converges in the trace norm
topology (i.e., in $\mathcal{L}^{1}(\mathfrak{h})$) to an operator $\Delta
_{\infty }\in \mathcal{L}^{1}(\mathfrak{h})$ satisfying 
\begin{equation*}
\Upsilon _{\infty }\doteq \Upsilon _{0}+\Delta _{\infty }\geq \left\vert \mu
\right\vert \mathbf{1}\ .
\end{equation*}
\end{theorem}

Compare the assumptions of this theorem with the ones of Theorem \ref{thm3}.

Note that \cite[Theorem 4]{EllipticFlow} gives more general results beyond
the Hilbert-Schmidt topology. For more details as well as additional
discussions on the above assumptions, see \cite[Section 2.3]{EllipticFlow}.

\subsubsection{Paradigmatic Example in Condensed Matter Physics\label%
{exemplesuperconductivity}}

Hamiltonians which are quadratic in terms of annihilation and creation
operators appear very often in theoretical physics. Here, we shortly present
the paradigmatic example of the BCS theory of superconductivity, proposed in
the late 1950s (1957) to explain conventional type I superconductors and
named after Bardeen, Cooper and Schrieffer \cite{BCS1,BCS2,BCS3}. Indeed,
this theory implies the first use of a fermionic quadratic Hamiltonian in
Physics, even if it is very elementary as compared to the general quadratic
Hamiltonians studied here.

An important model (of the lattice version) of the BCS theory is given by
the so-called (reduced) BCS\ Hamiltonian defined as follows: In solid-state
physics, quantum systems are usually done on crystalline structures. Having
in mind a traditional view of crystals as objects with a regular spatial
order, we consider fermions inside a cubic box $\Lambda _{L}\doteq \{\mathbb{%
Z}\cap \left[ -L,L\right] \}^{d}$ of volume $|\Lambda _{L}|$ for $L\in 
\mathbb{N}$, where $d\in \mathbb{N}$. It means that the separable
one-particle Hilbert space is $\mathfrak{h}=\ell ^{2}(\Lambda _{L}\times 
\mathrm{S})$, where $\mathrm{S}$ is some finite set of spins that is fixed
once and for all by $\mathrm{S}\doteq \left\{ \downarrow ,\uparrow \right\} $%
. The orthonormal basis of this Hilbert space we use is defined by the
functions 
\begin{equation}
\varphi _{\left( k,\mathrm{s}\right) }\left( x,\mathrm{t}\right) \doteq 
\frac{1}{\left\vert \Lambda _{L}\right\vert ^{1/2}}\mathrm{e}^{-ik\cdot
x}\delta _{\mathrm{s},\mathrm{t}}\ ,\qquad x\in \Lambda _{L},\ \mathrm{t}\in 
\mathrm{S}\ ,  \label{basis}
\end{equation}%
for any $k\in \Lambda _{L}^{\ast }$ and $\mathrm{s}\in \mathrm{S}$, where 
\begin{equation*}
\Lambda _{L}^{\ast }\doteq \frac{2\pi }{(2L+1)}\Lambda _{L}\subseteq \left[
-\pi ,\pi \right] ^{d}
\end{equation*}%
is the reciprocal lattice of quasi-momenta (periodic boundary conditions)
associated with $\Lambda _{L}$. The orthonormal basis taken in this example
is not real. Reality of orthonormal bases is a condition used in Section \ref%
{Diagonalization of Quadratic}. This assumption is however not necessary and
is only used for convenience, in order to slightly simplify our computations
by avoiding various conjugate functions in our expressions. See Equation (%
\ref{dddddd}). One could anyway use the real and imaginary parts of each
function (\ref{basis}) to define the real orthonormal basis, and thus obtain
a new form of the Hamiltonian presented below. But we stick to the complex
basis to perform the computations since it is the natural (particularly in
Physics) and original framework to describe the BCS theory.

The reduced BCS\ Hamiltonian is a self-adjoint operator acting on the
fermionic Fock space $\mathcal{F}$ (\ref{Fock}) and defined by 
\begin{equation}
\mathrm{H}_{L}^{BCS}\doteq \sum\limits_{k\in \Lambda _{L}^{\ast },\ \mathrm{s%
}\in \mathrm{S}}\left( \varepsilon _{k}-\kappa \right) a_{(k,\mathrm{s}%
)}^{\ast }a_{(k,\mathrm{s})}-\frac{1}{\left\vert \Lambda _{L}\right\vert }%
\sum_{k,q\in \Lambda _{L}^{\ast }}\gamma _{k,q}a_{(k,\uparrow )}^{\ast
}a_{(-k,\downarrow )}^{\ast }a_{(-q,\downarrow )}a_{(q,\uparrow )}\ ,
\label{BCS Hamilt}
\end{equation}%
where $\gamma _{k,q}$ is a positive function and $\kappa \in \mathbb{R}$ is
the chemical potential, while $\varepsilon _{k}=\varepsilon _{-k}\in \mathbb{%
R}_{0}^{+}$ is the kinetic energy\footnote{%
In the lattice $\mathbb{Z}^{d}$, usually, $\varepsilon _{k}=2\left( d-\left(
\cos (k_{1})+\cdots +\cos (k_{d})\right) \right) $ is the one-particle
energy spectrum in the quasi-momenta $k=(k_{1},\ldots ,k_{d})\in \Lambda
_{L}^{\ast }$ of free fermions.} of a fermion in the quasi-momenta $k\in
\Lambda _{L}^{\ast }$. In physics, one usually takes 
\begin{equation*}
\gamma _{k,q}\doteq \left\{ 
\begin{array}{l}
\gamma \geq 0 \\ 
0%
\end{array}%
\begin{array}{l}
\mathrm{for\ }\left\vert k-q\right\vert \leq \mathrm{C} \\ 
\mathrm{for\ }\left\vert k-q\right\vert >\mathrm{C}%
\end{array}%
\right.
\end{equation*}%
with $\mathrm{C}\in \left( 0,\infty \right] $. The simple choice $\mathrm{C}%
=\infty $, i.e., $\gamma _{k,q}=\gamma >0$ in (\ref{BCS Hamilt}), is still
physically very interesting since, even when $\varepsilon _{k}=0$, the BCS\
Hamiltonian qualitatively displays most of basic properties of real
conventional type I superconductors. See, e.g. \cite[Chapter VII, Section 4]%
{Thou} or \cite{BruPedra1} for a more recent mathematically rigorous study.

In fact, the thermodynamic and dynamical behaviors of the quantum system
described by the (mean-field) model (\ref{BCS Hamilt}) for $\gamma
_{k,q}=\gamma >0$ can be \emph{rigorously}\footnote{%
See \cite[Section 6.6.2]{Bru-Pedra-livre} (static case) and \cite%
{Bru-pedra-MF-IV} (dynamical case) for a first overview or \cite%
{BruPedra2,BruPedra-MFII,BruPedra-MFIII} for the full theory of mean-field
models for lattice fermions and quantum spin systems.} obtained in infinite
volume $L\rightarrow \infty $ by using the approximating model 
\begin{equation}
\mathrm{H}_{L}^{BCS}(c)=\sum\limits_{k\in \Lambda _{L}^{\ast },\ \mathrm{s}%
\in \mathrm{S}}\left( \varepsilon _{k}-\kappa \right) a_{(k,\mathrm{s}%
)}^{\ast }a_{(k,\mathrm{s})}-\gamma \sum_{k\in \Lambda _{L}^{\ast }}\left(
ca_{(k,\uparrow )}^{\ast }a_{(-k,\downarrow )}^{\ast }+\overline{c}%
a_{(k,\downarrow )}a_{(-k,\uparrow )}\right) +\gamma \left\vert \Lambda
_{L}\right\vert \left\vert c\right\vert ^{2}\mathbf{1}  \label{BCSlimit0}
\end{equation}%
for any complex number $c\in \mathbb{C}$. Note that the above Hamiltonian
can be written in the real orthonormal basis made of the real and imaginary
parts of $\varphi _{\left( k,\mathrm{s}\right) }$ for $k\in \Lambda
_{L}^{\ast }$ and $\mathrm{s}\in \{\uparrow ,\downarrow \}$, since the
mapping $\varphi \mapsto a\left( \varphi \right) $ is antilinear. It is a
quadratic model in terms of annihilation and creation operators that can be
diagonalized and written as 
\begin{equation}
H_{L}^{\mathrm{BCS}}(c)=\sum_{k\in \Lambda _{L}^{\ast },\ \mathrm{s}\in 
\mathrm{S}}\sqrt{\left( \varepsilon _{k}-\kappa \right) ^{2}+\gamma
^{2}|c|^{2}}\ b_{(k,\mathrm{s})}^{\ast }b_{(k,\mathrm{s})}+E\mathbf{1}
\label{BCSlimit}
\end{equation}%
for $c\in \mathbb{C}\backslash \{0\}$, where the ground state energy $E$ is
given by%
\begin{equation}
E\doteq \gamma |\Lambda _{L}||c|^{2}+\sum_{k\in \Lambda _{L}^{\ast }}\left(
\varepsilon _{k}-\kappa -\sqrt{\left( \varepsilon _{k}-\kappa \right)
^{2}+\gamma ^{2}|c|^{2}}\right) \ .  \label{BCSlimit2}
\end{equation}

This is done via a Bogoliubov $\ast $--automorphism uniquely defined by the
algebraic relation 
\begin{equation}
\begin{cases}
b_{(k,\uparrow )}=u_{k}a_{(k,\uparrow )}-v_{k}a_{(-k,\downarrow )}^{\ast }\ ,
\\ 
b_{(-k,\downarrow )}^{\ast }=v_{k}^{\ast }a_{(k,\uparrow )}+u_{k}^{\ast
}a_{(-k,\downarrow )}^{\ast }\ ,%
\end{cases}
\label{ssdsdsd}
\end{equation}%
where $u_{k}$ and $v_{k}$ are the complex numbers%
\begin{equation*}
u_{k}=\frac{\mathrm{e}^{i\arg c}}{\sqrt{2}}\sqrt{1+\frac{\varepsilon
_{k}-\kappa }{\sqrt{\left( \varepsilon _{k}-\kappa \right) ^{2}+\gamma
^{2}|c|^{2}}}}\quad \text{and}\quad v_{k}=\frac{\mathrm{e}^{i\arg c}}{\sqrt{2%
}}\sqrt{1-\frac{\varepsilon _{k}-\kappa }{\sqrt{\left( \varepsilon
_{k}-\kappa \right) ^{2}+\gamma ^{2}|c|^{2}}}}
\end{equation*}%
for each $k\in \Lambda _{L}^{\ast }$ and $c\in \mathbb{C}\backslash \{0\}$.
To get (\ref{BCSlimit}), it suffices to inverse the system (\ref{ssdsdsd})
and substitute the corresponding equalities into (\ref{BCSlimit0}). Such
kind of algebraic manipulation was originally invented for bosons by
Bogoliubov \cite{Bogoliubov1} in 1947 and it is a key ingredient in the
Bogoliubov theory of superfluidity for Helium 4. For fermions, such
quadratic Hamiltonians are at the origin of the BCS theory of
superconductivity done in 1957 \cite{BCS1,BCS2,BCS3} and their
diagonalization has been performed in 1958 \cite{BogoforBCS,ValatinBogo} via
a fermionic version of Bogoliubov's transformation. In modern words, the
above algebraic relations are nothing else than the implementation of a
Bogoliubov transformation, as given by Theorem \ref{Shale-Stinespring THM}.
Compare for instance (\ref{ssdsdsd}) with Theorem \ref{Shale-Stinespring THM}%
.

In our framework, it refers to applying Theorem \ref{thm3} and Equation (\ref%
{constant of motion eq2bis}) for $E_{0}=\gamma |\Lambda _{L}||c|^{2}$ and
operators $\Upsilon _{0}$,$D_{0}$ acting on the $4$-dimensional Hilbert
space with orthonormal basis 
\begin{equation*}
\{\varphi _{(-k,\downarrow )},\varphi _{(-k,\uparrow )},\varphi
_{(k,\downarrow )},\varphi _{(k,\uparrow )}\}\ ,
\end{equation*}%
in each fiber indexed by the set $\left\{ k,-k\right\} $ for $k\in \Lambda
_{L}^{\ast }$. In this orthonormal basis, $\Upsilon _{0}$ and $D_{0}$ take
the following matrix forms:%
\begin{equation}
\Upsilon _{0}=\left( 
\begin{array}{cccc}
\varepsilon _{k}-\kappa & 0 & 0 & 0 \\ 
0 & \varepsilon _{k}-\kappa & 0 & 0 \\ 
0 & 0 & \varepsilon _{k}-\kappa & 0 \\ 
0 & 0 & 0 & \varepsilon _{k}-\kappa%
\end{array}%
\right) \quad \text{and}\quad D_{0}=\frac{1}{2}\left( 
\begin{array}{cccc}
0 & 0 & 0 & \gamma c \\ 
0 & 0 & \gamma c & 0 \\ 
0 & -\gamma c & 0 & 0 \\ 
-\gamma c & 0 & 0 & 0%
\end{array}%
\right)  \label{seeeee}
\end{equation}%
for $\varepsilon _{k}\in \mathbb{R}_{0}^{+}$, $\kappa \in \mathbb{R}$, $%
\gamma \in \mathbb{R}^{+}$ and $c\in \mathbb{C}$. See (\ref{DefH0}) and (\ref%
{BCSlimit0}). Note in particular that $D_{0}=-D_{0}^{\top }$, $\Upsilon
_{0}=\Upsilon _{0}^{\top }=\Upsilon _{0}^{\ast }$ and $\Upsilon
_{0}D_{0}=D_{0}\Upsilon _{0}^{\top }$. Additionally, explicit computations
show that (\ref{condition plus}) is equivalent in this elementary case to
find $\mu \in \mathbb{R}^{+}$ such that 
\begin{equation*}
\kappa -\varepsilon _{k}<\mu \leq \sqrt{\left( \varepsilon _{k}-\kappa
\right) ^{2}+\left\vert c\right\vert ^{2}\gamma ^{2}}\ .
\end{equation*}%
Avoiding the trivial case $c=0$ for which $D_{0}=0$, such $\mu $ always
exists for $\varepsilon _{k}\in \mathbb{R}_{0}^{+}$, $\kappa \in \mathbb{R}$%
, $\gamma \in \mathbb{R}^{+}$ and $c\in \mathbb{C}\backslash \{0\}$. As a
consequence, one can satisfy all the assumptions of Theorem \ref{thm3} in
this case and use Equations (\ref{constant of motion eq2bis}) and (\ref%
{dfgdgdfgdfg}) to deduce in each fiber $\left\{ k,-k\right\} $ that 
\begin{equation*}
\Upsilon _{\infty }=\sqrt{\Upsilon _{0}^{2}+4D_{0}D_{0}^{\ast }}=\sqrt{%
\left( \varepsilon _{k}-\kappa \right) ^{2}+\gamma ^{2}|c|^{2}}\ \mathbf{1}
\end{equation*}%
as well as 
\begin{equation*}
-8\int_{0}^{\infty }\left\Vert D_{\tau }\right\Vert _{2}^{2}\mathrm{d}\tau =%
\frac{1}{2}\mathrm{tr}\left( \Upsilon _{0}-\sqrt{\Upsilon
_{0}^{2}+4D_{0}D_{0}^{\ast }}\right) =2\left( \varepsilon _{k}-\kappa -\sqrt{%
\left( \varepsilon _{k}-\kappa \right) ^{2}+\gamma ^{2}|c|^{2}}\right)
\end{equation*}%
for any $\varepsilon _{k}\in \mathbb{R}_{0}^{+}$, $\kappa \in \mathbb{R}$, $%
\gamma \in \mathbb{R}^{+}$ and $c\in \mathbb{C}\backslash \{0\}$. To recover
the results (\ref{BCSlimit})--(\ref{BCSlimit2}) it then suffices to sum up
over the set $\left\{ k,-k\right\} _{k\in \Lambda _{L}^{\ast }}$, keeping in
mind the explicit form of the Hamiltonian \textrm{H}$_{\infty }$ in Theorem %
\ref{thm3}.

Last but not least, observe that the asssumptions of Berezin's theorem
(Theorem \ref{thm 1}) do not always hold true even in this elementary
situation because the condition $\Upsilon _{0}\geq \alpha \mathbf{1}$ for a 
\emph{strictly positive} $\alpha \in \mathbb{R}^{+}$ in Theorem \ref{thm 1}
is broken as soon as $\kappa \geq \varepsilon _{k}$.

This concludes the section on the $\mathrm{N}$--diagonalization of quadratic
fermionic Hamiltonians via a non-trivial evolution equation for unbounded
operators. 

\section{Alternative Approaches on Quadratic Hamiltonians\label%
{Diagonalization of Quadraticbis}}

\subsection{Araki's Approach on the Level of $C^{\ast }$-Algebras\label%
{Araiki section}}

In 1968, Araki established a very general method for some \textquotedblleft $%
\mathrm{N}$--diagonalization\textquotedblright\ of bilinear Hamiltonians.
This refers to the well-known paper \cite{Araki}. Such a diagonalization
holds true in great generality. The Hilbert-Schmidt condition $D_{0}\in 
\mathcal{L}^{2}(\mathfrak{h})$ and the semiboundedness of $\Upsilon _{0}$
are not even required in \cite[Theorem 5.4]{Araki}. This is because the
concepts studied in Araki's paper \cite{Araki} are different than those
analyzed here, albeit still related. Araki's terminology and notation used
in \cite{Araki} are almost identical to the ones used in other works like in
Berezin's book \cite{Berezin}, but they refer to differents concepts or
frameworks. For instance, Araki's study \cite{Araki} uses a purely $C^{\ast
} $-algebraic formulation, while Berezin and we work on the Fock space
representation. See also \cite{Araki87} for more details on Araki's works in
relation with the Fock space representation.

In the sequel we give a brief synthesis of Araki's results \cite{Araki},
highlighting the similarities and differences with the work done here.
Hopefully, this section allows the non-expert reader to have an overview on
quadratic/bilinear fermionic models and prevents them from possible
misunderstandings. This section also allows us to introduce the celebrated
Shale-Stinespring condition and the standard concept of implementable
Bogoliubov transformations, which is useful afterwards and, in particular,
important to keep in mind in the more general (re)definition of quadratic
Hamiltonians in Section \ref{Section Bach}. See in particular Definition \ref%
{def quadratic}.

The paper \cite{Araki} has an algebraic approach based on self-dual CAR $%
C^{\ast }$-algebras defined as follows: Let $\mathcal{\mathcal{H}}$ be a
complex Hilbert space with either even or infinite dimension. Let $\mathfrak{%
A}$ be an antiunitary involution on $\mathcal{\mathcal{H}}$, see Remark \ref%
{remark idiote copy(1)}. A self-dual CAR algebra 
\begin{equation}
\mathcal{A}\equiv (\mathcal{A},+,\cdot ,\ast ,\Vert \cdot \Vert )  \label{A}
\end{equation}%
is the universal\footnote{%
For more details on the universal $C^{\ast }$-algebras of family of
polynomial relations, see \cite[Section 4.8]{Bru-Pedra-livre}. In the case
of self-dual CAR algebras, they are uniquely defined up to Bogoliubov $\ast $%
--automorphisms. See \cite[Section 4.8.3]{Bru-Pedra-livre}.} $C^{\ast }$%
-algebra generated by a unit $\mathfrak{1}$ and a family $\{\mathrm{B}(\psi
)\}_{\psi \in \mathcal{\mathcal{H}}}$ of elements satisfying the following
conditions:

\begin{enumerate}
\item[(a)] The mapping $\psi \mapsto \mathrm{B}\left( \psi \right) ^{\ast }$
is (complex) linear.

\item[(b)] $\mathrm{B}(\psi )^{\ast }=\mathrm{B}(\mathfrak{A}(\psi ))$ for
any $\psi \in \mathcal{H}$.

\item[(c)] The family $\{\mathrm{B}(\psi )\}_{\psi \in \mathcal{\mathcal{H}}%
} $ satisfies (a part of the) CAR\footnote{%
The use of \textquotedblleft CAR\textquotedblright\ is a slight misuse of
language since $\mathrm{B}(\varphi _{1})\mathrm{B}(\varphi _{2})+\mathrm{B}%
(\varphi _{2})\mathrm{B}(\varphi _{1})=0$ for $\varphi _{1},\varphi _{2}\in 
\mathcal{\mathcal{H}}$ does not hold true, in general.}: For any $\psi
_{1},\psi _{2}\in \mathcal{\mathcal{H}}$,%
\begin{equation}
\mathrm{B}(\psi _{1})\mathrm{B}(\psi _{2})^{\ast }+\mathrm{B}(\psi
_{2})^{\ast }\mathrm{B}(\psi _{1})=\left\langle \psi _{1},\psi
_{2}\right\rangle _{\mathcal{H}}\,\mathfrak{1}\ .  \label{CAR Grassmann III}
\end{equation}
\end{enumerate}

\noindent Note that the elements $\mathrm{B}(\psi )$, $\psi \in \mathcal{%
\mathcal{H}}$, are the so-called \emph{fields}, from which creation and
annihilation elements can be defined, see for instance (\ref{ghghghg2}) in
the Fock space representation. By the CAR\ (\ref{CAR Grassmann III}), the
antilinear mapping $\varphi \mapsto \mathrm{B}\left( \varphi \right) $ from $%
\mathcal{\mathcal{H}}$ to $\mathcal{A}$ is necessarily injective and
contractive.

For our setting (Section \ref{Diagonalization of Quadratic}), one takes%
\footnote{%
Note that it is common to use instead $\mathcal{H}=\mathfrak{h}\oplus 
\mathfrak{h}^{\ast }$, but we decide to keep Araki's choice \cite[Remark 5.3]%
{Araki} for better comparison. Here, $\oplus $ stands, as in all the paper,
for the (Hilbert) direct sums of Hilbert (sub)spaces (and not the algebraic
direct sum).} $\mathcal{H}=\mathfrak{h}\oplus \mathfrak{h}$ and the
canonical antiunitary involution of $\mathcal{H}$ is defined from the
complex conjugation $\mathcal{C}$ (Remark \ref{remark idiote copy(1)}) by 
\begin{equation}
\mathfrak{A}=\left( 
\begin{array}{cc}
0 & \mathcal{C} \\ 
\mathcal{C} & 0%
\end{array}%
\right) \ .  \label{ghghghg}
\end{equation}%
By contrast, we use in this paper the Fock space representation of such $%
C^{\ast }$-algebras, which is the representation\footnote{%
A representation on the Hilbert space $\mathcal{X}$ of a $C^{\ast }$%
--algebra $\mathcal{A}$ is, by definition \cite[Definition 2.3.2]%
{BratteliRobinsonI}, a $\ast $--homomorphism $\mathbf{\pi }$ from $\mathcal{A%
}$ to the unital $C^{\ast }$--algebra $\mathcal{B}(\mathcal{X})$ of all
bounded linear operators acting on $\mathcal{X}$.} $\mathbf{\pi }_{\mathcal{F%
}}$ on the fermionic Fock space $\mathcal{F}$ (\ref{Fock}) of the self-dual
CAR$\ $algebra $\mathcal{A}$, uniquely defined by 
\begin{equation}
\mathbf{\pi }_{\mathcal{F}}\left( \mathrm{B}\left( \psi \right) \right) =a(%
\mathrm{\psi }_{1})+a(\overline{\mathrm{\psi }}_{2})^{\ast }\ ,\qquad \psi =(%
\mathrm{\psi }_{1},\mathrm{\psi }_{2})\in \mathcal{\mathcal{H}}\ ,
\label{ghghghg2}
\end{equation}%
$a\left( \varphi \right) $ being the usual annihilation operator associated
with $\varphi \in \mathfrak{h}$ and acting on the (fermionic) Fock space $%
\mathcal{F}$.

Bilinear Hamiltonians used in \cite{Araki} are analogous to quadratic
operators studied here. Nonetheless, they are \textbf{no longer}
Hamiltonians, i.e., self-adjoint operators, acting on some Hilbert space,
like the Fock space $\mathcal{F}$. In \cite{Araki} a bilinear Hamiltonian is
a generator of a one-parameter automorphism group of the CAR $C^{\ast }$%
-algebra:

\begin{itemize}
\item A \emph{Bogoliubov transformation} is a unitary operator $\mathcal{U}%
\in \mathcal{B}(\mathcal{H})$ such that $\mathcal{U}\mathfrak{A=A}\mathcal{U}
$. A self-dual Hamiltonian $h$ on $\mathcal{H}$ is a self-adjoint operator
satisfying $h=-\mathfrak{A}h\mathfrak{A}$ with domain $\mathcal{D}(h)=%
\mathfrak{A}(\mathcal{D}(h))\subseteq \mathcal{\mathcal{H}}$. Basis
projections are orthogonal projections $P\in \mathcal{B}(\mathcal{H})$\
satisfying $\mathfrak{A}P\mathfrak{A}=P^{\bot }\doteq \mathbf{1}_{\mathcal{H}%
}-P$.

\item The Bogoliubov $\ast $--automorphism associated with a Bogoliubov
transformation $\mathcal{U}\in \mathcal{B}(\mathcal{H})$ is the unique $\ast 
$--automorphism $\mathbf{\chi }_{\mathcal{U}}$ of $\mathcal{A}$ satisfying 
\begin{equation}
\mathbf{\chi }_{\mathcal{U}}\left( \mathrm{B}(\varphi )\right) =\mathrm{B}(%
\mathcal{U}\varphi )\ ,\qquad \varphi \in \mathcal{\mathcal{H}}\ .
\label{def bogo}
\end{equation}%
See for instance \cite[Corollary 4.211]{Bru-Pedra-livre}. Given a self-dual
Hamiltonian $h$, $(\mathbf{\chi }_{\mathrm{e}^{ith}})_{t\in {\mathbb{R}}}$
is a strongly continuous group of Bogoliubov $\ast $--auto%
\-%
morphisms of $\mathcal{A}$.

\item It is known (see, e.g. \cite[Lemma 1]{EngelNagel} when $h\in \mathcal{B%
}(\mathcal{H})$) that the above group is generated by a symmetric derivation%
\footnote{%
Symmetric derivations $\delta $ refer to (linear) operators satisfying $%
\delta (A^{\ast })=\delta (A)^{\ast }$\ and $\delta (AB)=\delta (A)B+A\delta
(B)$ on $\mathcal{A}$.} $\delta _{h}$ acting on the $C^{\ast }$--algebra $%
\mathcal{A}$, \textbf{formally} denoted by 
\begin{equation}
\delta _{h}\left( \cdot \right) \equiv i\left[ \mathrm{H},\ \cdot \ \right]
\ ,  \label{symetric derivation}
\end{equation}%
keeping in mind that $[\cdot ,\cdot ]$ usually denotes a commutator.
Bilinear Hamiltonians are by definition such a derivation, or such a
\textquotedblleft $\mathrm{H}$\textquotedblright . The bilinear Hamiltonian
\textquotedblleft $\mathrm{H}$\textquotedblright\ is said hermitian or
self-adjoint because $h=h^{\ast }$. See \cite[Definition 5.1]{Araki}.

\item The diagonalization of bilinear Hamiltonians in Araki's sense \cite[%
Definition 5.1]{Araki} then only means the existence of a basis projection $%
P $ with range $\mathfrak{h}_{P}$, or a Bogoliubov transformation (cf. \cite[%
Lemma 3.6]{Araki}), such that the self-dual Hamiltonian $h$ has the form 
\begin{equation*}
\left( 
\begin{array}{cc}
h_{1,1} & 0 \\ 
0 & h_{2,2}%
\end{array}%
\right)
\end{equation*}%
on $\mathcal{H}=\mathfrak{h}_{P}\oplus \mathfrak{h}_{P^{\bot }}$, where $%
h_{1,1},h_{2,2}$ are self-adjoint operators acting on $\mathfrak{h}_{P},%
\mathfrak{h}_{P^{\bot }}$, respectively. Applying the spectral theorem to
the self-dual Hamiltonian $h$, Araki shows in \cite[Theorem 5.4]{Araki} that
this is always possible iff the kernel of $h$ has dimension even or
infinite, otherwise the same holds true but with a so-called extended
Bogoliubov transformation \cite[Theorem 5.6]{Araki}.
\end{itemize}

The motivation of these definitions comes from the fact that such a $\mathrm{%
H}$ in Araki's viewpoint can be seen in some very restricted situation as a
quadratic Hamiltonian in our sense. See, e.g., \cite[Remark 4.3]{Araki} and 
\cite[Sections 2.2-2.3]{LD1}, where $\mathrm{H}$ is shown to exist as a
self-adjoint element of the $C^{\ast }$-algebra $\mathcal{A}$ for
trace-class Hamiltonians $h$. However, in general, \textquotedblleft $%
\mathrm{H}$\textquotedblright\ is only a formal notation, without
necessarily another meaning than denoting the symmetric derivation via (\ref%
{symetric derivation}). In other words, \cite{Araki} analyzes another
situation and, in the general case, it is an open problem to show the
existence of some well-defined, self-adjoint, quadratic Hamiltonian that
leads to $\delta _{h}$ in some representation. In the Fock space
representation, one can answer this question provided (\ref{DefH0}) defines
a self-adjoint operator on the Fock space:\ 

\begin{lemma}[From quadratic Hamiltonians to quasi-free dynamics]
\label{Lemma quasi-free}\mbox{
}\newline
Fix $E_{0}\in \mathbb{R}$. Let $\Upsilon _{0}=\Upsilon _{0}^{\ast }$ and $%
D_{0}=-D_{0}^{\top }\in \mathcal{L}^{2}(\mathfrak{h})$ be two operators
acting on the Hilbert space $\mathfrak{h}$. Then, for any $t\in {\mathbb{R}}$
and $A\in \mathcal{A}$,%
\begin{equation*}
\mathbf{\pi }_{\mathcal{F}}\circ \mathbf{\chi }_{\mathrm{e}^{ith}}\left(
A\right) =\mathrm{e}^{it\mathrm{H}_{0}/2}\mathbf{\pi }_{\mathcal{F}}\left(
A\right) \mathrm{e}^{-it\mathrm{H}_{0}/2}\ ,
\end{equation*}%
where $h$ is the self-dual Hamiltonian defined on $\mathcal{H}=\mathfrak{h}%
\oplus \mathfrak{h}$ by 
\begin{equation}
h\doteq \frac{1}{2}\left( 
\begin{array}{cc}
\Upsilon _{0} & 2D_{0} \\ 
-2\overline{D}_{0} & -\Upsilon _{0}^{\top }%
\end{array}%
\right) =\frac{1}{2}\left( 
\begin{array}{cc}
\Upsilon _{0} & 2D_{0} \\ 
2D_{0}^{\ast } & -\Upsilon _{0}^{\top }%
\end{array}%
\right) \ .  \label{sdsdsdsdsdsdsdsd}
\end{equation}
\end{lemma}

\begin{proof}
Fix all parameters of the lemma. By Proposition \ref{Hamilselfadjoint},
Equations (\ref{DefH0})--(\ref{domain H0}) define a self-adjoint operator $%
\mathrm{H}_{0}$ with the dense subset $\mathcal{D}_{0}\subseteq \mathcal{F}$
as a core. In particular, $\mathrm{e}^{it\mathrm{H}_{0}/2}$ is well defined
for all times $t\in {\mathbb{R}}$ (via the functional calculus). The
corresponding self-dual Hamiltonian $h$ acting on $\mathcal{H}=\mathfrak{h}%
\oplus \mathfrak{h}$ which would formally lead to our quadratic Hamiltonian (%
\ref{DefH0}) (with $E_{0}=0$, the constant being irrelevant here) is equal
to (\ref{sdsdsdsdsdsdsdsd}), see (\ref{ghghghg2}) and \cite[Equation (5.6)]%
{Araki}. Note in this case that $h=h^{\ast }=-\mathfrak{A}h\mathfrak{A}$,
thanks to $\Upsilon _{0}=\Upsilon _{0}^{\ast }$, $D_{0}=-D_{0}^{\top }$, $%
\mathcal{C}^{2}=\mathbf{1}$, $X^{\top }\doteq \mathcal{C}X^{\ast }\mathcal{C}
$, $\overline{X}\doteq \mathcal{C}X\mathcal{C}$ (Remark \ref{remark idiote
copy(1)}) and Equation (\ref{ghghghg}). To prove now the assertion, it
suffices to show that 
\begin{equation}
\mathbf{\pi }_{\mathcal{F}}\circ \mathbf{\chi }_{\mathrm{e}^{ith}}\left( 
\mathrm{B}(\psi )\right) =\mathrm{e}^{it\mathrm{H}_{0}/2}\mathbf{\pi }_{%
\mathcal{F}}\left( \mathrm{B}\left( \psi \right) \right) \mathrm{e}^{-it%
\mathrm{H}_{0}/2}  \label{dfgdfgdffgd}
\end{equation}%
for any $\psi \in \mathcal{H}=\mathfrak{h}\oplus \mathfrak{h}$. The proof is
relatively standard in its principle. See for instance \cite[Lemma 2.8]{LD1}%
. We outline it for completeness. As usual, $[A,B]\doteq AB-BA$ is the
commutator of two elements $A,B\in \mathcal{A}$. Using the CAR (\ref{CAR})
(or (\ref{CAR Grassmann III})), Equation (\ref{ghghghg2}) together with the
properties of the antiunitary involution $\mathfrak{A}$, $h^{\ast }=-%
\mathfrak{A}h\mathfrak{A}$ and $\mathrm{B}(\psi )^{\ast }=\mathrm{B}(%
\mathfrak{A}(\psi ))$, we compute that 
\begin{equation}
\frac{1}{2}\left[ \mathrm{H}_{0},\mathbf{\pi }_{\mathcal{F}}\left( \mathrm{B}%
\left( \psi \right) \right) ^{\ast }\right] \varphi =\mathbf{\pi }_{\mathcal{%
F}}\left( \mathrm{B}\left( h\psi \right) \right) ^{\ast }\varphi \ ,
\label{tototototot}
\end{equation}%
for any $\psi \in \mathcal{D}\left( \Upsilon _{0}\right) \oplus \mathcal{D}%
\left( \Upsilon _{0}^{\top }\right) $ and $\varphi \in \mathcal{D}_{0}$,
with $\mathcal{D}\left( \Upsilon _{0}\right) ,\mathcal{D}\left( \Upsilon
_{0}^{\top }\right) \subseteq \mathfrak{h}$ being the dense domains of $%
\Upsilon _{0}$ and $\Upsilon _{0}^{\top }$, respectively, while $\mathcal{D}%
_{0}\subseteq \mathcal{F}$ is the dense subset defined by (\ref{domain H0}).
This statement is analogous to \cite[Equations (4.15)--(4.16)]{Araki} done
on the level of the algebra $\mathcal{A}$ for trace-class operators $h$ (to
obtain a true bilinear Hamiltonian $\mathrm{H}\in \mathcal{A}$). A
representation is automatically contractive \cite[Proposition 2.3.1]%
{BratteliRobinsonI} and even isometric when it is injective \cite[%
Proposition 2.3.3]{BratteliRobinsonI}. Meanwhile, the antilinear mapping $%
\varphi \mapsto \mathrm{B}\left( \varphi \right) $ on $\mathcal{H}$ is also
contractive. Together with (\ref{def bogo}) and (\ref{tototototot}), it
follows that, for any $t\in \mathbb{R}$, $\psi \in \mathcal{D}\left(
\Upsilon _{0}\right) \oplus \mathcal{D}\left( \Upsilon _{0}^{\top }\right) $
and $\varphi \in \mathcal{D}_{0}$, 
\begin{equation*}
\partial _{t}\left\{ \mathrm{e}^{-it\mathrm{H}_{0}/2}\mathbf{\pi }_{\mathcal{%
F}}\circ \mathbf{\chi }_{\mathrm{e}^{ith}}\left( \mathrm{B}\left( \psi
\right) \right) ^{\ast }\mathrm{e}^{it\mathrm{H}_{0}/2}\varphi \right\} =0\ ,
\end{equation*}%
which in turn implies that%
\begin{equation}
\mathbf{\pi }_{\mathcal{F}}\circ \mathbf{\chi }_{\mathrm{e}^{ith}}\left( 
\mathrm{B}(\psi )\right) \varphi =\mathrm{e}^{it\mathrm{H}_{0}/2}\mathbf{\pi 
}_{\mathcal{F}}\left( \mathrm{B}\left( \psi \right) \right) \mathrm{e}^{-it%
\mathrm{H}_{0}/2}\varphi  \label{sdsdsdsdsddffgfgfhg}
\end{equation}%
for $t\in \mathbb{R}$, $\psi \in \mathcal{D}\left( \Upsilon _{0}\right)
\oplus \mathcal{D}\left( \Upsilon _{0}^{\top }\right) $ and $\varphi \in 
\mathcal{D}_{0}$. Since $\mathcal{D}_{0}$ is dense in $\mathcal{F}$ and the
operators in (\ref{sdsdsdsdsddffgfgfhg}) are all bounded, we conclude that 
\begin{equation}
\mathbf{\pi }_{\mathcal{F}}\circ \mathbf{\chi }_{\mathrm{e}^{ith}}\left( 
\mathrm{B}(\psi )\right) =\mathrm{e}^{it\mathrm{H}_{0}/2}\mathbf{\pi }_{%
\mathcal{F}}\left( \mathrm{B}\left( \psi \right) \right) \mathrm{e}^{-it%
\mathrm{H}_{0}/2}  \label{dfgdfgdffgd2}
\end{equation}%
for any $\psi \in \mathcal{D}\left( \Upsilon _{0}\right) \oplus \mathcal{D}%
\left( \Upsilon _{0}^{\top }\right) $. Since $\mathbf{\pi }_{\mathcal{F}}$
and $\varphi \mapsto \mathrm{B}\left( \varphi \right) $ are contractive
mappings, we can extend by continuity (\ref{dfgdfgdffgd2}) to all $\psi \in 
\mathcal{H}=\mathfrak{h}\oplus \mathfrak{h}$ in order to get (\ref%
{dfgdfgdffgd}).
\end{proof}

Lemma \ref{Lemma quasi-free} implies that the (necessarily closed) generator
of the strongly continuous group $(\mathbf{\chi }_{\mathrm{e}^{ith}})_{t\in {%
\mathbb{R}}}$ of Bogoliubov $\ast $--auto%
\-%
morphisms of $\mathcal{A}$, $h$ being defined by (\ref{sdsdsdsdsdsdsdsd}),
can be seen in the Fock space representation as the commutator $\left[ 
\mathrm{H}_{0},\ \cdot \ \right] $ on some (dense) domain of $\mathcal{B}(%
\mathcal{F})$, making explicit the relation between Araki's bilinear
Hamiltonians and quadratic operators studied here.

\cite[Theorem 5.4]{Araki} would mean in this case that a quadratic
Hamiltonian in the sense of (\ref{DefH0}) is diagonalizable whenever (\ref%
{sdsdsdsdsdsdsdsd}) is essentially self-adjoint and such that the eigenspace
associated with the eigenvalue $0$ has dimension even or infinite. This
would be a very general result, even if $D_{0}$ is required to be
Hilbert-Schmidt to ensure the assertions of Lemma \ref{Lemma quasi-free}.
For instance, the operator $\Upsilon _{0}$ is not even semibounded in this
case. However, \cite[Definition 5.1]{Araki}, defining the diagonalization of
bilinear Hamiltonians, only involves the existence of a Bogoliubov
transformation $\mathcal{U}\in \mathcal{B}(\mathcal{H})$ such that $\mathcal{%
U}\mathfrak{A=A}\mathcal{U}$, which may \textbf{not} define a unitary
transformation in the Fock space representation. This is well known via the
celebrated Shale-Stinespring condition which refers to the Hilbert-Schmidt
assumption.

\begin{theorem}[Implementation of Bogoliubov transformations]
\label{Shale-Stinespring THM}\mbox{}\newline
Let $\mathcal{H}\doteq \mathfrak{h}\oplus \mathfrak{h}$ and $\mathfrak{A}$
be defined by (\ref{ghghghg}). Take any Bogoliubov transformation $\mathcal{U%
}$. In particular, $\mathcal{U}$ has the form\footnote{%
Use $\overline{X}=\mathcal{C}X\mathcal{C}$ (Remark \ref{remark idiote
copy(1)}) and $\mathcal{U}=\mathfrak{A}\mathcal{U}\mathfrak{A}$ together
with (\ref{ghghghg}). The fact that $\mathcal{U}\mathcal{U}^{\ast }=\mathcal{%
U}^{\ast }\mathcal{U}=1$ also implies other constraints on $\mathbf{u},%
\mathbf{v}\in \mathcal{B}\left( \mathfrak{h}\right) $, but this is not
important here.} 
\begin{equation*}
\mathcal{U}=\left( 
\begin{array}{cc}
\mathbf{u} & \mathbf{v} \\ 
\overline{\mathbf{v}} & \overline{\mathbf{u}}%
\end{array}%
\right)
\end{equation*}%
with $\mathbf{u},\mathbf{v}\in \mathcal{B}\left( \mathfrak{h}\right) $. Then
there is a unitary transformation $\mathrm{U}:\mathcal{F\rightarrow F}$ such
that $\mathrm{U}\left( \mathbf{\pi }_{\mathcal{F}}\left( A\right) \right) 
\mathrm{U}^{\ast }=\mathbf{\pi }_{\mathcal{F}}\circ \mathbf{\chi }_{\mathcal{%
U}}\left( A\right) $ for all $A\in \mathcal{A}$ iff $\mathbf{v}\in \mathcal{L%
}^{2}(\mathfrak{h})$. In particular, by Equation (\ref{ghghghg2}), when $%
\mathbf{v}\in \mathcal{L}^{2}(\mathfrak{h})$,%
\begin{equation*}
\mathrm{U}a\left( \varphi \right) \mathrm{U}^{\ast }=a(\mathbf{u}\varphi
)+a^{\ast }(\mathbf{v}\overline{\varphi })\ ,\qquad \varphi \in \mathfrak{h}%
\ .
\end{equation*}
\end{theorem}

\noindent See \cite[Theorems 3.1 and 6.2]{Ruijsenaars} or\ \cite[Theorem 7]%
{A70} for a complete proof. This theorem originally goes back to Shale and
Stinespring \cite{Shale-Strinespring}. For more recent or pedagogical
references, see \cite[Theorem 9.5]{Solovej} or \cite[Equation (2a.9) and
Theorem 2.2]{BLS}. Again the Hilbert-Schmidt condition is referred to as the 
\emph{Shale-Stinespring condition}. When it is satisfied, one says that $%
\mathcal{U}$ is (unitarily)\ implementable. Note that one can extend this
result to the case where $\mathrm{U}$ is not necessarily unitary. This
refers to \cite[Proposition 1.2]{Matsui87}.

Since Araki's construction of the Bogoliubov transformation, or the
corresponding basis projection, is only based on general considerations
applying the spectral theorem to the self-dual Hamiltonian $h$ (see \cite[%
Lemma 3.6 and 4.8]{Araki}), the main result \cite[Theorem 5.4]{Araki}
provides no information on the Shale-Stinespring condition. In other words,
we do not know whether the corresponding Bogoliubov transformation given by 
\cite[Theorem 5.4]{Araki} is implementable or not. This can certainly be
done in some general cases, but further studies are in any case required to
get a result like Theorem \ref{thm3}.

\begin{remark}[Implementation of non-autonomous flows of Bogoliubov
transformations]
\label{remark idiote copy(2)}\mbox{}\newline
In Theorems \ref{theorem Ht=00003Dunitary orbite copy(1)} and \ref{lemma H t
infinity copy(2)} as well as in Proposition \ref{lemma uv-Bogoliubov
transformation} and Corollary \ref{lemma uv-Bogoliubov transformation
copy(1)}, we provide a non-autonomous version\footnote{%
By non-autonomous, we refer to the fact that, instead of having a
one-parameter group of Bogoliubov transformations, we have a two-parameter
family that satisfies the cocycle property.} of implemented Bogoliubov
transformations, by using the theory of non-autonomous (Kato-hyperbolic)
evolution equations. This is used to find a solution to the Brockett-Wegner
flow on quadratic Hamiltonians as well as its asymptotics.
\end{remark}

\subsection{Bach, Lieb and Solovej's Approach to Quadratic Hamiltonians\label%
{Section Bach}}

Bach, Lieb and Solovej \cite{BLS} introduced a definition of quadratic
Hamiltonians that a priori differs from the one used here. This definition
is inspired by Araki's definition \cite[Definition 5.1]{Araki} and Theorem %
\ref{Shale-Stinespring THM}. For the reader's convenience, it is reproduced
below within the terminology and notation used in Section \ref{Araiki
section}. We take here $\mathcal{H}\doteq \mathfrak{h}\oplus \mathfrak{h}$
and a Bogoliubov transformation is a unitary operator $\mathcal{U}\in 
\mathcal{B}(\mathcal{H})$ such that $\mathcal{U}\mathfrak{A=A}\mathcal{U}$.

\begin{definition}[Quadratic Hamilonians \protect\cite{BLS}]
\label{def quadratic}\mbox{}\newline
A self-adjoint operator $\mathrm{H}$ on the fermionic Fock space $\mathcal{F}
$ is said to be a quadratic Hamiltonian if the unitary operator $\mathrm{U}%
_{t}=\mathrm{e}^{it\mathrm{H}/2}$ implements a Bogoliubov transformation $%
\mathcal{U}_{t}$ for any $t\in \mathbb{R}$, i.e., 
\begin{equation}
\mathrm{U}_{t}\left( \mathbf{\pi }_{\mathcal{F}}\left( A\right) \right) 
\mathrm{U}_{t}^{\ast }=\mathbf{\pi }_{\mathcal{F}}\circ \mathbf{\chi }_{%
\mathcal{U}_{t}}\left( A\right) \ ,\qquad A\in \mathcal{A},\ t\in \mathbb{R}%
\ ,  \label{defff}
\end{equation}%
where $\mathbf{\chi }_{\mathcal{U}_{t}}$ is the unique $\ast $--automorphism 
$\mathbf{\chi }_{\mathcal{U}}$ of $\mathcal{A}$ satisfying (\ref{def bogo})
for $\mathcal{U}=\mathcal{U}_{t}$ while $\mathbf{\pi }_{\mathcal{F}}$ is the
Fock space representation of the universal $C^{\ast }$-algebra $\mathcal{A}$
(\ref{A}).
\end{definition}

\noindent This point of view is also studied for bosonic systems, for
instance by Bruneau and Derezinski via their definition of Bogoliubov
Hamiltonians \cite{Bruneau-derezinski2007}. Observe also that, if $\mathrm{H}
$ is a quadratic Hamiltonian as in Definition \ref{def quadratic} then, for
any real constant $C\in \mathbb{R}$,\ $\mathrm{H}+C\mathbf{1}$ is of course
a quadratic Hamiltonian leading to the same Bogoliubov transformation.

Quadratic operators in Bach, Lieb and Solovej's sense are therefore \emph{%
automatically} self-adjoint. In Section \ref{Setup of the Problem2},
following Berezin's approach, we define quadratic operators on some
restricted domain by using explicit sums over annihilation and creation
operators acting on the fermionic Fock space. In this case, their
essentially self-adjointness is an open issue solved by Proposition \ref%
{Hamilselfadjoint}. It leads to self-adjoint operators that are always
quadratic Hamiltonians in Bach, Lieb and Solovej's sense:

\begin{corollary}[From Berezin's view point to BLS's approach]
\label{Corollaire sympa copy(1)}\mbox{}\newline
Fix $E_{0}\in \mathbb{R}$. Let $\Upsilon _{0}=\Upsilon _{0}^{\ast }$ and $%
D_{0}=-D_{0}^{\top }\in \mathcal{L}^{2}(\mathfrak{h})$ be two operators
acting on the Hilbert space $\mathfrak{h}$. Then, the self-adjoint extension 
$\mathrm{H}_{0}\equiv \mathrm{H}_{0}^{\ast \ast }$ of the quadratic operator
defined by (\ref{DefH0})--(\ref{domain H0}) on the domain (\ref{domain H0})
is a quadratic Hamiltonian in the sense of Definition \ref{def quadratic}.
\end{corollary}

\begin{proof}
Combine Lemma \ref{Lemma quasi-free} with Definition \ref{def quadratic}.
\end{proof}

We now discuss whether any quadratic Hamiltonian in the sense of Definition %
\ref{def quadratic} is the self-adjoint extension of a quadratic operators
defined by (\ref{DefH0}) on the domain (\ref{domain H0}). To this end, we
recall that $\mathfrak{A}$ is the antiunitary involution of $\mathcal{H}%
\doteq \mathfrak{h}\oplus \mathfrak{h}$ defined by (\ref{ghghghg}), while $\{%
\mathrm{B}(\psi )\}_{\psi \in \mathcal{\mathcal{H}}}$ is the family of
generators of the universal $C^{\ast }$-algebra $\mathcal{A}$ (\ref{A}). We
start with a first elementary, but important, lemma:\ 

\begin{lemma}[Generators of Bogoliubov transformations]
\label{Lemma Bach1}\mbox{
}\newline
Let $\mathrm{H}$ be a quadratic Hamiltonian in the sense of Definition \ref%
{def quadratic}. Then, (\ref{defff}) defines a strongly continuous group $(%
\mathcal{U}_{t}=\mathrm{e}^{ith})_{t\in \mathbb{R}}$ of Bogoliubov
transformations on $\mathcal{H}\doteq \mathfrak{h}\oplus \mathfrak{h}$,
where $h$ is a self-adjoint operator acting on $\mathcal{H}$ satisfying $%
h=h^{\ast }=-\mathfrak{A}h\mathfrak{A}$ with domain $\mathcal{D}(h)=%
\mathfrak{A}(\mathcal{D}(h))\subseteq \mathcal{\mathcal{H}}$.
\end{lemma}

\begin{proof}
As done in \cite[Equation (2b.36)]{BLS}, for any quadratic Hamiltonian $%
\mathrm{H}$ in the sense of Definition \ref{def quadratic}, one computes
from the CAR (\ref{CAR Grassmann III}) and $\mathcal{U}_{t}\mathfrak{A=A}%
\mathcal{U}_{t}$ that, for any $\varphi ,\psi \in \mathcal{H}$ and $t\in 
\mathbb{R}$,%
\begin{equation*}
\left\{ \mathbf{\pi }_{\mathcal{F}}\left( \mathrm{B}\left( \varphi \right)
\right) ,\mathrm{e}^{it\mathrm{H}/2}\mathbf{\pi }_{\mathcal{F}}\left( 
\mathrm{B}\left( \mathfrak{A}\psi \right) \right) \mathrm{e}^{-it\mathrm{H}%
/2}\right\} =\mathbf{\pi }_{\mathcal{F}}\left( \left\{ \mathrm{B}\left(
\varphi \right) ,\mathrm{B}\left( \mathcal{U}_{t}\mathfrak{A}\psi \right)
\right\} \right) =\left\langle \varphi ,\mathcal{U}_{t}\psi \right\rangle _{%
\mathcal{H}}\,\mathbf{1}\ ,
\end{equation*}%
where $\left\{ A,B\right\} \doteq AB+BA$ is the usual anticommutator. So, by
elementary computations using the strong continuity of the family $\left( 
\mathrm{e}^{it\mathrm{H}/2}\right) _{t\in \mathbb{R}}$ of unitary operators,
we deduce that $(\mathcal{U}_{t})_{t\in \mathbb{R}}$ is a strongly
continuous one-parameter group of Bogoliubov transformations, see \cite[%
Chapter I, Theorem 5.8]{EngelNagel}. In particular, we infer from Stone's
theorem \cite[Theorem 6.2]{Konrad} the existence of some self-adjoint
operator $h=h^{\ast }$ acting on $\mathcal{H}$ such that $\mathcal{U}_{t}=%
\mathrm{e}^{ith}$ for any $t\in \mathbb{R}$. The assertion then follows from
Lemma \ref{Lenma Bach-tech1}.
\end{proof}

A priori, one may represent on $\mathcal{H}\doteq \mathfrak{h}\oplus 
\mathfrak{h}$ the Hamiltonian $h$ of Lemma \ref{Lemma Bach1} by some $%
2\times 2$ matrix 
\begin{equation}
h=\left( 
\begin{array}{cc}
P_{\mathfrak{h}}hP_{\mathfrak{h}} & P_{\mathfrak{h}}hP_{\mathfrak{h}}^{\perp
} \\ 
P_{\mathfrak{h}}^{\perp }hP_{\mathfrak{h}} & P_{\mathfrak{h}}^{\perp }hP_{%
\mathfrak{h}}^{\perp }%
\end{array}%
\right)  \label{blck matrices}
\end{equation}%
with operator-valued coefficients\footnote{%
By a slight abuse of notation, we use in an appropriate way the
identification $\mathfrak{h}\oplus \{0\}\equiv \mathfrak{h}$ and $%
\{0\}\oplus \mathfrak{h}\equiv \mathfrak{h}$ to see $P_{\mathfrak{h}}hP_{%
\mathfrak{h}}$, $P_{\mathfrak{h}}hP_{\mathfrak{h}}^{\perp }$, $P_{\mathfrak{h%
}}^{\perp }hP_{\mathfrak{h}}$ and $P_{\mathfrak{h}}^{\perp }hP_{\mathfrak{h}%
}^{\perp }$ as four operators acting on $\mathfrak{h}$.}, $P_{\mathfrak{h}}$
being the orthogonal projection on $\mathfrak{h}\oplus \{0\}$ and $P_{%
\mathfrak{h}}^{\perp }\doteq \mathbf{1}-P_{\mathfrak{h}}$. If $h\in \mathcal{%
B}\left( \mathcal{H}\right) $ then one easily computes from $h=h^{\ast }=-%
\mathfrak{A}h\mathfrak{A}$ (Lemma \ref{Lemma Bach1}) together with (\ref%
{ghghghg}) and $X^{\top }\doteq \mathcal{C}X^{\ast }\mathcal{C}$, $\overline{%
X}\doteq \mathcal{C}X\mathcal{C}$ and $\overline{X}^{\top }=\overline{%
X^{\top }}=X^{\ast }$ (Remark \ref{remark idiote copy(1)}) that the block
operator matrix (\ref{blck matrices}) is well-defined and of the form%
\begin{equation}
h=\frac{1}{2}\left( 
\begin{array}{cc}
\Upsilon _{0} & 2D_{0} \\ 
2D_{0}^{\ast } & -\Upsilon _{0}^{\top }%
\end{array}%
\right) =\frac{1}{2}\left( 
\begin{array}{cc}
\Upsilon _{0} & 2D_{0} \\ 
-2\overline{D}_{0} & -\Upsilon _{0}^{\top }%
\end{array}%
\right) \ ,  \label{dfdfdfdfdfdfdf}
\end{equation}%
where $\Upsilon _{0}\doteq P_{\mathfrak{h}}hP_{\mathfrak{h}}=\Upsilon
_{0}^{\ast }$ and $D_{0}\doteq P_{\mathfrak{h}}hP_{\mathfrak{h}}^{\perp
}=-D_{0}^{\top }$ are two bounded operators acting on $\mathfrak{h}$. By
comparing Equation (\ref{dfdfdfdfdfdfdf}) with Lemma \ref{Lemma quasi-free},
one sees that quadratic Hamiltonians in the sense of Definition \ref{def
quadratic} or (\ref{DefH0})--(\ref{domain H0}) are similarly encoded on the
(dubbed) Hilbert space $\mathcal{H}$, at least when the Hamiltonian $h$ is
bounded.

Nevertheless, general block operator matrices like (\ref{blck matrices})
with unbounded operators are generally nontrivial to define. See for
instance \cite{block1,block2,block2bis,block3,block4,block5,block6} which
analyze properties of operators (e.g., self-adjointness, closeness) given
via a $2\times 2$ matrix of unbounded operator-valued coefficients. In the
present case, if the Hamiltonian $h$ of Lemma \ref{Lemma Bach1} is
unbounded, with domain $\mathcal{D}(h)\subseteq \mathcal{\mathcal{H}}$, then
it is for instance necessary to ensure at least that $\mathfrak{h}\cap 
\mathcal{D}(h)\neq 0$ in order to give a mathematically rigorous meaning of
the coefficients $P_{\mathfrak{h}}hP_{\mathfrak{h}}$, $P_{\mathfrak{h}}hP_{%
\mathfrak{h}}^{\perp }$, $P_{\mathfrak{h}}^{\perp }hP_{\mathfrak{h}}$ and $%
P_{\mathfrak{h}}^{\perp }hP_{\mathfrak{h}}^{\perp }$ of (\ref{blck matrices}%
) as well-defined operators acting on $\mathfrak{h\equiv h}\oplus
\{0\}\equiv \{0\}\oplus \mathfrak{h}$. Additionally, even if $\mathfrak{h}%
\cap \mathcal{D}(h)\neq 0$, it is not clear that $\Upsilon _{0}$ is a
(essentially) self-adjoint operator while one has to make sense of the sum $%
\Upsilon _{0}+2D_{0}$ with $\Upsilon _{0}\doteq P_{\mathfrak{h}}hP_{%
\mathfrak{h}}$ and $D_{0}\doteq P_{\mathfrak{h}}hP_{\mathfrak{h}}^{\perp }$
being possibly unbounded operators. There is in particular important domain
issues associated with these operators $\Upsilon _{0}$ and $D_{0}$.

For instance, in Lemma \ref{Lenma Bach-tech1 copy(1)} (ii), we show that, as
soon as the kernel of $h$ has even or infinite dimension, $h$ can be
represented on $\mathcal{H}\doteq \mathfrak{h}\oplus \mathfrak{h}$ by (\ref%
{dfdfdfdfdfdfdf}), where $\Upsilon _{0}$ and $D_{0}$ are two (possibly
unbounded) operators acting on $\mathfrak{h}$ with some dense domain 
\begin{equation}
\mathcal{Y}=\mathcal{D}\left( \Upsilon _{0}\right) \subseteq \mathfrak{%
h\qquad }\text{and}\mathfrak{\qquad }\mathcal{D}\left( D_{0}\right) =%
\mathcal{CY}\ ,  \label{domain h0}
\end{equation}%
where $\mathcal{C}$ is the complex conjugation of Remark \ref{remark idiote
copy(1)}. Note in this case that 
\begin{equation}
\mathcal{D}\left( h\right) =\mathcal{Y}\oplus \mathcal{CY}\ .
\label{domain h}
\end{equation}%
However, even in this situation, $\Upsilon _{0}$ is only a \emph{symmetric}
operator while $D_{0}=-D_{0}^{\top }$ on its domain of definition. To get a
self-adjoint operator $\Upsilon _{0}=\Upsilon _{0}^{\ast }$, it is necessary
to construct a self-adjoint extension of the previous symmetric operator.
This can be performed via the use of the Cayley and Krein transforms or the
notion of boundary triplets \cite[Part VI]{Konrad}. For example, as a
consequence of the Cayley transform, von Neumann's extension theorem \cite[%
Theorem 13.10]{Konrad} demonstrates the existence of self-adjoint extensions
of a densely defined symmetric operator $S$ iff the deficiency indices of $S$
are identical\footnote{%
If this is not the case, one may use larger Hilbert space than $\mathcal{H}$%
, by \cite[Corollary 13.11]{Konrad}.}. This property holds true whenever $S$
is semibounded, thanks to \cite[Proposition 3.3]{Konrad}. Applied to the
densely defined symmetric operator $\Upsilon _{0}$ of Lemma \ref{Lenma
Bach-tech1 copy(1)} (ii), these mathematical observations strongly highlight
our condition $\Upsilon _{0}\geq -\tilde{\mu}\mathbf{1}$ for some $\tilde{\mu%
}\in \mathbb{R}$ given in Theorem \ref{thm3}, as a sufficient condition to
obtain a self-adjoint operator $\Upsilon _{0}=\Upsilon _{0}^{\ast }$ from
Lemma \ref{Lenma Bach-tech1 copy(1)} (ii).

A general and exhaustive discussion of fully unbounded cases like in Lemma %
\ref{Lenma Bach-tech1 copy(1)} (ii) is therefore nontrivial with rather
technical arguments and proofs. This simply goes beyond the main objective
of the paper. In fact, even Araki in \cite[Theorems 5.4 and 5.6]{Araki} or
Bruneau and Derezinski in the bosonic case \cite{Bruneau-derezinski2007} use
(essentially) self-adjoint operators $h$ which can be written as a block
operator matrix like (\ref{dfdfdfdfdfdfdf}) (in the fermionic case) and we
thus restrict the definition of quadratic Hamiltonians given in Definition %
\ref{def quadratic} by using the notion of \emph{compatible} quadratic
Hamiltonians:

\begin{definition}[Compatible quadratic Hamilonians]
\label{def quadratic compatible}\mbox{}\newline
A quadratic Hamiltonian of Definition \ref{def quadratic} is said to be
compatible with the (one-particle) Hilbert space $\mathfrak{h}$ whenever its
associated Hamiltonian $h$, as given by Lemma \ref{Lemma Bach1}, can be
written as (\ref{dfdfdfdfdfdfdf}) with domain $\mathcal{D}\left( h\right) =%
\mathcal{Y}\oplus \mathcal{CY}$, where $\mathcal{Y}\subseteq \mathfrak{h}$
is a dense subspace while $\Upsilon _{0}=\Upsilon _{0}^{\ast }$ and $%
D_{0}=-D_{0}^{\top }$ are two operators with domains $\mathcal{D}\left(
\Upsilon _{0}\right) \supseteq \mathcal{Y}$ and $\mathcal{D}\left(
D_{0}\right) \supseteq \mathcal{C}\mathcal{Y}$.
\end{definition}

\noindent From now, we only study compatible quadratic Hamiltonians. This
case is one standard situation in Physics for which $\Upsilon _{0}$ is
interpreted as the Hamiltonian acting on the one-particle Hilbert space
describing free fermions while $D_{0}$ could result from effective theories,
as already explained in Section \ref{Setup of the Problem2}. We emphasize
again that a generator of a group of Bogoliubov transformations is \textbf{%
not} necessarily compatible, in spite of Lemma \ref{Lenma Bach-tech1 copy(1)}
(ii), which can only gives the existence of densely defined symmetric
operators $\Upsilon _{0}$, as explained above.

It turns out that the set of quadratic Hamiltonians defined via (\ref{DefH0}%
)--(\ref{domain H0}) includes \emph{all} compatible quadratic Hamiltonians
of Definition \ref{def quadratic} having a Hilbert-Schmidt operator $D_{0}$:

\begin{corollary}[From BLS's approach to Berezin's view point]
\label{Corollaire sympa copy(1)bis}\mbox{
}\newline
Take a compatible quadratic Hamiltonian $\mathrm{H}$ in the sense of
Definitions \ref{def quadratic} and \ref{def quadratic compatible}. If $%
D_{0}\in \mathcal{L}^{2}(\mathfrak{h})$ in (\ref{dfdfdfdfdfdfdf}) then, for
some constant $E_{0}\in \mathbb{R}$,%
\begin{equation*}
\mathrm{H}=\mathrm{H}_{0}^{\ast \ast }\equiv \mathrm{H}_{0}=\sum_{k,l\in 
\mathbb{N}}\{\Upsilon _{0}\}_{k,l}a_{k}^{\ast
}a_{l}+\{D_{0}\}_{k,l}a_{k}^{\ast }a_{l}^{\ast }+\{\bar{D}%
_{0}\}_{k,l}a_{l}a_{k}+E_{0}\mathbf{1}\ .
\end{equation*}
\end{corollary}

\begin{proof}
Fix all parameters of the corollary. Using the operators $\Upsilon
_{0}=\Upsilon _{0}^{\ast }$ and $D_{0}=-D_{0}^{\top }$ given by Definition %
\ref{def quadratic compatible} (see (\ref{dfdfdfdfdfdfdf})) together with $%
E_{0}=0$, we can define via (\ref{DefH0})--(\ref{domain H0}) a self-adjoint
operator $\mathrm{H}_{0}$ under the condition that $D_{0}\in \mathcal{L}^{2}(%
\mathfrak{h})$. This is a consequence of Proposition \ref{Hamilselfadjoint}.
By Lemma \ref{Lemma quasi-free} and Definition \ref{def quadratic}, the
(unitary) groups $\left( \mathrm{e}^{it\mathrm{H}/2}\right) _{t\in \mathbb{R}%
}$ and $\left( \mathrm{e}^{it\mathrm{H}_{0}/2}\right) _{t\in \mathbb{R}}$
implement the same Bogoliubov transformation and so, we deduce that 
\begin{equation*}
\mathrm{e}^{-it\mathrm{H}_{0}/2}\mathrm{e}^{it\mathrm{H}/2}\mathbf{\pi }_{%
\mathcal{F}}\left( A\right) =\mathbf{\pi }_{\mathcal{F}}\left( A\right) 
\mathrm{e}^{-it\mathrm{H}_{0}/2}\mathrm{e}^{it\mathrm{H}/2}\ ,\qquad t\in {%
\mathbb{R}},\text{ }A\in \mathcal{A}\ ,
\end{equation*}%
i.e., for any $t\in {\mathbb{R}}$, $\mathrm{e}^{-it\mathrm{H}_{0}/2}\mathrm{e%
}^{it\mathrm{H}/2}$ is in the commutant of $\pi _{\mathcal{F}}(A)$. Hence,
by irreducibility of the Fock representation, 
\begin{equation}
\mathrm{e}^{-it\mathrm{H}_{0}/2}\mathrm{e}^{it\mathrm{H}/2}=c_{t}\in \mathbb{%
C}\ ,  \label{opopop}
\end{equation}%
see Equations (\ref{commutant1})--(\ref{commutant2}) for more details. Since 
$(\mathrm{e}^{it\mathrm{H}/2})_{t\in \mathbb{R}}$ and $(\mathrm{e}^{it%
\mathrm{H}_{0}/2})_{t\in \mathbb{R}}$\ are strongly continuous groups of
unitary operators, using (\ref{opopop}) and elementary estimates, one shows%
\footnote{%
For more details, see \cite{Nathan}.} that $(c_{t})_{t\in \mathbb{R}}\in
C\left( \mathbb{R};\mathbb{C}\right) $ is a continuous complex-valued
function satisfying $\left\vert c_{t}\right\vert =1$ and $c_{t+s}=c_{t}c_{s}$
for any $s,t\in {\mathbb{R}}$. As is well-known, there is therefore a real
constant $\theta \in \mathbb{R}$ such that $c_{t}=\mathrm{e}^{it\theta /2}$
for any $t\in {\mathbb{R}}$. Combining this last property with (\ref{opopop}%
) and the uniqueness of the generator of the strongly continuous group $%
\left( \mathrm{e}^{it\mathrm{H}/2}=\mathrm{e}^{it\left( \mathrm{H}%
_{0}+\theta \right) /2}\right) _{t\in \mathbb{R}}$ of unitary operators (see 
\cite[Theorem 6.2]{Konrad}), we arrive at the assertion.
\end{proof}

Corollary \ref{Corollaire sympa copy(1)bis} is of course the converse of
Corollary \ref{Corollaire sympa copy(1)} in the Hilbert-Schmidt case. Note,
however, that it is not clear whether the off-diagonal operator $D_{0}$ in
Definition \ref{def quadratic compatible} (see (\ref{dfdfdfdfdfdfdf})) is a
bounded operator.

On the one hand, the Hilbert-Schmidt condition for $D_{0}$ results from the
(natural) use of the domain $\mathcal{D}_{0}$ (\ref{domain H0}), which is a
core for all our quadratic Hamiltonians. See Proposition \ref%
{Hamilselfadjoint}. In particular, the vacuum state $\Psi \doteq \left(
1,0,\ldots \right) $ of the fermionic Fock space $\mathcal{F}$ always
belongs to the domain of our quadratic Hamiltonians, see Equation (\ref%
{quadratic opereators vaccum}). By contrast, $\Psi $ may not be in the
domain of a quadratic Hamiltonian in the sense of Definition \ref{def
quadratic}. But, a quadratic Hamiltonian $\mathrm{H}$ of Definition \ref{def
quadratic} that does not include $\Psi $ in its domain would be
intrinsically different from our quadratic Hamiltonians and would refer to
an non-traditional (albeit possibly very interesting) model in Physics. We
strongly suspect that the vacuum state would not belong to the domain of 
\textrm{H} when the operator $D_{0}$ of Definition \ref{def quadratic
compatible} (see (\ref{dfdfdfdfdfdfdf})) is not a Hilbert-Schmidt operator.
This conjecture is based on our next theorems.

\begin{theorem}[Shale-Stinespring-like condition for generators -- I]
\label{prop bach2 copy(1)}\mbox{
}\newline
Take any compatible quadratic Hamiltonian $\mathrm{H}$ in the sense of
Definitions \ref{def quadratic} and \ref{def quadratic compatible}, the
domain of which is $\mathcal{D}\left( \mathrm{H}\right) \subseteq \mathcal{F}
$. Assume that, for some $c\in \mathbb{R}^{+}\cup \{\infty \}$,%
\begin{equation}
\mathrm{e}^{it\Upsilon _{0}}\mathcal{Y}\subseteq \mathcal{Y}\ ,\qquad t\in
\left( -c,c\right) \ ,  \label{assumptionassumption}
\end{equation}%
with the operator family $(\overline{D}_{0}\mathrm{e}^{it\Upsilon
_{0}})_{t\in (-c,c)}$ being strongly continuous on $\mathcal{Y}$. Then, $%
\Psi \doteq \left( 1,0,\ldots \right) \in \mathcal{D}\left( \mathrm{H}%
\right) $ iff $D_{0}\in \mathcal{L}^{2}(\mathfrak{h})$\footnote{%
By a slight abuse of notation, we mean here the original $D_{0}$ or its
(well-defined) continuous extension to the whole Hilbert space $\mathfrak{h}$%
.}.
\end{theorem}

\begin{proof}
Combine Corollaries \ref{Corollaire sympa copy(1)bis} and \ref{coro
Bach-tech9} (ii), keeping in mind Definitions \ref{def quadratic} and \ref%
{def quadratic compatible}.
\end{proof}

\begin{remark}
\mbox{}\newline
Condition (\ref{assumptionassumption}) implies that $\mathcal{Y}$ must be a
core for the self-adjoint operator $\Upsilon _{0}$, by \cite[Proposition 6.3]%
{Konrad}. In other words, $\Upsilon _{0}$ restricted to the subspace $%
\mathcal{Y}\subseteq \mathcal{D}\left( \Upsilon _{0}\right) $ is essentially
self-adjoint.
\end{remark}

In applications, it is very natural to have an off-diagonal operator-valued
coefficient $\overline{D}_{0}$ that is relatively bounded with respect to
the diagonal one $\Upsilon _{0}$. For the block operator matrix (\ref%
{dfdfdfdfdfdfdf}) with densely defined linear operators $\Upsilon _{0}$ and $%
\overline{D}_{0}$ acting on $\mathfrak{h}$, it means that $\mathcal{D}(%
\overline{D}_{0})\supseteq \mathcal{D}(\Upsilon _{0})$ and that there exist
constants $\alpha _{1},\alpha _{2}\in \mathbb{R}^{+}$ such that, for any $%
\varphi \in \mathcal{D}\left( \Upsilon _{0}\right) $,%
\begin{equation}
\left\Vert \overline{D}_{0}\varphi \right\Vert _{\mathfrak{h}}\leq \alpha
_{1}\left\Vert \varphi \right\Vert _{\mathfrak{h}}+\alpha _{2}\left\Vert
\Upsilon _{0}\varphi \right\Vert _{\mathfrak{h}}\ .  \label{rel}
\end{equation}%
See \cite[Equation (X.19a)]{ReedSimonII}. It is an usual assumption which
enables well-defined block operator matrices like (\ref{dfdfdfdfdfdfdf}).
See for instance \cite{block1,block6,block7}. In Physics, it means that the
off-diagonal terms $D_{0}$,$\overline{D}_{0}$ can be viewed as a bounded
perturbation of the free system, characterized \ by the free (one particle)
Hamiltonian $\Upsilon _{0}$. On the level of quadratic Hamiltonians, it
means that the non-$\mathrm{N}$--diagonal part (see (\ref{non-N--diagonal
part})) should be relatively bounded with respect to the $\mathrm{N}$%
--diagonal part of the given quadratic Hamiltonian. The latter is in fact
true whenever $D_{0}$ is not only relatively bounded with respect $\Upsilon
_{0}$ in the operator norm as in (\ref{rel}) but in the Hilbert-Schmidt
norm, in the sense of Lemma \ref{lemmfermion} with $\theta =\Upsilon
_{0}+\mu \mathbf{1}\geq 0$ for some appropriate $\mu \in \mathbb{R}$. The
Shale-Stinespring-like condition for such generators can be studied by
applying directly Theorem \ref{prop bach2 copy(1)}:

\begin{corollary}[Shale-Stinespring-like condition for generators -- II]
\label{Corollaire sympa copy(2)}\mbox{
}\newline
Take any compatible quadratic Hamiltonian $\mathrm{H}$ in the sense of
Definitions \ref{def quadratic} and \ref{def quadratic compatible}, the
domain of which is $\mathcal{D}\left( \mathrm{H}\right) \subseteq \mathcal{F}
$. Assume that $D_{0}$ is relatively bounded with respect to $\Upsilon _{0}$%
. Then, $\Psi \doteq \left( 1,0,\ldots \right) \in \mathcal{D}\left( \mathrm{%
H}\right) $ iff $D_{0}\in \mathcal{L}^{2}(\mathfrak{h})$.
\end{corollary}

\begin{proof}
Combine Theorem \ref{prop bach2 copy(1)} with Lemma \ref{Lenma Bach-tech2
copy(2)}.
\end{proof}

Similar to the usual Shale-Stinespring condition given by Theorem \ref%
{Shale-Stinespring THM}, the property $D_{0}\in \mathcal{L}^{2}(\mathfrak{h}%
) $ in Theorem \ref{prop bach2 copy(1)} and Corollary \ref{Corollaire sympa
copy(2)} can be seen as a Shale-Stinespring condition\footnote{%
The Shale-Stinespring condition on Bogoliubov transformations does not
necessarily imply a Hilbert-Schmidt condition on the off-diagonal elements
of its generator. See, however, Theorem \ref{prop bach2} (i).} for
generators of implementable groups of Bogoliubov transformations. This
condition is directly related with the question of whether the vacuum state
belongs to the domain of the quadratic Hamiltonian or not. When it is the
case, we obtain the quadratic Hamiltonians diagonalized in the present paper.

We now conclude this section by studying the specific case for which the
off-diagonal operator-valued coefficient $D_{0}$ of Definition \ref{def
quadratic compatible} (see (\ref{dfdfdfdfdfdfdf})) is not only relatively
bounded with respect to $\Upsilon _{0}$, but already bounded. In this
situation, we lurch from Corollary \ref{Corollaire sympa copy(2)} by not
assuming that the vacuum state $\Psi \doteq \left( 1,0,\ldots \right) \in 
\mathcal{F}$ belongs to the domain of the compatible quadratic Hamiltonians.
Even in this case, the operator $D_{0}$ of Definition \ref{def quadratic
compatible} turns out to be very close to a Hilbert-Schmidt operator. To see
this, we define the continuous operator family $(\varkappa _{t})_{t\in 
\mathbb{R}}\subseteq \mathcal{B}(\mathcal{B}(\mathfrak{h}))$ by%
\begin{equation}
\varkappa _{t}\left( A\right) \doteq \int_{0}^{t}\mathrm{e}^{-i\tau \Upsilon
_{0}/2}A\mathrm{e}^{-i\tau \Upsilon _{0}^{\top }/2}\mathrm{d}\tau \ ,\qquad
A\in \mathcal{B}(\mathfrak{h})\ ,  \label{mapping familybis}
\end{equation}%
for each time $t\in \mathbb{R}$. The integral is well-defined for any vector 
$\varphi \in \mathcal{H}$ since $(\mathrm{e}^{it\Upsilon _{0}})_{t\in 
\mathbb{R}}$ and $(\mathrm{e}^{it\Upsilon _{0}^{\top }})_{t\in \mathbb{R}}$
are strongly continuous groups of unitary operators.

\begin{theorem}[Shale-Stinespring-like condition for generators -- III]
\label{prop bach2}\mbox{
}\newline
Take any compatible quadratic Hamiltonian $\mathrm{H}$ in the sense of
Definitions \ref{def quadratic} and \ref{def quadratic compatible}, the
domain of which is $\mathcal{D}\left( \mathrm{H}\right) \subseteq \mathcal{F}
$. Assume $D_{0}\in \mathcal{B}(\mathfrak{h})$. Then, the following
assertions hold true:\ 

\begin{enumerate}
\item[\emph{(i)}] $(\varkappa _{t}(D_{0}))_{t\in \mathbb{R}}\subseteq 
\mathcal{B}(\mathfrak{h})$ belongs to $\mathcal{L}^{2}(\mathfrak{h})$ for
times $t$ in a neighborhood of $0$ and is even continuous in $\mathcal{L}%
^{2}(\mathfrak{h})$ at $t=0$.

\item[\emph{(ii)}] If $D_{0}\Upsilon _{0}=-\Upsilon _{0}^{\top }D_{0}$ then $%
D_{0}\in \mathcal{L}^{2}(\mathfrak{h})$ and $\Psi \in \mathcal{D}\left( 
\mathrm{H}\right) $.

\item[\emph{(iii)}] If $\Upsilon _{0}\in \mathcal{B}(\mathfrak{h})$ and $%
D_{0}\Upsilon _{0}=\Upsilon _{0}^{\top }D_{0}$ then $D_{0}\in \mathcal{L}%
^{2}(\mathfrak{h})$ and $\Psi \in \mathcal{D}\left( \mathrm{H}\right) $.
\end{enumerate}
\end{theorem}

\begin{proof}
Assertion (i) results from Corollary \ref{coro Bach-tech9} combined with
Definitions \ref{def quadratic} and \ref{def quadratic compatible}. We
deduce Assertions (ii)--(iii) by means of Theorem \ref{Shale-Stinespring THM}%
, Corollary \ref{Corollaire sympa copy(2)} and Lemma \ref{Lenma Bach-tech7},
keeping in mind again Definitions \ref{def quadratic} and \ref{def quadratic
compatible}.
\end{proof}

Note that, for simplicity, we restrict the last theorem to bounded
off-diagonal elements $D_{0}\in \mathcal{B}(\mathfrak{h})$ since a general
and exhaustive discussion of the fully unbounded cases would a priori
deserve much longer discussions with rather technical arguments and new
proofs. Here, the most general case is given by Theorem \ref{prop bach2}
(i), which shows that implementable strongly continuous groups of Bogoliubov
transformations lead to operators $D_{0}$ that are close to a
Hilbert-Schmidt operator (if it is not always a Hilbert-Schmidt): By the
functional calculus, for any $\varphi \in \mathfrak{h}$, and $A\in \mathcal{B%
}(\mathfrak{h})$,%
\begin{equation}
\varkappa _{t}\left( A\right) \varphi =f_{\varkappa _{t}}\left( \Upsilon
_{0}\right) A\varphi +B_{\varkappa _{t}}\varphi \ ,\qquad t\in \mathbb{R}\ ,
\label{F-B-1}
\end{equation}%
with $f_{\varkappa _{t}}:\mathbb{R}\rightarrow \mathbb{R}$ and $B_{\varkappa
_{t}}\in \mathcal{B}(\mathfrak{h})$ respectively defined by 
\begin{equation}
f_{\varkappa _{t}}\left( x\right) \doteq 2ix^{-1}\left( \mathrm{e}%
^{-itx/2}-1\right) \qquad \text{and}\qquad B_{\varkappa _{t}}\doteq \int_{0}^{t}\mathrm{e}^{-i\tau \Upsilon _{0}/2}A\left( \mathrm{e}^{-i\tau
\Upsilon _{0}^{\top }/2}-\mathbf{1}\right) \mathrm{d}\tau \ .
\label{F-B}
\end{equation}%
In particular, for times $t$ in a neighborhood $\mathcal{V}$ of $0$ (cf.
Theorem \ref{prop bach2} (i)), 
\begin{equation}
f_{\varkappa _{t}}\left( x\right) =t-\frac{1}{4}it^{2}x+\mathcal{O}\left(
t^{3}x^{2}\right) \qquad \text{and}\qquad \left\Vert B_{\varkappa
_{t}}\varphi \right\Vert _{\mathrm{op}}\leq t\left\Vert A\right\Vert _{%
\mathrm{op}}\sup_{t\in \mathcal{V}}\left\Vert \left( \mathbf{1}-\mathrm{e}%
^{-i\tau \Upsilon _{0}^{\top }/2}\right) \varphi \right\Vert _{\mathfrak{h}%
}=o\left( t\right) \ ,  \label{F-B+1}
\end{equation}%
which demonstrate that $\varkappa _{t}\left( A\right) \varphi \simeq
tA\varphi $ as $t\rightarrow 0$. Applying this to $D_{0}\in \mathcal{B}(%
\mathfrak{h})$ and having in mind Theorem \ref{prop bach2} (i), this
heuristic observation suggests that $D_{0}$ is close to be a Hilbert-Schmidt
operator $\mathcal{L}^{2}(\mathfrak{h})$, as shown in the special cases of
Theorem \ref{prop bach2} (ii)--(iii). This concludes our study of Bach, Lieb
and Solovej's approach to quadratic Hamiltonians in relation with Berezin's
approach used here.

\subsection{Technical Results\label{Section tech7 bis}\label%
{Shale-Stinespring Condition for Generators}}

In this subsection, we explicit the relation between Bogoliubov
transformations and quadratic Hamiltonians and finally prove Theorems \ref%
{prop bach2 copy(1)} and \ref{prop bach2}. Indeed, as explained above,
recall that Bach, Lieb and Solovej \cite{BLS} introduced a general
definition of quadratic Hamiltonians on fermionic Fock spaces. This refers
to Definition \ref{def quadratic}. It leads to a strongly continuous
one-parameter group of Bogoliubov transformations, as explained in the proof
of Lemma \ref{Lemma Bach1}. Here, we study such kind of groups and discuss
their generators in relation with the Shale-Stinespring condition (see
Theorem \ref{Shale-Stinespring THM}).

We take $\mathcal{H}\doteq \mathfrak{h}\oplus \mathfrak{h}$ and recall that
a Bogoliubov transformation is a unitary operator $\mathcal{U}\in \mathcal{B}%
(\mathcal{H})$ such that $\mathcal{U}\mathfrak{A}=\mathfrak{A}\mathcal{U}$,
as explained in Section \ref{Araiki section}. We thus start by relating
these properties with the generator of a group of Bogoliubov transformations:

\begin{lemma}[Self-dual operators and groups of Bogoliubov transformations]
\label{Lenma Bach-tech1}\mbox{
}\newline
Take any self-adjoint operator $h=h^{\ast }$ acting on $\mathcal{H}$. Then, $%
\mathrm{e}^{ith}$ is a Bogoliubov transformation (in the sense of \cite%
{Araki}) for any $t\in \mathbb{R}$ iff $h=-\mathfrak{A}h\mathfrak{A}$ with
domain $\mathcal{D}(h)=\mathfrak{A}(\mathcal{D}(h))\subseteq \mathcal{%
\mathcal{H}}$.
\end{lemma}

\begin{proof}
Let $h=h^{\ast }$ acting on $\mathcal{H}$. Recall that $\mathfrak{A}$ is an
involution, i.e., $\mathfrak{A}^{2}=\mathbf{1}$, and it is in particular
one-to-one. So, the family $(\mathfrak{A}\mathrm{e}^{ith}\mathfrak{A})_{t\in 
\mathbb{R}}$ is a strongly continuous one-parameter group of unitary
operators on $\mathcal{H}$. By Stone's theorem \cite[Theorem 6.2]{Konrad},
there is a unique self-adjoint operator $g=g^{\ast }$ such that $\mathfrak{A}%
\mathrm{e}^{ith}\mathfrak{A}=\mathrm{e}^{itg}$ for any $t\in \mathbb{R}$.
Using the antiunitarity of $\mathfrak{A}$, $\mathfrak{A}^{2}=\mathbf{1}$ and
the theory of strongly continuous groups, one easily checks that $g=-%
\mathfrak{A}h\mathfrak{A}$ with domain $\mathcal{D}(g)=\mathfrak{A}(\mathcal{%
D}(h))$. Now, if $\mathrm{e}^{ith}$ is a Bogoliubov transformation for any $%
t\in \mathbb{R}$ then 
\begin{equation*}
\mathrm{e}^{ith}=\mathfrak{A}\mathrm{e}^{ith}\mathfrak{A}=\mathrm{e}^{itg}\
\qquad t\in \mathbb{R}\ ,
\end{equation*}%
and so, $h=g=-\mathfrak{A}h\mathfrak{A}$ with domain $\mathcal{D}(h)=%
\mathfrak{A}(\mathcal{D}(h))$, by unicity of the generator $g$ (see again 
\cite[Theorem 6.2]{Konrad}). Conversely, if $h=-\mathfrak{A}h\mathfrak{A}$
with domain $\mathcal{D}(h)=\mathfrak{A}(\mathcal{D}(h))\subseteq \mathcal{%
\mathcal{H}}$ then 
\begin{equation*}
\mathrm{e}^{ith}=\mathrm{e}^{-it\mathfrak{A}h\mathfrak{A}}=\mathfrak{A}%
\mathrm{e}^{ith}\mathfrak{A}\ ,\qquad t\in \mathbb{R}\ ,
\end{equation*}%
because $ig=-i\mathfrak{A}h\mathfrak{A}$ is the (uniquely defined) generator
of the strongly continuous one-parameter group $(\mathfrak{A}\mathrm{e}^{ith}%
\mathfrak{A})_{t\in \mathbb{R}}$ of unitary operators, as explained above.
In particular, $\mathrm{e}^{ith}$ is a Bogoliubov transformation for any $%
t\in \mathbb{R}$.
\end{proof}

This lemma shows that a group of Bogoliubov transformations is constructed
from a special class of self-adjoint operators defined on the self-dual
Hilbert space, which are named \emph{self-dual }Hamiltonians \cite[%
Definition 2.7]{LD1}:\ A self-dual Hamiltonian is a self-adjoint operator $%
h=h^{\ast }$ acting on $\mathcal{H}$ and satisfying $h=-\mathfrak{A}h%
\mathfrak{A}$ with domain $\mathcal{D}(h)=\mathfrak{A}(\mathcal{D}%
(h))\subseteq \mathcal{\mathcal{H}}$.

The next step is to represent groups of Bogoliubov transformations and their
generator on $\mathcal{H}\doteq \mathfrak{h}\oplus \mathfrak{h}$ as a $%
2\times 2$ matrix of operator-valued coefficients. As already explained
after Lemma \ref{Lemma Bach1} in Section \ref{Section Bach}, such a block
operator matrix 
\begin{equation*}
A=\left( 
\begin{array}{cc}
A_{1,1} & A_{1,2} \\ 
A_{2,1} & A_{2,2}%
\end{array}%
\right) \ ,\qquad A_{1,1},A_{1,2},A_{2,1},A_{2,2}\in \mathcal{L}\left( 
\mathfrak{h}\right) \ ,
\end{equation*}%
is trivial to obtain on $\mathcal{H}\doteq \mathfrak{h}\oplus \mathfrak{h}$
when $A\in \mathcal{B}\left( \mathcal{H}\right) $ is a bounded (linear)
operator. Here, $\mathcal{L}\left( \mathfrak{h}\right) $ denotes the set of
linear operators acting on $\mathfrak{h}$. However, this is not obvious
anymore for unbounded operators $A$ acting on $\mathcal{H}$, as for some
generators of groups of Bogoliubov transformations. Below, using Araki's
study \cite{Araki}, we give a general result on groups of Bogoliubov
transformations and their generator as block operator matrices on $\mathcal{H%
}\doteq \mathfrak{h}\oplus \mathfrak{h}$.

\begin{lemma}[Representation of Bogoliubov transformations and their
generator]
\label{Lenma Bach-tech1 copy(1)}\mbox{
}\newline
A group $(\mathrm{e}^{ith})_{t\in \mathbb{R}}$ of Bogoliubov transformations
and its generator $ih$ acting on $\mathcal{H}$, $h$ being a self-adjoint
operator, can be represented on $\mathcal{H}\doteq \mathfrak{h}\oplus 
\mathfrak{h}$ as follows:

\begin{enumerate}
\item[\emph{(i)}] Bogoliubov transformations: There are two strongly
continuous families $(\mathbf{u}_{t})_{t\in \mathbb{R}},(\mathbf{v}%
_{t})_{t\in \mathbb{R}}\in \mathcal{B}\left( \mathfrak{h}\right) $ of
bounded operators such that 
\begin{equation*}
\mathrm{e}^{ith}=\left( 
\begin{array}{cc}
\mathbf{u}_{t} & \mathbf{v}_{t} \\ 
\overline{\mathbf{v}}_{t} & \overline{\mathbf{u}}_{t}%
\end{array}%
\right) \ ,\qquad t\in \mathbb{R}\ .
\end{equation*}

\item[\emph{(ii)}] Generators: If $\dim \ker h\in 2\mathbb{N}\cup \{\infty
\} $ then there are a dense subspace $\mathcal{Y}\subseteq \mathfrak{h}$, a
symmetric operator $\Upsilon _{0}$ defined on $\mathcal{Y}$ and another
operator $D_{0}=-D_{0}^{\top }$ acting on $\mathcal{CY}$ such that%
\begin{equation*}
h=\frac{1}{2}\left( 
\begin{array}{cc}
\Upsilon _{0} & 2D_{0} \\ 
-2\overline{D}_{0} & -\Upsilon _{0}^{\top }%
\end{array}%
\right) \ ,\qquad \mathcal{D}\left( h\right) =\mathcal{Y}\oplus \mathcal{CY}%
\ ,
\end{equation*}%
where $\mathcal{C}$ is the complex conjugation of Remark \ref{remark idiote
copy(1)}.
\end{enumerate}
\end{lemma}

\begin{proof}
Any bounded operator $\mathcal{U}$ acting on $\mathcal{H}$ can always be
represented on $\mathcal{H}\doteq \mathfrak{h}\oplus \mathfrak{h}$ by 
\begin{equation}
\mathcal{U}=\left( 
\begin{array}{cc}
P_{\mathfrak{h}}\mathcal{U}P_{\mathfrak{h}} & P_{\mathfrak{h}}\mathcal{U}P_{%
\mathfrak{h}}^{\perp } \\ 
P_{\mathfrak{h}}^{\perp }\mathcal{U}P_{\mathfrak{h}} & P_{\mathfrak{h}%
}^{\perp }\mathcal{U}P_{\mathfrak{h}}^{\perp }%
\end{array}%
\right)  \label{blck matrices0}
\end{equation}%
with $P_{\mathfrak{h}}$ being the orthogonal projection on $\mathfrak{h}%
\oplus \{0\}$ and $P_{\mathfrak{h}}^{\perp }\doteq \mathbf{1}-P_{\mathfrak{h}%
}$. (By a slight abuse of notation, we can use in an appropriate way the
identification $\mathfrak{h}\oplus \{0\}\equiv \mathfrak{h}$ and $%
\{0\}\oplus \mathfrak{h}\equiv \mathfrak{h}$ to see $P_{\mathfrak{h}}%
\mathcal{U}P_{\mathfrak{h}}$, $P_{\mathfrak{h}}\mathcal{U}P_{\mathfrak{h}%
}^{\perp }$, $P_{\mathfrak{h}}^{\perp }\mathcal{U}P_{\mathfrak{h}}$ and $P_{%
\mathfrak{h}}^{\perp }\mathcal{U}P_{\mathfrak{h}}^{\perp }$ as four
operators acting on $\mathfrak{h}$.) If $\mathcal{U}$ is a Bogoliubov
transformation, i.e., $\mathcal{U}\mathcal{U}^{\ast }=\mathcal{U}^{\ast }%
\mathcal{U}=\mathbf{1}$ and $\mathcal{U}\mathfrak{A}=\mathfrak{A}\mathcal{U}$
then, using (\ref{ghghghg}) together with $\overline{X}\doteq \mathcal{C}X%
\mathcal{C}$ (Remark \ref{remark idiote copy(1)}) one computes that the
above representation can be reduced to%
\begin{equation}
\mathcal{U}=\left( 
\begin{array}{cc}
\mathbf{u} & \mathbf{v} \\ 
\overline{\mathbf{v}} & \overline{\mathbf{u}}%
\end{array}%
\right) \ ,\qquad \mathbf{u},\mathbf{v}\in \mathcal{B}\left( \mathfrak{h}%
\right) \ .  \label{rrrr}
\end{equation}%
Applied to the group $(\mathrm{e}^{ith})_{t\in \mathbb{R}}$ of Bogoliubov
transformations, we get Assertion (i). It remains to represent the
self-adjoint Hamiltonian $h$ leading to the group $(\mathrm{e}^{ith})_{t\in 
\mathbb{R}}$. A priori, one can represent $h$ on $\mathcal{H}\doteq 
\mathfrak{h}\oplus \mathfrak{h}$ like (\ref{blck matrices0}) for $\mathcal{U}%
=h$, but as explained after Equation (\ref{blck matrices}), there are
nontrivial domain issues. To have some control on the domain of the
corresponding operators, we first use Lemma \ref{Lenma Bach-tech1} to deduce
that $h$ is a self-dual Hamiltonian, i.e., $h=h^{\ast }=-\mathfrak{A}h%
\mathfrak{A}$ with domain $\mathcal{D}(h)=\mathfrak{A}(\mathcal{D}%
(h))\subseteq \mathcal{\mathcal{H}}$. Secondly, recall that basis
projections are orthogonal projections $P\in \mathcal{B}(\mathcal{H})$\
satisfying $\mathfrak{A}P\mathfrak{A}=P^{\bot }\doteq \mathbf{1}_{\mathcal{H}%
}-P$. If $\dim \ker h\in 2\mathbb{N}\cup \{\infty \}$ and $h=-\mathfrak{A}h%
\mathfrak{A}$, then \cite[Lemma 4.8 and its proof]{Araki} leads to the
existence of a pair of basis projections $P$ and $P^{\bot }\doteq \mathbf{1}%
-P$ with range $\mathfrak{h}_{P}$ and $\mathfrak{h}_{P^{\bot }}$,
respectively, such that $h$ can be represented on $\mathcal{H}=\mathfrak{h}%
_{P}\oplus \mathfrak{h}_{P^{\bot }}$ by 
\begin{equation*}
h=\left( 
\begin{array}{cc}
\mathbb{A} & 0 \\ 
0 & -\overline{\mathbb{A}}%
\end{array}%
\right) =\left( 
\begin{array}{cc}
\mathbb{A} & 0 \\ 
0 & -\mathbb{A}^{\top }%
\end{array}%
\right) \ ,
\end{equation*}%
where $\mathbb{A}=\mathbb{A}^{\ast }$ is a (possibility unbounded)
self-adjoint operator with domain $\mathcal{D}\left( \mathbb{A}\right)
\subseteq \mathfrak{h}_{P}$. Even in the finite dimension situation, one can
use the identification $\mathfrak{h}_{P}\equiv \mathfrak{A}\mathfrak{h}_{P}=%
\mathfrak{h}_{P^{\bot }}$ and see $-\mathbb{A}^{\top }$ as an operator
acting on $\mathfrak{h}_{P^{\bot }}$. We meanwhile observe from \cite[Lemma
4.8]{Araki} that 
\begin{equation}
\mathcal{D}(h)=\mathcal{D}(\mathbb{A})\oplus \mathcal{D}(\mathbb{A}^{\top
})\ .  \label{ssssss}
\end{equation}%
In the above matrix representation of $h$, note that we used $X^{\top
}\doteq \mathcal{C}X^{\ast }\mathcal{C}$ (Remark \ref{remark idiote copy(1)}%
), which yields $\mathbb{A}^{\top }=\overline{\mathbb{A}}=\mathcal{C\mathbb{A%
}C}$. By \cite[Lemma 3.6]{Araki}, there is a Bogoliubov transformation $%
\mathcal{U}$ such that $h$ can be represented on $\mathcal{H}\doteq 
\mathfrak{h}\oplus \mathfrak{h}$ via the following product 
\begin{equation*}
h=\mathcal{U}^{\ast }\left( 
\begin{array}{cc}
\mathbb{A} & 0 \\ 
0 & -\mathbb{A}^{\top }%
\end{array}%
\right) \mathcal{U}\ .
\end{equation*}%
In this case, by \cite[Proof of Lemma 3.6]{Araki} and $\mathcal{D}(h)=%
\mathfrak{A}(\mathcal{D}(h))$, one checks from (\ref{ssssss}) the existence
of one subspace $\mathcal{Y}\subseteq \mathfrak{h}$ such that 
\begin{equation}
\mathcal{D}\left( h\right) =\mathcal{Y}\oplus \mathcal{CY}\ ,  \label{yyyyy}
\end{equation}%
where $\oplus $ stands (as in all the paper) for the (Hilbert) direct sums
of Hilbert (sub)spaces (and not the algebraic direct sum). In fact, using (%
\ref{rrrr}), $h$ can be represented on $\mathcal{H}\doteq \mathfrak{h}\oplus 
\mathfrak{h}$ by the following block operator matrix: 
\begin{eqnarray}
h &=&\left( 
\begin{array}{cc}
\mathbf{u}^{\ast } & \mathbf{v}^{\top } \\ 
\mathbf{v}^{\ast } & \mathbf{u}^{\top }%
\end{array}%
\right) \left( 
\begin{array}{cc}
\mathbb{A} & 0 \\ 
0 & -\mathbb{A}^{\top }%
\end{array}%
\right) \left( 
\begin{array}{cc}
\mathbf{u} & \mathbf{v} \\ 
\overline{\mathbf{v}} & \overline{\mathbf{u}}%
\end{array}%
\right)  \notag \\
&=&\left( 
\begin{array}{cc}
\mathbf{u}^{\ast } & \mathbf{v}^{\top } \\ 
\mathbf{v}^{\ast } & \mathbf{u}^{\top }%
\end{array}%
\right) \left( 
\begin{array}{cc}
\mathbb{A}\mathbf{u} & \mathbb{A}\mathbf{v} \\ 
-\mathbb{A}^{\top }\overline{\mathbf{v}} & -\mathbb{A}^{\top }\overline{%
\mathbf{u}}%
\end{array}%
\right)  \label{ssdsdsdsdsds} \\
&=&\left( 
\begin{array}{cc}
\mathbf{u}^{\ast }\mathbb{A}\mathbf{u}-\mathbf{v}^{\top }\mathbb{A}^{\top }%
\overline{\mathbf{v}} & \mathbf{u}^{\ast }\mathbb{A}\mathbf{v}-\mathbf{v}%
^{\top }\mathbb{A}^{\top }\overline{\mathbf{u}} \\ 
\mathbf{v}^{\ast }\mathbb{A}\mathbf{u}-\mathbf{u}^{\top }\mathbb{A}^{\top }%
\overline{\mathbf{v}} & \mathbf{v}^{\ast }\mathbb{A}\mathbf{v}-\mathbf{u}%
^{\top }\mathbb{A}^{\top }\overline{\mathbf{u}}%
\end{array}%
\right)  \label{ssdssdsds}
\end{eqnarray}%
and the subspace $\mathcal{Y}$ of Equation (\ref{yyyyy}) can be defined by 
\begin{equation*}
\mathcal{Y}\doteq \left\{ \varphi \in \mathfrak{h}:\mathbf{u}\varphi \in 
\mathcal{D}(\mathbb{A}),\ \overline{\mathbf{v}}\varphi \in \mathcal{D}(%
\mathbb{A}^{\top })\right\} =\left\{ \varphi \in \mathfrak{h}:\mathbf{u}%
\varphi ,\mathbf{v}\varphi \in \mathcal{D}(\mathbb{A})\right\} \subseteq 
\mathfrak{h}\ .
\end{equation*}%
If $\mathcal{D}\left( h\right) $ is a dense set in $\mathcal{H}$ then $%
\mathcal{Y}$ must\footnote{%
By (\ref{yyyyy}) and density of $\mathcal{D}\left( h\right) \subseteq 
\mathcal{H}\doteq \mathfrak{h}\oplus \mathfrak{h}$, for any $\varepsilon >0$
and $\varphi \in \mathfrak{h}$, there are $\psi _{1}\in \mathcal{Y}$ and $%
\psi _{2}\in \mathcal{CY}$ such that $\Vert \psi _{1}-\varphi \Vert _{%
\mathfrak{h}}\leq \Vert \psi _{1}-\varphi \Vert _{\mathfrak{h}}+\Vert \psi
_{2}-\varphi \Vert _{\mathfrak{h}}=\Vert \left( \psi _{1},\psi _{2}\right)
-\left( \varphi ,0\right) \Vert _{\mathcal{H}}<\varepsilon $.} be a dense
set of $\mathfrak{h}$. Define 
\begin{equation*}
\Upsilon _{0}\doteq 2\left( \mathbf{u}^{\ast }\mathbb{A}\mathbf{u}-\mathbf{v}%
^{\top }\mathbb{A}^{\top }\overline{\mathbf{v}}\right) =P_{\mathfrak{h}}hP_{%
\mathfrak{h}}\qquad \text{and}\qquad D_{0}\doteq \mathbf{u}^{\ast }\mathbb{A}%
\mathbf{v}-\mathbf{v}^{\top }\mathbb{A}^{\top }\overline{\mathbf{u}}=P_{%
\mathfrak{h}}hP_{\mathfrak{h}}^{\perp }\ ,
\end{equation*}%
which are both well-defined operators acting on $\mathfrak{h}$, with domains 
$\mathcal{Y}$ and $\mathcal{CY}$ respectively. Note that $\Upsilon
_{0}^{\top }\mathcal{C=C}\Upsilon _{0}$ and $\Upsilon _{0}^{\top }$ is
therefore defined on $\mathcal{CY}$, like $D_{0}$. Using $X^{\top }\doteq 
\mathcal{C}X^{\ast }\mathcal{C}$, $\overline{X}\doteq \mathcal{C}X\mathcal{C}
$ and $\overline{X}^{\top }=\overline{X^{\top }}=X^{\ast }$, we observe that 
$D_{0}=-D_{0}^{\top }$ on the subspace $\mathcal{CY}$ and $\Upsilon _{0}$ is
clearly a symmetric operator on $\mathcal{Y}$. Then, we obtain Assertion
(ii).
\end{proof}

Note that Lemma \ref{Lenma Bach-tech1 copy(1)} (ii) does not assert that
every self-dual Hamiltonian is compatible when $\dim \ker h\in 2\mathbb{N}%
\cup \{\infty \}$. Indeed, it does not necessarily imply that the densely
defined symmetric operator $\Upsilon _{0}$ is essentially self-adjoint, even
if $h$ is itself self-adjoint. For instance, if $h$ is self-adjoint then $%
\ker \left( h\pm i\right) =\left\{ 0\right\} $, see \cite[Proposition 3.8]%
{Konrad}. In particular, using elements $\left( \varphi ,0\right) ,\left( 0,%
\mathcal{C}\varphi \right) \in \mathcal{D}(h)$ for $\varphi \in \mathcal{Y}$
we only deduce that 
\begin{equation*}
\ker \left( \Upsilon _{0}\pm i\right) \cap \ker \left( \overline{D}%
_{0}\right) =\ker \left( \Upsilon _{0}^{\top }\pm i\right) \cap \ker \left(
D_{0}\right) =\left\{ 0\right\} \ .
\end{equation*}%
Meanwhile, $\Upsilon _{0}$ is essentially self-adjoint iff $\ker \left(
\Upsilon _{0}^{\ast }\pm i\right) =\left\{ 0\right\} $, by \cite[Proposition
3.8]{Konrad}. Since we have a priori no special control on the kernels $\ker
\left( D_{0}\right) $ and $\ker \left( \Upsilon _{0}\pm i\right) $, the
essential self-adjointness of $\Upsilon _{0}$ stays unclear, even when we
have the self-adjointness of $h$ at our disposal. Then, one may try to make
a self-adjoint extension of the densely defined symmetric operator $\Upsilon
_{0}$ , provided the deficiency indices of $\Upsilon _{0}$ are identical%
\footnote{%
If this is not the case, one may use larger Hilbert space than $\mathcal{H}$%
, by \cite[Corollary 13.11]{Konrad}.} (cf. \cite[Theorem 13.10]{Konrad}), as
explained after Equation (\ref{domain h}) in Section \ref{Section Bach}.

An updated description of available methods to extend densely defined
symmetric operators can be found in \cite[Part VI]{Konrad}. Nevertheless,
such methods are rather technical and further discussions on this subject
simply go beyond the main objective of the paper, beside the fact that
additional conditions on $h$ should probably be imposed. In fact, even Araki
in \cite[Theorems 5.4 and 5.6]{Araki} uses essentially self-adjoint,
self-dual operators $h$ which can be written as well-defined block operator
matrices and from now we consider the following assumptions on the self-dual
Hamiltonians used below:

\begin{assumption}[Self-dual Hamiltonians as Block operator matrices]
\label{Assumption1}\mbox{}\newline
$h$ is a self-dual Hamiltonian with domain $\mathcal{D}\left( h\right) =%
\mathcal{Y}\oplus \mathcal{CY}$, $\mathcal{Y}\subseteq \mathfrak{h}$ being a
dense vector subspace, which can be represented on $\mathcal{H}\doteq 
\mathfrak{h}\oplus \mathfrak{h}$ by the block operator matrix 
\begin{equation*}
h\doteq \frac{1}{2}\left( 
\begin{array}{cc}
\Upsilon _{0} & 2D_{0} \\ 
2D_{0}^{\ast } & -\Upsilon _{0}^{\top }%
\end{array}%
\right) =\frac{1}{2}\left( 
\begin{array}{cc}
\Upsilon _{0} & 2D_{0} \\ 
-2\overline{D}_{0} & -\Upsilon _{0}^{\top }%
\end{array}%
\right) =\frac{1}{2}\left( 
\begin{array}{cc}
\Upsilon _{0} & 2D_{0} \\ 
-2\overline{D}_{0} & -\overline{\Upsilon }_{0}%
\end{array}%
\right) \ ,
\end{equation*}%
where $\Upsilon _{0}=\Upsilon _{0}^{\ast }$ and $D_{0}=-D_{0}^{\top }$ are
two (possibly unbounded) operators with domains $\mathcal{D}\left( \Upsilon
_{0}\right) \supseteq \mathcal{Y}$ and $\mathcal{D}\left( D_{0}\right)
\supseteq \mathcal{C}\mathcal{Y}$. (Note that $\mathcal{D}(\Upsilon
_{0}^{\top })=\mathcal{CD}(\Upsilon _{0})$ since $\Upsilon _{0}^{\top }=%
\overline{\Upsilon }_{0}=\mathcal{C}\Upsilon _{0}\mathcal{C}$.)
\end{assumption}

A group $(\mathrm{e}^{ith})_{t\in \mathbb{R}}$ of Bogoliubov
transformations, represented by Lemma \ref{Lenma Bach-tech1 copy(1)} (i) on $%
\mathcal{H}\doteq \mathfrak{h}\oplus \mathfrak{h}$, satisfies the
Shale-Stinespring condition at some time $t\in \mathbb{R}$ whenever its
off-diagonal operator $\mathbf{v}_{t}$ is Hilbert-Schmidt, i.e., $\mathbf{v}%
_{t}\in \mathcal{L}^{2}(\mathfrak{h})$. See Theorem \ref{Shale-Stinespring
THM}. Meanwhile, if $h$ satisfies Assumption \ref{Assumption1}, with
off-diagonal operator $D_{0}$ acting on the Hilbert space $\mathfrak{h}$,
then we would like to understand the relations between the Hilbert-Schmidt
properties of $\mathbf{v}_{t}$ and $D_{0}$. To analyze this, it is very
convenient to formulate the problem by means of the theory of non-autonomous
evolution equations \cite{Katobis,Caps,Schnaubelt1,Pazy,Neidhardt-zagrebnov}.

\begin{lemma}[Interaction picture of groups of Bogoliubov transformations]
\label{Lenma Bach-tech2 copy(1)}\mbox{
}\newline
Take any self-adjoint operator $h$ satisfying Assumption \ref{Assumption1}
and use Lemmata \ref{Lenma Bach-tech1} and \ref{Lenma Bach-tech1 copy(1)}
(i) to represent the group $(\mathrm{e}^{ith})_{t\in \mathbb{R}}$ of
Bogoliubov transformations. Let $c\in \mathbb{R}^{+}\cup \{\infty \}$ and
assume that 
\begin{equation}
\mathrm{e}^{it\Upsilon _{0}}\mathcal{Y}\subseteq \mathcal{Y}\ ,\qquad t\in
\left( -c,c\right) \ ,  \label{assumption}
\end{equation}%
with the operator family $(\overline{D}_{0}\mathrm{e}^{it\Upsilon
_{0}})_{t\in (-c,c)}$ being strongly continuous on $\mathcal{Y}$. Then, the
block operator matrix%
\begin{equation*}
\left( 
\begin{array}{cc}
\mathbf{x}_{t,s} & \mathbf{y}_{t,s} \\ 
\mathbf{\tilde{y}}_{t,s} & \mathbf{\tilde{x}}_{t,s}%
\end{array}%
\right) \doteq \left( 
\begin{array}{cc}
\mathrm{e}^{-it\Upsilon _{0}/2}\mathbf{u}_{t-s}\mathrm{e}^{is\Upsilon _{0}/2}
& \mathrm{e}^{-it\Upsilon _{0}/2}\mathbf{v}_{t-s}\mathrm{e}^{-is\Upsilon
_{0}^{\top }/2} \\ 
\mathrm{e}^{it\Upsilon _{0}^{\top }/2}\overline{\mathbf{v}}_{t-s}\mathrm{e}%
^{is\Upsilon _{0}/2} & \mathrm{e}^{it\Upsilon _{0}^{\top }/2}\overline{%
\mathbf{u}}_{t-s}\mathrm{e}^{-is\Upsilon _{0}^{\top }/2}%
\end{array}%
\right) \ ,\qquad s,t\in \mathbb{R}\ ,
\end{equation*}%
on $\mathcal{H}\doteq \mathfrak{h}\oplus \mathfrak{h}$ defines a strongly
continuous two-parameter family $(V_{t,s})_{s,t\in \mathbb{R}}$ of unitary
operators satisfying, for any $s,t\in (-c,c)$,%
\begin{equation*}
\left( 
\begin{array}{cc}
\partial _{t}\mathbf{x}_{t,s} & \partial _{t}\mathbf{y}_{t,s} \\ 
\partial _{t}\mathbf{\tilde{y}}_{t,s} & \partial _{t}\mathbf{\tilde{x}}_{t,s}%
\end{array}%
\right) =i\left( 
\begin{array}{cc}
C_{t}\mathbf{\tilde{y}}_{t,s} & C_{t}\mathbf{\tilde{x}}_{t,s} \\ 
C_{t}^{\ast }\mathbf{x}_{t,s} & C_{t}^{\ast }\mathbf{y}_{t,s}%
\end{array}%
\right) \quad \text{and}\quad \left( 
\begin{array}{cc}
\partial _{s}\mathbf{x}_{t,s} & \partial _{s}\mathbf{y}_{t,s} \\ 
\partial _{s}\mathbf{\tilde{y}}_{t,s} & \partial _{s}\mathbf{\tilde{x}}_{t,s}%
\end{array}%
\right) =-i\left( 
\begin{array}{cc}
\mathbf{y}_{t,s}C_{s}^{\ast } & \mathbf{x}_{t,s}C_{s} \\ 
\mathbf{\tilde{x}}_{t,s}C_{s}^{\ast } & \mathbf{\tilde{y}}_{t,s}C_{s}%
\end{array}%
\right)
\end{equation*}%
on the dense set $\mathcal{D}\left( h\right) \subseteq \mathcal{H}$, where%
\begin{equation}
C_{t}\doteq \mathrm{e}^{-it\Upsilon _{0}/2}D_{0}\mathrm{e}^{-it\Upsilon
_{0}^{\top }/2}=-C_{t}^{\top }\ ,\qquad t\in \left( -c,c\right) \ .
\label{sssss4}
\end{equation}
\end{lemma}

\begin{proof}
Assume all conditions of the lemma. The proof is rather elementary. For
pedagogical reasons, it is divided in two steps: \medskip

\noindent \underline{Step 1:} Define on $\mathcal{H}\doteq \mathfrak{h}%
\oplus \mathfrak{h}$ the self-adjoint operators 
\begin{equation*}
\gamma _{0}\doteq \frac{1}{2}\left( 
\begin{array}{cc}
\Upsilon _{0} & 0 \\ 
0 & -\Upsilon _{0}^{\top }%
\end{array}%
\right) =\gamma _{0}^{\ast }\qquad \text{and}\qquad d_{0}\doteq \left( 
\begin{array}{cc}
0 & D_{0} \\ 
D_{0}^{\ast } & 0%
\end{array}%
\right) =\left( 
\begin{array}{cc}
0 & D_{0} \\ 
-\overline{D}_{0} & 0%
\end{array}%
\right) =d_{0}^{\ast }\ ,
\end{equation*}%
the domains of which equal 
\begin{equation*}
\mathcal{D}\left( \gamma _{0}\right) =\mathcal{D}\left( \Upsilon _{0}\right)
\oplus \mathcal{D}\left( \Upsilon _{0}^{\top }\right) \qquad \text{and}%
\qquad \mathcal{D}\left( d_{0}\right) =\mathcal{D}\left( \overline{D}%
_{0}\right) \oplus \mathcal{D}\left( D_{0}\right) \ .
\end{equation*}%
Note from the self-adjointness of $\Upsilon _{0}$ that $\mathcal{D}\left(
\Upsilon _{0}^{\top }\right) =\mathcal{C}\mathcal{D}\left( \Upsilon
_{0}\right) $. Compare with (\ref{yyyyy}). Their sum is the self-adjoint
operator $h=\gamma _{0}+d_{0}$ on the domain 
\begin{equation*}
\mathcal{D}\left( h\right) =\mathcal{Y}\oplus \mathcal{CY\subseteq D}\left(
\gamma _{0}\right) \cap \mathcal{D}\left( d_{0}\right) =\mathcal{D}\left(
\Upsilon _{0}\right) \cap \mathcal{D}\left( \overline{D}_{0}\right) \oplus 
\mathcal{D}\left( \Upsilon _{0}^{\top }\right) \cap \mathcal{D}\left(
D_{0}\right) \ .
\end{equation*}%
Note that the family $(\mathcal{C}\mathrm{e}^{it\Upsilon _{0}^{\top }}%
\mathcal{C})_{t\in \mathbb{R}}$ is a strongly continuous one-parameter group
of unitary operators on $\mathcal{H}$ and we deduce from the theory of
strongly continuous unitary groups (in particular Stone's theorem \cite[%
Theorem 6.2]{Konrad}) together with the antiunitarity of $\mathcal{C}$, $%
\mathcal{C}^{2}=\mathbf{1}$ and $X^{\top }\doteq \mathcal{C}X^{\ast }%
\mathcal{C}$ (Remark \ref{remark idiote copy(1)}) that 
\begin{equation}
\mathcal{C}\mathrm{e}^{it\Upsilon _{0}^{\top }}\mathcal{C}=\mathrm{e}^{-it%
\mathcal{C}\Upsilon _{0}^{\top }\mathcal{C}}=\mathrm{e}^{-it\Upsilon _{0}}\
,\qquad t\in \mathbb{R}\ .  \label{sdsdsdsdd}
\end{equation}%
By Condition (\ref{assumption}) and recalling again that $\mathcal{C}$ is an
involution, it follows that 
\begin{equation}
\mathrm{e}^{it\Upsilon _{0}^{\top }}\mathcal{CY}\subseteq \mathcal{CY}\
,\qquad t\in \left( -c,c\right) \ ,  \label{assumption2}
\end{equation}%
and therefore, 
\begin{equation}
\mathrm{e}^{it\gamma _{0}}\mathcal{D}\left( h\right) \subseteq \mathcal{D}%
\left( \gamma _{0}\right) \cap \mathcal{D}\left( d_{0}\right) \ ,\qquad t\in
\left( -c,c\right) \ .  \label{assumption 3}
\end{equation}%
Meanwhile, $\overline{X}\doteq \mathcal{C}X\mathcal{C}$, 
\begin{equation}
\mathcal{C}\mathrm{e}^{it\Upsilon _{0}}\mathcal{C}=\mathrm{e}^{-it\mathcal{C}%
\Upsilon _{0}\mathcal{C}}=\mathrm{e}^{-it\Upsilon _{0}^{\top }}\ ,\qquad
t\in \mathbb{R}\ .  \label{sdsdsdsdd2}
\end{equation}%
(cf. (\ref{sdsdsdsdd}))\ and $(\overline{D}_{0}\mathrm{e}^{it\Upsilon
_{0}})_{t\in (-c,c)}$ is by assumption strongly continuous on $\mathcal{Y}$.
Hence, the collection $(D_{0}\mathrm{e}^{it\Upsilon _{0}^{\top }})_{t\in
(-c,c)}$ is a well-defined strongly continuous family of operators on $%
\mathcal{CY}$. As a consequence $(C_{t})_{t\in (-c,c)}$ is also a strongly
continuous family of operators on $\mathcal{CY}$, because $(\mathrm{e}%
^{it\Upsilon _{0}})_{t\in \mathbb{R}}$ is a strongly continuous group of
unitary operators as $\Upsilon _{0}=\Upsilon _{0}^{\ast }$. Note
additionally from $D_{0}=-D_{0}^{\top }$, $\mathcal{C}^{2}=\mathbf{1}$, (\ref%
{sdsdsdsdd}) and (\ref{sdsdsdsdd2}) that 
\begin{equation*}
C_{t}^{\top }\doteq \mathcal{C}C_{t}^{\ast }\mathcal{C}=(\mathcal{C}\mathrm{e%
}^{it\Upsilon _{0}^{\top }/2}\mathcal{C})(\mathcal{C}D_{0}^{\ast }\mathcal{C}%
)(\mathcal{C}\mathrm{e}^{it\Upsilon _{0}/2}\mathcal{C})=\mathrm{e}%
^{-it\Upsilon _{0}/2}(-D_{0})\mathrm{e}^{-it\Upsilon _{0}^{\top }/2}=-C_{t}
\end{equation*}%
for all times $t\in \mathbb{R}$. \medskip

\noindent \underline{Step 2:} We are now in a position to use evolution
equations to prove the lemma. We compute that, for any $s,t\in (-c,c)$,%
\begin{equation*}
\partial _{t}\left\{ \mathrm{e}^{-it\gamma _{0}}\mathrm{e}^{i\left(
t-s\right) h}\mathrm{e}^{is\gamma _{0}}\right\} =i\left( \mathrm{e}%
^{-it\gamma _{0}}d_{0}\mathrm{e}^{it\gamma _{0}}\right) \left( \mathrm{e}%
^{-it\gamma _{0}}\mathrm{e}^{i\left( t-s\right) h}\mathrm{e}^{is\gamma
_{0}}\right)
\end{equation*}%
as well as 
\begin{equation*}
\partial _{s}\left\{ \mathrm{e}^{-it\gamma _{0}}\mathrm{e}^{i\left(
t-s\right) h}\mathrm{e}^{is\gamma _{0}}\right\} =\left( \mathrm{e}%
^{-it\gamma _{0}}\mathrm{e}^{i\left( t-s\right) h}\mathrm{e}^{is\gamma
_{0}}\right) \left( -i\mathrm{e}^{-is\gamma _{0}}d_{0}\mathrm{e}^{is\gamma
_{0}}\right)
\end{equation*}%
both in the strong sense on the dense set $\mathcal{Y}\oplus \mathcal{CY}%
\subseteq \mathcal{H}$. In other words, the family 
\begin{equation}
V_{t,s}\doteq \mathrm{e}^{-it\gamma _{0}}\mathrm{e}^{i\left( t-s\right) h}%
\mathrm{e}^{is\gamma _{0}}\ ,\qquad s,t\in \mathbb{R}\ ,  \label{ssssssss}
\end{equation}%
of (uniformly) bounded operators is a strongly continuous two-parameter
family of unitary operators solving the non-autonomous evolution equations%
\begin{equation}
\forall s,t\in (-c,c):\qquad \partial _{t}Z_{t,s}=id_{t}Z_{t,s}\ ,\qquad
\partial _{s}Z_{t,s}=-iZ_{t,s}d_{s}\ ,\qquad Z_{s,s}=\mathbf{1}\ ,
\label{sssss1}
\end{equation}%
in the strong sense on the dense set $\mathcal{Y}\oplus \mathcal{CY}%
\subseteq \mathcal{H}$, where 
\begin{equation*}
d_{t}\doteq \mathrm{e}^{-it\gamma _{0}}d_{0}\mathrm{e}^{it\gamma
_{0}}=d_{t}^{\ast }\ ,\qquad t\in (-c,c)\ .
\end{equation*}%
This formulation of the strongly continuous one-parameter group $(\mathrm{e}%
^{ith})_{t\in \mathbb{R}}$ of Bogoliubov transformations via $%
(V_{t,s})_{s,t\in \mathbb{R}}$ is advantageous because the operator family $%
(d_{t})_{t\in (-c,c)}$ can be represented in a simple way on the Hilbert
space $\mathcal{H}\doteq \mathfrak{h}\oplus \mathfrak{h}$:%
\begin{equation}
d_{t}=\left( 
\begin{array}{cc}
\mathrm{e}^{-it\Upsilon _{0}/2} & 0 \\ 
0 & \mathrm{e}^{it\Upsilon _{0}^{\top }/2}%
\end{array}%
\right) \left( 
\begin{array}{cc}
0 & D_{0} \\ 
D_{0}^{\ast } & 0%
\end{array}%
\right) \left( 
\begin{array}{cc}
\mathrm{e}^{it\Upsilon _{0}/2} & 0 \\ 
0 & \mathrm{e}^{-it\Upsilon _{0}^{\top }/2}%
\end{array}%
\right) =\left( 
\begin{array}{cc}
0 & C_{t} \\ 
C_{t}^{\ast } & 0%
\end{array}%
\right)  \label{sssss3}
\end{equation}%
for each time $t\in \mathbb{R}$, where $C_{t}$ is the operator defined by (%
\ref{sssss4}). Note that $(C_{t})_{t\in (-c,c)}$ is a strongly continuous
family of operators on $\mathcal{CY}$ and so, $(d_{t})_{t\in (-c,c)}$ is a
strongly continuous family of operators on $\mathcal{D}\left( h\right) =%
\mathcal{Y}\oplus \mathcal{CY}$. The lemma is then obtained by writing all
operators on $\mathcal{H}$ as well as the non-autonomous evolution equations
in terms of block operator matrices. For instance, one computes from Lemma %
\ref{Lenma Bach-tech1 copy(1)} (i) that 
\begin{equation*}
V_{t,s}\doteq \left( 
\begin{array}{cc}
\mathbf{x}_{t,s} & \mathbf{y}_{t,s} \\ 
\mathbf{\tilde{y}}_{t,s} & \mathbf{\tilde{x}}_{t,s}%
\end{array}%
\right) =\left( 
\begin{array}{cc}
\mathrm{e}^{-it\Upsilon _{0}/2}\mathbf{u}_{t-s}\mathrm{e}^{is\Upsilon _{0}/2}
& \mathrm{e}^{-it\Upsilon _{0}/2}\mathbf{v}_{t-s}\mathrm{e}^{-is\Upsilon
_{0}^{\top }/2} \\ 
\mathrm{e}^{it\Upsilon _{0}^{\top }/2}\overline{\mathbf{v}}_{t-s}\mathrm{e}%
^{is\Upsilon _{0}/2} & \mathrm{e}^{it\Upsilon _{0}^{\top }/2}\overline{%
\mathbf{u}}_{t-s}\mathrm{e}^{-is\Upsilon _{0}^{\top }/2}%
\end{array}%
\right)
\end{equation*}%
for any $s,t\in \mathbb{R}$.
\end{proof}

Observe that Assumption \ref{Assumption1} and Condition (\ref{assumption})
imply that $\mathcal{Y}$ must be a core for the self-adjoint operator $%
\Upsilon _{0}$, by \cite[Proposition 6.3]{Konrad}. In other words, $\Upsilon
_{0}$ restricted to the subspace $\mathcal{Y}\subseteq \mathcal{D}\left(
\Upsilon _{0}\right) $ is essentially self-adjoint. Note also that Lemma \ref%
{Lenma Bach-tech2 copy(1)} can always be applied whenever the off-diagonal
operator-valued coefficient $\overline{D}_{0}$ is relatively bounded with
respect to the diagonal one $\Upsilon _{0}$ (see (\ref{rel})). This is a
direct consequence of the following lemma:

\begin{lemma}[Relative boundedness of off-diagonal coefficients]
\label{Lenma Bach-tech2 copy(2)}\mbox{
}\newline
Let $\Upsilon _{0}=\Upsilon _{0}^{\ast }$ and $D_{0}$ be two (possibly
unbounded) operators with domains $\mathcal{Y}\doteq \mathcal{D}\left(
\Upsilon _{0}\right) $ and $\mathcal{D}\left( \overline{D}_{0}\right)
\supseteq \mathcal{D}\left( \Upsilon _{0}\right) $. If 
\begin{equation}
\left\Vert \overline{D}_{0}\left( \Upsilon _{0}+i\mathbf{1}\right)
^{-1}\right\Vert _{\mathrm{op}}<\infty \ ,  \label{relative boundedness}
\end{equation}%
then $\mathrm{e}^{it\Upsilon _{0}}\mathcal{Y}\subseteq \mathcal{Y}$ for any $%
t\in \mathbb{R}$ and the operator family $(\overline{D}_{0}\mathrm{e}%
^{it\Upsilon _{0}})_{t\in \mathbb{R}}$ is strongly continuous on the dense
vector subspace $\mathcal{Y}\subseteq \mathfrak{h}$.
\end{lemma}

\begin{proof}
As is well-known, if $\Upsilon _{0}$ is a self-adjoint operator, then $%
\mathcal{Y\doteq D}\left( \Upsilon _{0}\right) $ is a dense vector subspace
of $\mathfrak{h}$ which is preserved by $\mathrm{e}^{it\Upsilon _{0}}$, i.e. 
$\mathrm{e}^{it\Upsilon _{0}}\mathcal{Y}\subseteq \mathcal{Y}$ for any $t\in 
\mathbb{R}$ and $\mathrm{e}^{it\Upsilon _{0}}\Upsilon _{0}=\Upsilon _{0}%
\mathrm{e}^{it\Upsilon _{0}}$ on $\mathcal{Y}$. See for instance \cite[%
Chapter II, Lemma 1.3 (ii)]{EngelNagel}. Moreover, for any $\varphi \in 
\mathcal{Y}$ and $s,t\in \mathbb{R}$,%
\begin{equation*}
\left\Vert \overline{D}_{0}\left( \mathrm{e}^{it\Upsilon _{0}}-\mathrm{e}%
^{is\Upsilon _{0}}\right) \varphi \right\Vert _{\mathfrak{h}}\leq \left\Vert 
\overline{D}_{0}\left( \Upsilon _{0}+i\mathbf{1}\right) ^{-1}\right\Vert _{%
\mathrm{op}}\left\Vert \left( \mathrm{e}^{it\Upsilon _{0}}-\mathrm{e}%
^{is\Upsilon _{0}}\right) \left( \Upsilon _{0}+i\mathbf{1}\right) \varphi
\right\Vert _{\mathfrak{h}}\ .
\end{equation*}%
Therefore, if (\ref{relative boundedness}) holds true, then the operator
family $(\overline{D}_{0}\mathrm{e}^{it\Upsilon _{0}})_{t\in \mathbb{R}}$ is
strongly continuous on $\mathcal{Y}$, because $(\mathrm{e}^{it\Upsilon
_{0}})_{t\in \mathbb{R}}$ is a strongly continuous group of unitary
operators as $\Upsilon _{0}=\Upsilon _{0}^{\ast }$.
\end{proof}

\begin{remark}
Under the same assumptions as Lemma \ref{Lenma Bach-tech2 copy(2)} by
replacing $\overline{D}_{0}$ with $D_{0}$, the family $(D_{0}\mathrm{e}%
^{it\Upsilon _{0}^{\top }})_{t\in \mathbb{R}}$ is strongly continuous on $%
\mathcal{C}\mathcal{Y}$.
\end{remark}

Condition (\ref{relative boundedness}) is a very natural condition to
consider in order to be able to define block operator matrices as at least
closable operators, like in Assumption \ref{Assumption1}. See for instance 
\cite{block1,block6,block7}. The best situation is when the off-diagonal
operator-valued coefficient is directly bounded, i.e., $D_{0}\in \mathcal{B}(%
\mathfrak{h})$. In the following lemma, we give a sufficient condition in
terms of groups of Bogoliubov transformations to have their generator $ih$
with bounded off-diagonal term $D_{0}\in \mathcal{B}(\mathfrak{h})$:

\begin{lemma}[Boundedness of off-diagonal coefficients]
\label{Lenma Bach-tech4 copy(1)}\mbox{
}\newline
Under conditions of Lemma \ref{Lenma Bach-tech2 copy(1)}, if 
\begin{equation}
\liminf_{s\rightarrow 0^{-}}s^{-1}\left\Vert \mathbf{v}_{s}\right\Vert _{%
\mathrm{op}}<\infty \ ,  \label{limit bounded}
\end{equation}%
then $D_{0}$ can be extended by continuity to a bounded operator (again
denoted by $D_{0}\in \mathcal{B}(\mathfrak{h})$, by a slight abuse of
notation).
\end{lemma}

\begin{proof}
Using Lemma \ref{Lenma Bach-tech2 copy(1)} and the triangle inequality, for
any $\varphi \in \mathcal{C}\mathcal{Y}$ and $s\in \left( -c,0\right) $, we
have 
\begin{equation*}
\left\Vert D_{0}\varphi \right\Vert _{\mathfrak{h}}\leq \left\vert
s\right\vert ^{-1}\int_{s}^{0}\left\Vert \left( D_{0}-C_{\tau }\right)
\varphi \right\Vert _{\mathfrak{h}}\mathrm{d}\tau +\left\Vert
s^{-1}\int_{s}^{0}C_{\tau }\varphi \mathrm{d}\tau \right\Vert _{\mathfrak{h}}
\end{equation*}%
while 
\begin{equation*}
\mathbf{y}_{0,s}\varphi =i\int_{s}^{0}\mathbf{x}_{0,\tau }C_{\tau }\varphi 
\mathrm{d}\tau \ .
\end{equation*}%
It follows that, for any $\varphi \in \mathcal{C}\mathcal{Y}$ and $s\in
\left( -c,0\right) $,%
\begin{equation}
\left\Vert D_{0}\varphi \right\Vert _{\mathfrak{h}}\leq \left\vert
s\right\vert ^{-1}\left\Vert \mathbf{y}_{0,s}\right\Vert _{\mathrm{op}%
}\left\Vert \varphi \right\Vert _{\mathfrak{h}}+\left\vert s\right\vert
^{-1}\int_{s}^{0}\left\Vert \left( D_{0}-C_{\tau }\right) \varphi
\right\Vert _{\mathfrak{h}}\mathrm{d}\tau +\left\vert s\right\vert
^{-1}\int_{s}^{0}\left\Vert \left( \mathbf{x}_{0,\tau }-\mathbf{1}\right)
C_{\tau }\varphi \right\Vert _{\mathfrak{h}}\mathrm{d}\tau \ .  \label{dddd0}
\end{equation}%
Recall that $(C_{t})_{t\in (-c,c)}$ is a strongly continuous family of
operators on $\mathcal{CY}$ with $C_{0}=D_{0}$. Therefore, for any $\varphi
\in \mathcal{C}\mathcal{Y}$, 
\begin{equation}
\lim_{s\rightarrow 0^{-}}\left\vert s\right\vert ^{-1}\int_{s}^{0}\left\Vert
\left( D_{0}-C_{\tau }\right) \varphi \right\Vert _{\mathfrak{h}}\mathrm{d}%
\tau =0\ .  \label{dddd}
\end{equation}%
Since $\mathrm{e}^{ith}$ is a Bogoliubov transformation for all times $t\in 
\mathbb{R}$, one checks that 
\begin{equation}
\mathbf{u}_{t}^{\ast }\mathbf{u}_{t}+\mathbf{v}_{t}^{\top }\overline{\mathbf{%
v}}_{t}=\mathbf{1}\ ,  \label{eqdfdf}
\end{equation}%
see for instance Equation (\ref{CCRu,v}). It follows that $\mathbf{u}%
_{t}^{\ast }\mathbf{u}_{t},\mathbf{v}_{t}^{\top }\overline{\mathbf{v}}%
_{t}\leq \mathbf{1}$ and therefore, 
\begin{equation}
\left\Vert \mathbf{u}_{t}\right\Vert _{\mathrm{op}},\left\Vert \mathbf{v}%
_{t}\right\Vert _{\mathrm{op}}\leq 1\ ,\qquad t\in \mathbb{R}\ .
\label{sssssssss}
\end{equation}%
In particular, through Lemma \ref{Lenma Bach-tech2 copy(1)}, we deduce that 
\begin{equation*}
\sup_{s,t\in \mathbb{R}}\left\Vert \mathbf{x}_{t,s}\right\Vert _{\mathrm{op}%
}=\sup_{s,t\in \mathbb{R}}\left\Vert \mathrm{e}^{-it\Upsilon _{0}/2}\mathbf{u%
}_{t-s}\mathrm{e}^{is\Upsilon _{0}/2}\right\Vert _{\mathrm{op}}\leq 1
\end{equation*}%
and, by the triangle inequality, 
\begin{equation}
\left\vert s\right\vert ^{-1}\int_{s}^{0}\left\Vert \left( \mathbf{x}%
_{0,\tau }-\mathbf{1}\right) C_{\tau }\varphi \right\Vert _{\mathfrak{h}}%
\mathrm{d}\tau \leq 2\left\vert s\right\vert ^{-1}\int_{s}^{0}\left\Vert
\left( C_{\tau }-D_{0}\right) \varphi \right\Vert _{\mathfrak{h}}\mathrm{d}%
\tau +\left\vert s\right\vert ^{-1}\int_{s}^{0}\left\Vert \left( \mathbf{x}%
_{0,\tau }-\mathbf{1}\right) D_{0}\varphi \right\Vert _{\mathfrak{h}}\mathrm{%
d}\tau \ .  \label{check2}
\end{equation}%
The family $(\mathrm{e}^{ith})_{t\in \mathbb{R}}$ and $(\mathrm{e}%
^{it\Upsilon _{0}})_{t\in \mathbb{R}}$ are both strongly continuous group of
unitary operators. In particular, $(\mathbf{x}_{0,\tau }=\mathbf{u}_{-\tau }%
\mathrm{e}^{i\tau \Upsilon _{0}})_{t\in \mathbb{R}}$ is a strongly
continuous family of bounded operators satisfying $\mathbf{x}_{0,0}=\mathbf{1%
}$, while $(C_{t})_{t\in (-c,c)}$ is a strongly continuous family of
operators on $\mathcal{CY}$ with $C_{0}=D_{0}$. Therefore, we infer from (%
\ref{check2}) that, for any $\varphi \in \mathcal{C}\mathcal{Y}$, 
\begin{equation}
\lim_{s\rightarrow 0^{-}}\left\vert s\right\vert ^{-1}\int_{s}^{0}\left\Vert
\left( \mathbf{x}_{0,\tau }-\mathbf{1}\right) C_{\tau }\varphi \right\Vert _{%
\mathfrak{h}}\mathrm{d}\tau =0\ .  \label{dddd2}
\end{equation}%
Now, we combine (\ref{dddd0}) with the limits (\ref{dddd}) and (\ref{dddd2})
to arrive at 
\begin{equation}
\left\Vert D_{0}\varphi \right\Vert _{\mathfrak{h}}\leq \left\Vert \varphi
\right\Vert _{\mathfrak{h}}\liminf_{s\rightarrow 0^{-}}s^{-1}\left\Vert 
\mathbf{v}_{s}\right\Vert _{\mathrm{op}}  \label{dddd3}
\end{equation}%
for any $\varphi \in \mathcal{C}\mathcal{Y}$. Note that $\mathcal{Y}%
\subseteq \mathfrak{h}$ is by assumption a dense vector subspace and $%
\mathcal{C}^{2}=\mathbf{1}$. Therefore, $\mathcal{C}\mathcal{Y}$ is also a
dense vector space and if (\ref{limit bounded}) holds true then we deduce
from (\ref{dddd3}) that $D_{0}$ can be extended by continuity to a bounded
operator acting on $\mathfrak{h}$.
\end{proof}

We focus now on bounded off-diagonal operator-valued coefficients, i.e., $%
D_{0}\in \mathcal{B}(\mathfrak{h})$. It is worth mentioning that the
diagonal element $\Upsilon _{0}$ is still taken as a possibly unbounded
operator. In this case, using very standard arguments with Lemmata \ref%
{Lenma Bach-tech2 copy(1)} and \ref{Lenma Bach-tech2 copy(2)}, we easily
write all operator-valued coefficients of groups of Bogoliubov
transformations via Dyson series involving operator-valued coefficients of
their generator.

\begin{lemma}[Coefficients of Bogoliubov transformations as Dyson series]
\label{Lenma Bach-tech2}\mbox{
}\newline
Take any self-adjoint operator $h$ satisfying Assumption \ref{Assumption1}
with $D_{0}\in \mathcal{B}(\mathfrak{h})$ and use Lemmata \ref{Lenma
Bach-tech1} and \ref{Lenma Bach-tech1 copy(1)} (i) to represent the group $(%
\mathrm{e}^{ith})_{t\in \mathbb{R}}$ of Bogoliubov transformations as 
\begin{equation*}
\mathrm{e}^{ith}=\left( 
\begin{array}{cc}
\mathbf{u}_{t} & \mathbf{v}_{t} \\ 
\overline{\mathbf{v}}_{t} & \overline{\mathbf{u}}_{t}%
\end{array}%
\right) \ ,\qquad t\in \mathbb{R}\ .
\end{equation*}%
Then, for any $t\in \mathbb{R}$, 
\begin{eqnarray*}
\mathrm{e}^{-it\Upsilon _{0}/2}\mathbf{u}_{t} &=&\mathbf{1}%
+\sum_{p=1}^{\infty }\left( -1\right) ^{p}\int_{0}^{t}\mathrm{d}\tau
_{1}\cdots \int_{0}^{\tau _{2p-1}}\mathrm{d}\tau _{2p}(C_{\tau _{1}}C_{\tau
_{2}}^{\ast })\cdots (C_{\tau _{2p-1}}C_{\tau _{2p}}^{\ast })\ , \\
\mathrm{e}^{-it\Upsilon _{0}/2}\mathbf{v}_{t} &=&i\int_{0}^{t}\mathrm{d}\tau
C_{\tau }+i\sum_{p=1}^{\infty }\left( -1\right) ^{p}\int_{0}^{t}\mathrm{d}%
\tau _{1}\cdots \int_{0}^{\tau _{2p}}\mathrm{d}\tau _{2p+1}C_{\tau
_{1}}\left( (C_{\tau _{2}}^{\ast }C_{\tau _{3}})\cdots (C_{\tau _{2p}}^{\ast
}C_{\tau _{2p+1}})\right) \ ,
\end{eqnarray*}%
the series being absolutely summable in $\mathcal{B}(\mathfrak{h})$, where $%
\left( C_{t}\right) _{t\in \mathbb{R}}\subseteq \mathcal{B}(\mathfrak{h})$
is the strongly continuous family of bounded operators defined by (\ref%
{sssss4}).
\end{lemma}

\begin{proof}
Lemma \ref{Lenma Bach-tech2 copy(1)} can always be applied for any bounded $%
D_{0}\in \mathcal{B}(\mathfrak{h})$. In this situation, the parameter $c$ of
Lemma \ref{Lenma Bach-tech2 copy(1)} is $c=\infty $, as one can see from
Lemma \ref{Lenma Bach-tech2 copy(2)}. For instance, under all the
assumptions of the lemma, by Equation (\ref{sssss3}), $(d_{t})_{t\in \mathbb{%
R}}\subseteq \mathcal{B}(\mathcal{H})$ is a strongly continuous family of
bounded operators, since $(\mathrm{e}^{it\Upsilon _{0}})_{t\in \mathbb{R}}$
and $(\mathrm{e}^{it\Upsilon _{0}^{\top }})_{t\in \mathbb{R}}$ are strongly
continuous group of unitary operators, $\Upsilon _{0}$ and $\Upsilon
_{0}^{\top }$ being self-adjoint operators. As is well-known, we can
therefore define the Dyson series 
\begin{equation}
\tilde{V}_{t,s}\doteq \mathbf{1}+\sum_{n=1}^{\infty }i^{n}\int_{s}^{t}%
\mathrm{d}\tau _{1}\cdots \int_{s}^{\tau _{n-1}}\mathrm{d}\tau _{n}d_{\tau
_{1}}\cdots d_{\tau _{n}}\mathrm{\ },\qquad s,t\in \mathbb{R}\ .
\label{dyson series}
\end{equation}%
This series is of course well-defined for any $\varphi \in \mathcal{H}$
since $\tilde{V}_{t,s}\varphi $ becomes an absolutely summable series in $%
\mathcal{B}(\mathfrak{h})$. In fact, $(\tilde{V}_{t,s})_{s,t\in \mathbb{R}}$
is another strongly continuous two-parameter operator family solving the
non-autonomous evolution equations (\ref{sssss1}) and is therefore nothing
else that $(V_{t,s})_{s,t\in \mathbb{R}}$, by uniqueness\footnote{%
Compute that, for any $s,t\in \mathbb{R}$ and $\varphi \in \mathcal{H}$, $(%
\tilde{V}_{t,s}-V_{t,s})\varphi =\int_{s}^{t}\partial _{\tau }\{V_{t,\tau }%
\tilde{V}_{\tau ,s}\}\varphi \ \mathrm{d}\tau =0$.} of the solution to (\ref%
{sssss1}). This is completely standard. Using the representation (\ref%
{sssss3}) of the operator family $(d_{t})_{t\in \mathbb{R}}$ and 
\begin{equation*}
V_{t,s}=\left( 
\begin{array}{cc}
\mathbf{x}_{t,s} & \mathbf{y}_{t,s} \\ 
\mathbf{\tilde{y}}_{t,s} & \mathbf{\tilde{x}}_{t,s}%
\end{array}%
\right) =\tilde{V}_{t,s}\ ,\qquad s,t\in \mathbb{R}\ ,
\end{equation*}%
on the Hilbert space $\mathcal{H}\doteq \mathfrak{h}\oplus \mathfrak{h}$, we
compute from direct computations of the Dyson series (\ref{dyson series})
that%
\begin{eqnarray}
\mathbf{x}_{t,s} &=&\mathbf{1}+\sum_{p=1}^{\infty }\left( -1\right)
^{p}\int_{s}^{t}\mathrm{d}\tau _{1}\cdots \int_{s}^{\tau _{2p-1}}\mathrm{d}%
\tau _{2p}(C_{\tau _{1}}C_{\tau _{2}}^{\ast })\cdots (C_{\tau
_{2p-1}}C_{\tau _{2p}}^{\ast })  \label{x11} \\
\mathbf{\tilde{x}}_{t,s} &=&\mathbf{1}+\sum_{p=1}^{\infty }\left( -1\right)
^{p}\int_{s}^{t}\mathrm{d}\tau _{1}\cdots \int_{s}^{\tau _{2p-1}}\mathrm{d}%
\tau _{2p}(C_{\tau _{1}}^{\ast }C_{\tau _{2}})\cdots (C_{\tau _{2p-1}}^{\ast
}C_{\tau _{2p}})  \label{x22} \\
\mathbf{y}_{t,s} &=&i\int_{s}^{t}\mathrm{d}\tau C_{\tau
}+i\sum_{p=1}^{\infty }\left( -1\right) ^{p}\int_{s}^{t}\mathrm{d}\tau
_{1}\cdots \int_{s}^{\tau _{2p}}\mathrm{d}\tau _{2p+1}C_{\tau _{1}}\left(
(C_{\tau _{2}}^{\ast }C_{\tau _{3}})\cdots (C_{\tau _{2p}}^{\ast }C_{\tau
_{2p+1}})\right)  \label{x12} \\
\mathbf{\tilde{y}}_{t,s} &=&i\int_{s}^{t}\mathrm{d}\tau C_{\tau }^{\ast
}+i\sum_{p=1}^{\infty }\left( -1\right) ^{p}\int_{s}^{t}\mathrm{d}\tau
_{1}\cdots \int_{s}^{\tau _{2p}}\mathrm{d}\tau _{2p+1}C_{\tau _{1}}^{\ast
}\left( (C_{\tau _{2}}C_{\tau _{3}}^{\ast })\cdots (C_{\tau _{2p}}C_{\tau
_{2p+1}}^{\ast })\right)  \label{x21}
\end{eqnarray}%
for all times $s,t\in \mathbb{R}$. Observe that $\mathbf{\tilde{x}}_{t,s}=%
\overline{\mathbf{x}}_{t,s}$ and $\mathbf{\tilde{y}}_{t,s}=\overline{\mathbf{%
y}}_{t,s}$ for any $s,t\in \mathbb{R}$. Since Lemma \ref{Lenma Bach-tech2
copy(1)} shows that%
\begin{equation*}
\left( 
\begin{array}{cc}
\mathbf{x}_{t,s} & \mathbf{y}_{t,s} \\ 
\mathbf{\tilde{y}}_{t,s} & \mathbf{\tilde{x}}_{t,s}%
\end{array}%
\right) \doteq \left( 
\begin{array}{cc}
\mathrm{e}^{-it\Upsilon _{0}/2}\mathbf{u}_{t-s}\mathrm{e}^{is\Upsilon _{0}/2}
& \mathrm{e}^{-it\Upsilon _{0}/2}\mathbf{v}_{t-s}\mathrm{e}^{-is\Upsilon
_{0}^{\top }/2} \\ 
\mathrm{e}^{it\Upsilon _{0}^{\top }/2}\overline{\mathbf{v}}_{t-s}\mathrm{e}%
^{is\Upsilon _{0}/2} & \mathrm{e}^{it\Upsilon _{0}^{\top }/2}\overline{%
\mathbf{u}}_{t-s}\mathrm{e}^{-is\Upsilon _{0}^{\top }/2}%
\end{array}%
\right) \ ,\qquad s,t\in \mathbb{R}\ ,
\end{equation*}%
we arrive at the two equalities of the lemma from (\ref{x11}) and (\ref{x12}%
) for $s=0$.
\end{proof}

Lemma \ref{Lenma Bach-tech2}\ is a key result to understand the relation
between the Hilbert-Schmidt property of $\mathbf{v}_{t}$ and $D_{0}$
because, at the cost of time-dependent generators, it allows us to disregard
in some sense the possibly unbounded diagonal element $\Upsilon _{0}$ in
Assumption \ref{Assumption1}, which only appears via the strongly continuous
one-parameter group $(\mathrm{e}^{it\Upsilon _{0}})_{t\in \mathbb{R}}$ of
unitary operators. Using such a result, it is straightforward to prove that
Hilbert-Schmidt off-diagonal coefficients $D_{0}$ lead to the
Shale-Stinespring condition at all times:

\begin{lemma}[Sufficient condition for the Shale-Stinespring condition]
\label{Lenma Bach-tech3}\mbox{
}\newline
Take any self-adjoint operator $h$ satisfying Assumption \ref{Assumption1}
and use Lemmata \ref{Lenma Bach-tech1} and \ref{Lenma Bach-tech1 copy(1)}
(i) to represent the group $(\mathrm{e}^{ith})_{t\in \mathbb{R}}$ of
Bogoliubov transformations. If $D_{0}\in \mathcal{L}^{2}(\mathfrak{h})$ then 
$\mathbf{v}_{t}\in \mathcal{L}^{2}(\mathfrak{h})$ for all $t\in \mathbb{R}$.
\end{lemma}

\begin{proof}
If $D_{0}\in \mathcal{L}^{2}(\mathfrak{h})$ is Hilbert-Schmidt then one can
combine Proposition \ref{Hamilselfadjoint} and Lemma \ref{Lemma quasi-free}
with Theorem \ref{Shale-Stinespring THM} to deduce that $\mathbf{v}_{t}\in 
\mathcal{L}^{2}(\mathfrak{h})$ for all times $t\in \mathbb{R}$, proving in
this way\ this lemma. However, these arguments are far from being natural.
It is indeed somehow artificial to use the second quantization formalism in
order to analyze a property only related to operators on the (dubbed)
one-particle Hilbert space $\mathcal{H}\doteq \mathfrak{h}\oplus \mathfrak{h}
$. As a result, we present here another (direct) proof of this lemma that
can be obtained from Lemma \ref{Lenma Bach-tech2} via an explicit upper
bound: If $D_{0}\in \mathcal{L}^{2}(\mathfrak{h})$ then we deduce from Lemma %
\ref{Lenma Bach-tech2} together with the triangle inequality and properties
of the trace that%
\begin{equation*}
\left\Vert \mathbf{v}_{t}\right\Vert _{2}\leq \sum_{p=0}^{\infty }\frac{%
\left( \left\vert t\right\vert \left\Vert D_{0}\right\Vert _{2}\right)
^{2p+1}}{(2p+1)!}=\sinh \left( \left\vert t\right\vert \left\Vert
D_{0}\right\Vert _{2}\right) <\infty
\end{equation*}%
for all times $t\in \mathbb{R}$.
\end{proof}

We now study the converse of Lemma \ref{Lenma Bach-tech3}. In other words,
we would like to know whether $\mathbf{v}_{t}\in \mathcal{L}^{2}(\mathfrak{h}%
)$ for some times $t\in \mathbb{R}$ can imply $D_{0}\in \mathcal{L}^{2}(%
\mathfrak{h})$. This problem is much more delicate. We first give a general
result dependent on the continuous operator family $(\varkappa _{t})_{t\in 
\mathbb{R}}\subseteq \mathcal{B}(\mathcal{B}(\mathfrak{h}))$ defined by (\ref%
{mapping familybis}), that is, for each time $t\in \mathbb{R}$,%
\begin{equation}
\varkappa _{t}\left( A\right) \doteq \int_{0}^{t}\mathrm{e}^{-i\tau \Upsilon
_{0}/2}A\mathrm{e}^{-i\tau \Upsilon _{0}^{\top }/2}\mathrm{d}\tau \ ,\qquad
A\in \mathcal{B}(\mathfrak{h})\ .  \label{mapping family}
\end{equation}%
Note again that the integral is well-defined for any vector $\varphi \in 
\mathfrak{h}$ since $(\mathrm{e}^{it\Upsilon _{0}})_{t\in \mathbb{R}}$ and $(%
\mathrm{e}^{it\Upsilon _{0}^{\top }})_{t\in \mathbb{R}}$ are strongly
continuous groups of unitary operators.

\begin{lemma}[Hilbert-Schmidt upper bound of twisted off-diagonal elements]
\label{Lenma Bach-tech4}\mbox{
}\newline
Take any self-adjoint operator $h$ acting on $\mathcal{H}=\mathfrak{h}\oplus 
\mathfrak{h}$ satisfying Assumption \ref{Assumption1} and use Lemmata \ref%
{Lenma Bach-tech1} and \ref{Lenma Bach-tech1 copy(1)} (i) to represent the
group $(\mathrm{e}^{ith})_{t\in \mathbb{R}}$ of Bogoliubov transformations.
If $D_{0}\in \mathcal{B}(\mathfrak{h})$ then, for any $t\in \mathbb{R}$, 
\begin{equation*}
\left\Vert \varkappa _{t}\left( D_{0}\right) \right\Vert _{2}\leq \left\Vert 
\mathbf{v}_{t}\right\Vert _{2}+\frac{\left\vert t\right\vert ^{2}}{2}%
\left\Vert D_{0}\right\Vert _{\mathrm{op}}^{2}\sup_{\tau \in \left[
-\left\vert t\right\vert ,\left\vert t\right\vert \right] }\left\Vert 
\mathbf{v}_{\tau }\right\Vert _{2}\ .
\end{equation*}
\end{lemma}

\begin{proof}
Using Lemmata \ref{Lenma Bach-tech2 copy(1)} and \ref{Lenma Bach-tech2
copy(2)} with $D_{0}\in \mathcal{B}(\mathfrak{h})$ and $c=\infty $ as well
as the mapping (\ref{mapping family}), we obtain that 
\begin{equation}
\mathbf{y}_{t,0}=i\varkappa _{t}\left( D_{0}\right) -\int_{0}^{t}\mathrm{d}%
\tau _{1}C_{\tau _{1}}\int_{0}^{\tau _{1}}\mathrm{d}\tau _{2}C_{\tau
_{2}}^{\ast }\mathbf{y}_{\tau _{2},0}\ ,  \label{hjk}
\end{equation}%
from which we deduce that 
\begin{equation*}
\left\Vert \varkappa _{t}\left( D_{0}\right) \right\Vert _{2}\leq \left\Vert 
\mathbf{y}_{t,0}\right\Vert _{2}+\int_{0}^{\left\vert t\right\vert }\mathrm{d%
}\tau _{1}\int_{0}^{\tau _{1}}\mathrm{d}\tau _{2}\left\Vert C_{\tau
_{1}}C_{\tau _{2}}^{\ast }\mathbf{y}_{\tau _{2},0}\right\Vert _{2}\leq
\left\Vert \mathbf{y}_{t,0}\right\Vert _{2}+\frac{\left\vert t\right\vert
^{2}}{2}\left\Vert D_{0}\right\Vert _{\mathrm{op}}^{2}\sup_{\tau \in \left[
-\left\vert t\right\vert ,\left\vert t\right\vert \right] }\left\Vert 
\mathbf{y}_{\tau ,0}\right\Vert _{2}\ ,
\end{equation*}%
where $C_{t}$ is the operator defined by (\ref{sssss4}) for any $t\in 
\mathbb{R}$. Using now Lemma \ref{Lenma Bach-tech2 copy(1)} (with $c=\infty $%
), more precisely, 
\begin{equation*}
\mathbf{y}_{t,0}=\mathrm{e}^{-it\Upsilon _{0}/2}\mathbf{v}_{t}\ ,\qquad t\in 
\mathbb{R}\ ,
\end{equation*}%
one arrives at the desired upper bound.
\end{proof}

\begin{corollary}[Hilbert-Schmidt continuity of twisted off-diagonal elements%
]
\label{coro Bach-tech5}\mbox{}\newline
Under the conditions of Lemma \ref{Lenma Bach-tech4}, if $(\mathbf{v}%
_{t})_{t\in \left[ -c,c\right] }\subseteq \mathcal{L}^{2}(\mathfrak{h})$ for
some time $c\in \mathbb{R}^{+}$ and is continuous in $\mathcal{L}^{2}(%
\mathfrak{h})$ at $t=0$, i.e., 
\begin{equation*}
\lim_{t\rightarrow 0}\left\Vert \mathbf{v}_{t}\right\Vert _{2}=0\ ,
\end{equation*}%
then there is $\delta \in \mathbb{R}^{+}$ such that $(\varkappa
_{t}(D_{0}))_{t\in \left[ -\delta ,\delta \right] }\subseteq \mathcal{L}^{2}(%
\mathfrak{h})$ and this operator family is continuous in $\mathcal{L}^{2}(%
\mathfrak{h})$ at $t=0$, i.e., 
\begin{equation*}
\lim_{t\rightarrow 0}\left\Vert \varkappa _{t}\left( D_{0}\right)
\right\Vert _{2}=0\ .
\end{equation*}
\end{corollary}

\begin{proof}
This is an obvious consequence of Lemma \ref{Lenma Bach-tech4}. For more
details, see \cite{Nathan}.
\end{proof}

\begin{corollary}[Hilbert-Schmidt property of off-diagonal elements]
\label{coro Bach-tech6}\mbox{}\newline
Under the conditions of Lemma \ref{Lenma Bach-tech4}, if $(\mathbf{v}%
_{t})_{t\in \left[ 0,c\right] }\subseteq \mathcal{L}^{2}(\mathfrak{h})$ for
some time $c\in \mathbb{R}^{+}$ and%
\begin{equation}
\liminf_{t\rightarrow 0^{+}}t^{-1}\left\Vert \mathbf{v}_{t}\right\Vert
_{2}<\infty \ ,  \label{wwww}
\end{equation}%
then $D_{0}\in \mathcal{L}^{2}(\mathfrak{h})$.
\end{corollary}

\begin{proof}
By Lemma \ref{Lenma Bach-tech4}, for any $t\in \mathbb{R}^{+}$, 
\begin{equation}
\left\Vert t^{-1}\varkappa _{t}\left( D_{0}\right) \right\Vert _{2}\leq
t^{-1}\left\Vert \mathbf{v}_{t}\right\Vert _{2}+\frac{t}{2}\left\Vert
D_{0}\right\Vert _{\mathrm{op}}^{2}\sup_{\tau \in \left[ -\left\vert
t\right\vert ,\left\vert t\right\vert \right] }\left\Vert \mathbf{v}_{\tau
}\right\Vert _{2}\ .  \label{ssswww}
\end{equation}%
Under the conditions of Corollary \ref{coro Bach-tech6}, note that $(\mathbf{%
v}_{t})_{t\in \left[ 0,c\right] }\subseteq \mathcal{L}^{2}(\mathfrak{h})$ is
in particular continuous in $\mathcal{L}^{2}(\mathfrak{h})$ at $t=0$ along
the subsequence $(t_{n})_{n\in \mathbb{N}}$ satisfying 
\begin{equation*}
\lim_{n\rightarrow \infty }t_{n}^{-1}\left\Vert \mathbf{v}%
_{t_{n}}\right\Vert _{2}=\liminf_{t\rightarrow 0^{+}}t^{-1}\left\Vert 
\mathbf{v}_{t}\right\Vert _{2}<\infty
\end{equation*}%
and there is $N\in \mathbb{R}^{+}$ such that 
\begin{equation}
\sup_{n\in \{N,N+1,\ldots ,\infty \}}\left\Vert \mathbf{v}%
_{t_{n}}\right\Vert _{2}<+\infty \ .  \label{extra1}
\end{equation}%
Additionally, for any $\varphi \in \mathfrak{h}$ and $t\in \mathbb{R}^{+}$, 
\begin{equation*}
\left\Vert D_{0}\varphi \right\Vert _{\mathfrak{h}}\leq \left\Vert \left(
D_{0}-t^{-1}\varkappa _{t}\left( D_{0}\right) \right) \varphi \right\Vert _{%
\mathfrak{h}}+\left\Vert t^{-1}\varkappa _{t}\left( D_{0}\right) \varphi
\right\Vert _{\mathfrak{h}}\ ,
\end{equation*}%
by the triangle inequality. Since $(\mathrm{e}^{it\Upsilon _{0}})_{t\in 
\mathbb{R}}$ and $(\mathrm{e}^{it\Upsilon _{0}^{\top }})_{t\in \mathbb{R}}$
are strongly continuous groups of unitary operators and $D_{0}\in \mathcal{B}%
(\mathfrak{h})$, we deduce from the last inequality and the continuity of
the function $x\mapsto x^{2}$ that, for any $\varphi \in \mathfrak{h}$, 
\begin{equation*}
\left\Vert D_{0}\varphi \right\Vert _{\mathfrak{h}}^{2}\leq
\liminf_{t\rightarrow 0^{+}}\left\Vert t^{-1}\varkappa _{t}\left(
D_{0}\right) \varphi \right\Vert _{\mathfrak{h}}^{2}\ .
\end{equation*}%
As a consequence, taking any orthonormal basis $\left\{ \psi _{k}\right\}
_{k=1}^{\infty }\subseteq \mathfrak{h}$, for any $D_{0}\in \mathcal{B}(%
\mathfrak{h})$, we have to evaluate the limit $k_{0}\rightarrow \infty $ of
the quantity 
\begin{equation*}
\sum_{k=1}^{k_{0}}\left\Vert D_{0}\psi _{k}\right\Vert _{\mathfrak{h}%
}^{2}\leq \liminf_{t\rightarrow 0^{+}}\sum_{k=1}^{k_{0}}\left\Vert
t^{-1}\varkappa _{t}\left( D_{0}\right) \psi _{k}\right\Vert _{\mathfrak{h}%
}^{2}\leq \liminf_{t\rightarrow 0^{+}}\left\Vert t^{-1}\varkappa _{t}\left(
D_{0}\right) \right\Vert _{2}^{2}\ .
\end{equation*}%
By (\ref{ssswww})--(\ref{extra1}) it follows that, for any $\varepsilon \in 
\mathbb{R}^{+}$, 
\begin{equation*}
\left\Vert D_{0}\right\Vert _{2}\leq \liminf_{t\rightarrow 0^{+}}\left\Vert
t^{-1}\varkappa _{t}\left( D_{0}\right) \right\Vert _{2}\leq
\liminf_{t\rightarrow 0^{+}}t^{-1}\left\Vert \mathbf{v}_{t}\right\Vert
_{2}+\varepsilon \ .
\end{equation*}%
In particular, if (\ref{wwww}) holds true then $D_{0}\in \mathcal{L}^{2}(%
\mathfrak{h})$.
\end{proof}

The continuity properties expressed in Corollary \ref{coro Bach-tech5} are
satisfied for all groups $(\mathrm{e}^{ith})_{t\in \mathbb{R}}$ of
Bogoliubov transformations that can be implemented in the Fock space. This
is proven in Proposition \ref{prop Bach-tech8}. However, the sufficient
condition of Corollary \ref{coro Bach-tech6} to get a Hilbert-Schmidt
off-diagonal element $D_{0}$ is a priori unclear for such implementable
groups of Bogoliubov transformations, except if the vacuum state of the
fermionic Fock space belongs to the domain of the Hamiltonian implementing
the group of Bogoliubov transformations. See Proposition \ref{prop
Bach-tech8} and Corollary \ref{coro Bach-tech9}.

In particular situations, Corollary \ref{coro Bach-tech6} is not necessary.
In fact, heuristically, under the conditions of Corollary \ref{coro
Bach-tech5}, $\varkappa _{t}\left( D_{0}\right) \simeq tD_{0}$ when $%
t\rightarrow 0$, as heuristically observed in Equations (\ref{F-B-1})--(\ref%
{F-B+1}). For instance, $\varkappa _{t}\left( D_{0}\right) =tD_{0}$ when $%
D_{0}\mathrm{e}^{-i\tau \Upsilon _{0}^{\top }/2}=\mathrm{e}^{i\tau \Upsilon
_{0}/2}D_{0}$ for any $\tau \in \lbrack 0,t]$. In the following lemma, we
give the simple case for which the operator $D_{0}$ intertwines\footnote{%
We say that an operator $A$ intertwines between two operators $B$ and $C$
when $AB=CA$.} between $\Upsilon _{0}$ and $\Upsilon _{0}^{\top }$ (resp. $%
-\Upsilon _{0}^{\top }$).

\begin{lemma}[Hilbert-Schmidt property -- commuting case]
\label{Lenma Bach-tech7}\mbox{
}\newline
Take any self-adjoint operator $h$ acting on $\mathcal{H}=\mathfrak{h}\oplus 
\mathfrak{h}$ satisfying Assumption \ref{Assumption1} with $D_{0}\in 
\mathcal{B}(\mathfrak{h})$ and use Lemmata \ref{Lenma Bach-tech1} and \ref%
{Lenma Bach-tech1 copy(1)} (i) to represent the group $(\mathrm{e}%
^{ith})_{t\in \mathbb{R}}$ of Bogoliubov transformations.

\begin{enumerate}
\item[\emph{(i)}] Assume $D_{0}\Upsilon _{0}=-\Upsilon _{0}^{\top }D_{0}$
and $\mathbf{v}_{t}\in \mathcal{L}^{2}(\mathfrak{h})$ for positive times $t$
in a neighborhood of $0$. Then, $D_{0}\in \mathcal{L}^{2}(\mathfrak{h})$ and 
$\mathbf{v}_{t}\in \mathcal{L}^{2}(\mathfrak{h})$ for all $t\in \mathbb{R}$.

\item[\emph{(ii)}] Assume $D_{0}\Upsilon _{0}=\Upsilon _{0}^{\top }D_{0}$
and $\mathbf{v}_{t}\in \mathcal{L}^{2}(\mathfrak{h})$ for positive times $t$
in a neighborhood of $0$. If additionally $\Upsilon _{0}\in \mathcal{B}(%
\mathfrak{h})$, then $D_{0}\in \mathcal{L}^{2}(\mathfrak{h})$ and $\mathbf{v}%
_{t}\in \mathcal{L}^{2}(\mathfrak{h})$ for all $t\in \mathbb{R}$.
\end{enumerate}
\end{lemma}

\begin{proof}
We separate the proof in three short steps, the first one being a
preliminary study while Step 2 and 3 prove Assertions (i) and (ii),
respectively.\ \medskip

\noindent \underline{Step 1:} Assume all conditions of the lemma as well as $%
\Upsilon _{0}D_{0}=\pm D_{0}\Upsilon _{0}^{\top }$. We show here that $D_{0}$
also intertwines between the groups generated by $i\Upsilon _{0}$ and $\pm
i\Upsilon _{0}^{\top }$. Since $D_{0}\in \mathcal{B}(\mathfrak{h})$, the
equality $\Upsilon _{0}D_{0}=\pm D_{0}\Upsilon _{0}^{\top }$ means in
particular that 
\begin{equation*}
D_{0}\mathcal{D}\left( \Upsilon _{0}^{\top }\right) \subseteq \mathcal{D}%
\left( \Upsilon _{0}\right) \ .
\end{equation*}%
Then, observing that 
\begin{equation*}
\partial _{t}\left\{ \mathrm{e}^{it\Upsilon _{0}}D_{0}\mathrm{e}^{\mp
it\Upsilon _{0}^{\top }}\right\} =\mathrm{e}^{it\Upsilon _{0}}\left(
\Upsilon _{0}D_{0}\mp D_{0}\Upsilon _{0}^{\top }\right) \mathrm{e}^{\mp
it\Upsilon _{0}^{\top }}=0
\end{equation*}%
in the strong sense on the domain $\mathcal{D}(\Upsilon _{0}^{\top })$, we
deduce that 
\begin{equation}
\mathrm{e}^{it\Upsilon _{0}}D_{0}=D_{0}\mathrm{e}^{\pm it\Upsilon _{0}^{\top
}}\ ,\qquad t\in \mathbb{R}\ .  \label{ffffffffffffff}
\end{equation}%
Note that we use above the density of the domain $\mathcal{D}(\Upsilon
_{0}^{\top })$ and the unitarity of operators $\mathrm{e}^{it\Upsilon
_{0}^{\top }}$ for $t\in \mathbb{R}$, $\Upsilon _{0}^{\top }$ being
self-adjoint.\medskip

\noindent \underline{Step 2:} Assume in this step that $D_{0}\Upsilon
_{0}=-\Upsilon _{0}^{\top }D_{0}$. Then, we find from Lemma \ref{Lenma
Bach-tech2} and Equality (\ref{ffffffffffffff}) that 
\begin{equation}
\mathrm{e}^{-it\Upsilon _{0}/2}\mathbf{v}_{t}=itD_{0}\left( \mathbf{1}%
+R\left( t\right) \right) \ ,\qquad t\in \mathbb{R}_{0}^{+}\ ,  \label{d1}
\end{equation}%
with 
\begin{equation}
R\left( t\right) \doteq \sum_{p=1}^{\infty }\left( -1\right) ^{p}\frac{%
t^{2p}(D_{0}^{\ast }D_{0})^{p}}{(2p+1)!}\ ,\qquad t\in \mathbb{R}_{0}^{+}\ .
\label{d2}
\end{equation}%
Since $D_{0}\in \mathcal{B}(\mathfrak{h})$ and $\lim_{t\rightarrow 0}\cosh
(t)-1=0$, there is $\delta \in \mathbb{R}^{+}$ such that 
\begin{equation}
\max_{t\in \left[ 0,\delta \right] }\left\Vert R\left( t\right) \right\Vert
_{\mathrm{op}}\leq \sum_{p=1}^{\infty }\frac{\delta ^{2p}\left\Vert
D_{0}\right\Vert _{\mathrm{op}}^{2p}}{(2p+1)!}\leq \cosh \left( \delta
\left\Vert D_{0}\right\Vert _{\mathrm{op}}\right) -1\leq \frac{1}{4}\ .
\label{d3}
\end{equation}%
Therefore, we infer from (\ref{d1})--(\ref{d3}) that 
\begin{equation}
\left\Vert \mathbf{v}_{t}\right\Vert _{2}^{2}=\left\Vert \mathrm{e}%
^{-it\Upsilon _{0}/2}\mathbf{v}_{t}\right\Vert _{2}^{2}\geq t^{2}\left(
1-2\left\Vert R\left( t\right) \right\Vert _{\mathrm{op}}\right)
^{2}\left\Vert D_{0}\right\Vert _{2}^{2}\geq \frac{t^{2}}{4}\left\Vert
D_{0}\right\Vert _{2}^{2}  \label{sssssghhj}
\end{equation}%
for small times $t\in \left[ 0,\delta \right] $. In particular, if $%
D_{0}\Upsilon _{0}=-\Upsilon _{0}^{\top }D_{0}$ and $\mathbf{v}_{t}\in 
\mathcal{L}^{2}(\mathfrak{h})$ for positive times $t$ in a neighborhood of $%
0 $, then $D_{0}\in \mathcal{L}^{2}(\mathfrak{h})$, which, combined with
Lemma \ref{Lenma Bach-tech3}, leads to Assertion (i). \medskip

\noindent \underline{Step 3:} Assume now that $\Upsilon
_{0}D_{0}=D_{0}\Upsilon _{0}^{\top }$. If additionally $\Upsilon _{0}\in 
\mathcal{B}(\mathfrak{h})$, then 
\begin{equation}
2\sup_{t\in \left[ 0,c\right] }\left\Vert \mathrm{e}^{-2it\Upsilon _{0}}-%
\mathbf{1}\right\Vert _{\mathrm{op}}<1\ ,  \label{sdsdsdsd}
\end{equation}%
for some $c\in \mathbb{R}^{+}$, because $\Upsilon _{0}\in \mathcal{B}(%
\mathfrak{h})$ implies the norm continuity of the group $(\mathrm{e}%
^{it\Upsilon _{0}})_{t\in \mathbb{R}}$ of unitary operators, see, e.g., \cite%
[Chapter I, Proposition 3.5 and Theorem 3.7]{EngelNagel}. Then, again via
Lemma \ref{Lenma Bach-tech2} and Equality (\ref{ffffffffffffff}),%
\begin{equation}
\mathbf{y}_{t,0}=it\left( \mathbf{1}+S\left( t\right) \right) D_{0}\ ,\qquad
t\in \mathbb{R}^{+}\ ,  \label{d4}
\end{equation}%
where, for any $t\in \mathbb{R}^{+}$, 
\begin{equation*}
S\left( t\right) \doteq t^{-1}\int_{0}^{t}\mathrm{d}\tau \left( \mathrm{e}%
^{-i\tau \Upsilon _{0}}-\mathbf{1}\right) +t^{-1}\sum_{p=1}^{\infty }\left(
-1\right) ^{p}\int_{s}^{t}\mathrm{d}\tau _{1}\cdots \int_{s}^{\tau _{2p}}%
\mathrm{d}\tau _{2p+1}\mathrm{e}^{i\sum_{k=1}^{2p+1}\left( -1\right)
^{k}\tau _{k}\Upsilon _{0}}\left( D_{0}D_{0}^{\ast }\right) ^{p}\ .
\end{equation*}%
We now use the following estimate 
\begin{equation*}
\left\Vert S\left( t\right) \right\Vert _{\mathrm{op}}\leq \sup_{\tau \in %
\left[ 0,t\right] }\left\Vert \mathrm{e}^{-i\tau \Upsilon _{0}}-\mathbf{1}%
\right\Vert _{\mathrm{op}}+\sum_{p=1}^{\infty }\frac{t^{2p}\left\Vert
D_{0}\right\Vert _{\mathrm{op}}^{2p}}{(2p+1)!}\leq \sup_{\tau \in \left[ 0,t%
\right] }\left\Vert \mathrm{e}^{-it\Upsilon _{0}}-\mathbf{1}\right\Vert _{%
\mathrm{op}}+\cosh \left( t\left\Vert D_{0}\right\Vert _{\mathrm{op}}\right)
-1
\end{equation*}%
for any $t\in \mathbb{R}^{+}$\ to show the existence of $\delta \in \mathbb{R%
}^{+}$ such that\ 
\begin{equation}
2\max_{t\in \left[ 0,\delta \right] }\left\Vert S\left( t\right) \right\Vert
_{\mathrm{op}}<1\ ,  \label{d6}
\end{equation}%
provided that (\ref{sdsdsdsd}) holds true. By (\ref{d4})--(\ref{d6}), we
then get 
\begin{equation*}
\left\Vert \mathbf{v}_{t}\right\Vert _{2}^{2}=\left\Vert \mathrm{e}%
^{-it\Upsilon _{0}/2}\mathbf{v}_{t}\right\Vert _{2}^{2}\geq t^{2}\left(
1-2\max_{t\in \left[ 0,\delta \right] }\left\Vert S\left( t\right)
\right\Vert _{\mathrm{op}}\right) ^{2}\left\Vert D_{0}\right\Vert _{2}^{2}\ .
\end{equation*}%
In particular, if $\Upsilon _{0}\in \mathcal{B}(\mathfrak{h})$ as well as $%
D_{0}\Upsilon _{0}=\Upsilon _{0}^{\top }D_{0}$ and $\mathbf{v}_{t}\in 
\mathcal{L}^{2}(\mathfrak{h})$ for positive times $t$ in a neighborhood of $%
0 $, then $D_{0}\in \mathcal{L}^{2}(\mathfrak{h})$, which, combined with
Lemma \ref{Lenma Bach-tech3}, leads to Assertion (ii).
\end{proof}

\begin{remark}
\label{Lenma Bach-tech7 extra}\mbox{}\newline
Equation \eqref{sdsdsdsd} is equivalent to $\Upsilon _{0}\in \mathcal{B}(%
\mathfrak{h})$. Indeed, for any $c\in \mathbb{R}^{+}$ and $\Upsilon
_{0}\notin \mathcal{B}(\mathfrak{h})$, we compute using the spectral
theorem that
\begin{equation*}
\sup_{t\in \lbrack 0,c)}\left\Vert \mathrm{e}^{-2it\Upsilon _{0}}-\mathbf{1}%
\right\Vert _{\mathrm{op}}=\sup_{t\in \lbrack 0,c)}\sup_{\lambda \in \sigma
(\Upsilon _{0})}\left\vert \mathrm{e}^{-2it\lambda }-1\right\vert
=\sup_{t\in \lbrack 0,c)}\sup_{\lambda \in \sigma (\Upsilon _{0})}\sqrt{%
2+2\cos \left( 2t\lambda \right) }=2\ .
\end{equation*}
\end{remark}

We conclude this section by proving that the continuity properties given by
Corollary \ref{coro Bach-tech5} are satisfied for all groups $(\mathrm{e}%
^{ith})_{t\in \mathbb{R}}$ of Bogoliubov transformations that can be
implemented in the fermionic Fock space $\mathcal{F}$. In fact, Bruneau and
Derezinski prove in \cite[Theorem 4.2]{Bruneau-derezinski2007} that a
strongly continuous one-parameter symplectic group, or a group of Bogoliubov
transformations in the bosonic case\footnote{%
It corresponds to the framework presented in section \ref{Araiki section},
except we consider a CCR $C^{\ast }$ algebra instead of a CAR $C^{\ast }$
algebra, and the involution $\mathfrak{A}$ is replaced by an
anti-involution, i.e. $\mathfrak{A}^{2}=-\mathfrak{A}$. see \cite%
{Bruneau-derezinski2007}.}, is implementable on the bosonic Fock space if
and only if the off-diagonal part of this group, corresponding here to $%
\mathbf{v}_{t}$, is not only Hilbert-Schmidt for all times but also
continuous at time $t=0$ in the Hilbert-Schmidt topology. We partially show
this fact for the fermionic case\ in Proposition \ref{prop Bach-tech8} (i)
while Proposition \ref{prop Bach-tech8} (ii) is a new result extending the
previous one.

\begin{proposition}[Hilbert-Schmidt continuity of the Shale-Stinespring
condition]
\label{prop Bach-tech8}\mbox{
}\newline
Take any self-adjoint operator $h$ satisfying Assumption \ref{Assumption1}
and use Lemmata \ref{Lenma Bach-tech1} and \ref{Lenma Bach-tech1 copy(1)}
(i) to represent the group $(\mathrm{e}^{ith})_{t\in \mathbb{R}}$ of
Bogoliubov transformations. Assume that $(\mathrm{e}^{ith})_{t\in \mathbb{R}%
} $ can be implemented in the Fock space $\mathcal{F}$ via a strongly
continuous group $(\mathrm{e}^{it\mathrm{H}/2})_{t\in \mathbb{R}}$, $\mathrm{%
H}=\mathrm{H}^{\ast }$ being a self-adjoint operator acting on $\mathcal{F}$%
, as in Definition \ref{def quadratic}. Then the following assertions hold
true:

\begin{enumerate}
\item[\emph{(i)}] The Shale-Stinespring condition is continuous at $t=0$: 
\begin{equation*}
\lim_{t\rightarrow 0}\left\Vert \mathbf{v}_{t}\right\Vert
_{2}=\lim_{t\rightarrow 0}\left\Vert \mathbf{1}-\mathbf{u}_{t}^{\ast }%
\mathbf{u}_{t}\right\Vert _{1}=0\ .
\end{equation*}

\item[\emph{(ii)}] If the vacuum state $\Psi \doteq \left( 1,0,\ldots
\right) $ belongs to the domain $\mathcal{D}\left( \mathrm{H}\right) $ then 
\begin{equation*}
\limsup_{t\rightarrow 0}t^{-1}\left\Vert \mathbf{v}_{t}\right\Vert _{2}\in 
\mathbb{R}_{0}^{+}\ .
\end{equation*}
\end{enumerate}
\end{proposition}

\begin{proof}
Fix all conditions of the proposition. The proof is done in several steps.
Note that the proof of Assertion (i) is similar to the one of \cite[Theorem
4.2]{Bruneau-derezinski2007} done in the bosonic case, but the fermionic
case is a bit more complicated. \medskip

\noindent \underline{Step 1:} Recall that $\mathcal{A}$ is the universal
(self-dual CAR) $C^{\ast }$-algebra $\mathcal{A}$ (\ref{A}) generated by
fields $\mathrm{B}(\psi )$, $\psi \in \mathcal{\mathcal{H}}$, and an unit $%
\mathfrak{1}$, while $\mathbf{\pi }_{\mathcal{F}}$ denotes its Fock space
representation, as uniquely defined by Equation (\ref{ghghghg2}). Since $%
\mathrm{e}^{ith}$ is an implementable Bogoliubov transformation for all
times, as defined in \cite[Equations (6.4)--(6.5)]{Ruijsenaars}, for any $%
t\in \mathbb{R}$, there is a unitary operator $\mathbb{U}_{t}$ acting on the
fermionic Fock space $\mathcal{F}$ and satisfying, for any $t\in \mathbb{R}$%
, 
\begin{equation}
\mathbb{U}_{t}\left( \mathbf{\pi }_{\mathcal{F}}\left( A\right) \right) 
\mathbb{U}_{t}^{\ast }=\mathbf{\pi }_{\mathcal{F}}\circ \mathbf{\chi }_{%
\mathrm{e}^{ith}}\left( A\right) =\mathrm{e}^{it\mathrm{H}/2}\left( \mathbf{%
\pi }_{\mathcal{F}}\left( A\right) \right) \mathrm{e}^{-it\mathrm{H}/2}\
,\qquad A\in \mathcal{A}\ ,  \label{sdsddsddsdsd}
\end{equation}%
as well as 
\begin{equation}
\mathbb{U}_{t}\Psi =\left\{ 
\begin{array}{lll}
\det \left( 1+K_{t}^{\ast }K_{t}\right) ^{-1/4}\exp \left( \mathbb{K}%
_{t}\right) \Psi & \text{if} & L_{t}=0\ , \\ 
\det \left( 1+K_{t}^{\ast }K_{t}\right) ^{-1/4}\prod_{\ell
=1}^{L_{t}}a^{\ast }\left( \psi _{\ell ,t}\right) \exp \left( \mathbb{K}%
_{t}\right) \Psi & \text{if} & L_{t}\in \mathbb{N}\ ,%
\end{array}%
\right.  \label{sdsddsddsdsdsdsddsddsdsd}
\end{equation}%
where $\Psi \doteq \left( 1,0,\ldots \right) \in \mathcal{F}$ is the vacuum
state of the Fock space, $a\left( \varphi \right) $ is the usual creation
operator associated with $\varphi \in \mathfrak{h}$ and acting on $\mathcal{F%
}$, $L_{t}\doteq \dim \ker (\mathbf{u}_{t}^{\ast })<\infty $, $\{\psi _{\ell
,t}\}_{\ell =1}^{L_{t}}$ is an orthonormal basis of $\ker (\mathbf{u}%
_{t}^{\ast })$ when $L_{t}\in \mathbb{N}$, 
\begin{equation}
K_{t}=\mathbf{v}_{t}\overline{\mathbf{u}}_{t}^{-1}\in \mathcal{L}^{2}(%
\mathfrak{h})\ ,\qquad t\in \mathbb{R}\ ,  \label{ddddd}
\end{equation}%
and $\mathbb{K}_{t}$ is an operator formally represented from the kernel $%
k_{t}\left( p,q\right) $ of the Hilbert-Schmidt operator $K_{t}^{\ast }K_{t}$
as 
\begin{equation}
\mathbb{K}_{t}\doteq \frac{1}{2}\int_{\mathcal{M}}\mathrm{d}\mathfrak{m}%
\left( p\right) \mathrm{d}\mathfrak{m}\left( q\right) k_{t}\left( p,q\right)
a^{\ast }\left( p\right) a^{\ast }\left( q\right)  \label{ddddd2}
\end{equation}%
keeping in mind that $\mathfrak{h}\doteq L^{2}(\mathcal{M})$ is assumed (for
simplicity) to be realized as a space of square-integrable (complex-valued)
functions on a measure space $(\mathcal{M},\mathfrak{m})$. Equations (\ref%
{sdsddsddsdsd})--(\ref{ddddd2}) requires additional imputs to be
well-defined: Here, the operator $\overline{\mathbf{u}}_{t}$ is not
necessarly invertible. However, because of Lemma \ref{Lenma Bach-tech1
copy(1)} (i) and the fact that $\mathrm{e}^{ith}$ is always a Bogoliubov
transformation, one invokes Ruijsenaars' arguments of \cite{Ruijsenaars} to
show in the present case that the restriction of $\overline{\mathbf{u}}_{t}$
to $\ker (\overline{\mathbf{u}}_{t})^{\bot }$ has a bounded inverse from $%
\ker (\mathbf{u}_{t}^{\top })^{\bot }$ to $\ker (\overline{\mathbf{u}}%
_{t})^{\bot }$, which is by definition extended by setting it equal to zero
on $\ker (\mathbf{u}_{t}^{\top })$. The result of this extension is denoted
by $\overline{\mathbf{u}}_{t}^{-1}$. This is used for instance in Equation (%
\ref{ddddd}) and it shows that $K_{t}\in \mathcal{L}^{2}(\mathfrak{h})$ and $%
\mathbb{K}_{t}$ can be well-defined. Note that $\ker (\overline{\mathbf{u}}%
_{t})$ and $\ker (\mathbf{u}_{t}^{\ast })$ have always finite dimension
because 
\begin{equation*}
L_{t}\doteq \dim \ker (\mathbf{u}_{t}^{\ast })=\dim \ker (\mathbf{u}%
_{t})=\dim \ker (\overline{\mathbf{u}}_{t})=\mathrm{tr}\left( \left. (%
\mathbf{1}-\mathbf{u}_{t}\mathbf{u}_{t}^{\ast })\right\vert _{\ker (\mathbf{u%
}_{t}^{\ast })}\right)
\end{equation*}%
thanks to the equalities $\ker (\overline{\mathbf{u}}_{t})=\mathcal{C}\ker (%
\mathbf{u}_{t})$ together with \cite[Equation (6.1)]{Ruijsenaars}. So, by
using (\ref{eqdfdf}), one obtains that 
\begin{equation}
L_{t}\leq \mathrm{tr}(\mathbf{1}-\mathbf{u}_{t}\mathbf{u}_{t}^{\ast })=%
\mathrm{tr}\left( \mathbf{v}_{t}^{\top }\overline{\mathbf{v}}_{t}\right) =%
\mathrm{tr}\left( \mathbf{v}_{t}^{\ast }\mathbf{v}_{t}\right) =\left\Vert 
\mathbf{v}_{t}\right\Vert _{2}^{2}<\infty \ .  \label{eqdfdf201}
\end{equation}%
\medskip

\noindent \underline{Step 2:} Note that $\mathbb{U}_{t}$ is a priori
different from $\mathrm{e}^{it\mathrm{H}/2}$. In fact, (\ref{sdsddsddsdsd})
means that 
\begin{equation}
\mathbb{U}_{t}^{\ast }\mathrm{e}^{it\mathrm{H}/2}\in \mathbf{\pi }_{\mathcal{%
F}}\left( \mathcal{A}\right) ^{\prime }\doteq \left\{ B\in \mathcal{B}\left( 
\mathcal{F}\right) :\forall A\in \mathcal{A}\ ,\left[ B,\mathbf{\pi }_{%
\mathcal{F}}\left( A\right) \right] =0\right\} \ .  \label{commutant1}
\end{equation}%
The Fock space representation $\mathbf{\pi }_{\mathcal{F}}$ is irreducible,
see, e.g. \cite[Definition 2.3.7]{BratteliRobinsonI} and \cite[Proposition
5.2.2]{BratteliRobinson}. It follows from Schur's lemma \cite[Proposition
9.20]{TakesakiI} that 
\begin{equation}
\mathbf{\pi }_{\mathcal{F}}\left( \mathcal{A}\right) ^{\prime }=\mathbb{C}\,%
\mathbf{1}\ ,  \label{commutant2}
\end{equation}%
which, combined with (\ref{commutant1}), implies in turn the existence of
constants $b_{t}\in \mathbb{C}$, $t\in {\mathbb{R}}$, such that 
\begin{equation}
\mathrm{e}^{it\mathrm{H}/2}=b_{t}\mathbb{U}_{t}\ ,\qquad t\in {\mathbb{R}}\ .
\label{sssssss0}
\end{equation}%
Note that, for any $t\in {\mathbb{R}}$ and every normalized vector $\varphi
\in \mathcal{F}$, i.e., vectors $\varphi $ in the Fock space $\mathcal{F}$
satisfying $\left\Vert \varphi \right\Vert _{\mathcal{F}}=1$, 
\begin{equation}
\left\vert b_{t}\right\vert =\left\Vert b_{t}\varphi \right\Vert _{\mathcal{F%
}}=\left\Vert \mathbb{U}_{t}^{\ast }\mathrm{e}^{it\mathrm{H}/2}\varphi
\right\Vert _{\mathcal{F}}=\left\Vert \varphi \right\Vert _{\mathcal{F}}=1\ .
\label{sssssss00}
\end{equation}%
Since $(\mathrm{e}^{it\mathrm{H}})_{t\in \mathbb{R}}$ is a strongly
continous group of unitary operators, it follows from (\ref{sssssss0})--(\ref%
{sssssss00}) that 
\begin{equation}
\Psi =\lim_{t\rightarrow 0}\mathrm{e}^{it\mathrm{H}/2}\Psi
=\lim_{t\rightarrow 0}b_{t}\mathbb{U}_{t}\Psi \ ,  \label{sdsdsdsddsd}
\end{equation}%
keeping in mind that $\Psi \doteq \left( 1,0,\ldots \right) \in \mathcal{F}$
is the vacuum state. Assume that, for any $\delta \in \mathbb{R}^{+}$ there
is $t\in \left[ -\delta ,\delta \right] $ such that $L_{t}\geq 1$. Then, we
deduce from (\ref{sdsddsddsdsdsdsddsddsdsd}) and (\ref{sssssss00}) the
existence of a subsequence $(t_{n})_{n\in \mathbb{N}}\subseteq \mathbb{R} $
converging to zero such that%
\begin{equation*}
\left\langle \Psi ,b_{t_{n}}\mathbb{U}_{t_{n}}\Psi \right\rangle _{\mathcal{F%
}}=0\ ,\qquad n\in \mathbb{N}\ .
\end{equation*}%
This is not possible because of (\ref{sdsdsdsddsd}). Therefore, there is $%
\delta \in \mathbb{R}^{+}$ such that $L_{t}=0$ for any $t\in \left[ -\delta
,\delta \right] $. It follows from (\ref{eqdfdf201}) that $\overline{\mathbf{%
u}}_{t}$ is invertible for $t\in \left[ -\delta ,\delta \right] $ and from (%
\ref{sdsddsddsdsdsdsddsddsdsd}) that%
\begin{equation}
\left\langle \Psi ,\mathbb{U}_{t}\Psi \right\rangle _{\mathfrak{h}}=\det
\left( 1+K_{t}^{\ast }K_{t}\right) ^{-1/4}>0\ ,\qquad t\in \left[ -\delta
,\delta \right] \ .  \label{sdsddsddsdsd2}
\end{equation}%
In particular, by Equation (\ref{ddddd}), $\mathbf{v}_{t}=K_{t}\overline{%
\mathbf{u}}_{t}$ and 
\begin{equation}
\left\Vert \mathbf{v}_{t}\right\Vert _{2}^{2}=\mathrm{tr}\left( \mathbf{v}%
_{t}\mathbf{v}_{t}^{\ast }\right) =\mathrm{tr}\left( K_{t}\overline{\mathbf{u%
}}_{t}\mathbf{u}_{t}^{\top }K_{t}^{\ast }\right) \leq \left\Vert
K_{t}\right\Vert _{2}^{2}  \label{ddd20}
\end{equation}%
for any $t\in \left[ -\delta ,\delta \right] $, $\delta \in \mathbb{R}^{+}$
being a sufficient small but fixed parameter.\medskip

\noindent \underline{Step 3:} By (\ref{sssssss0})--(\ref{sssssss00}), the
mappings $\iota _{t}\in \mathcal{B}\left( \mathcal{B}\left( \mathcal{F}%
\right) \right) $, $t\in \mathbb{R}$, defined by%
\begin{equation}
\iota _{t}\left( B\right) \doteq \mathbb{U}_{t}B\mathbb{U}_{t}^{\ast }=%
\mathrm{e}^{it\mathrm{H}/2}B\mathrm{e}^{-it\mathrm{H}/2}\ ,\qquad t\in {%
\mathbb{R}},\text{ }B\in \mathcal{B}\left( \mathcal{F}\right) \ ,
\label{sdsddsddsdsd3}
\end{equation}%
form a weak$^{\ast }$ continuous one-parameter group of automorphisms of $%
\mathcal{B}\left( \mathcal{F}\right) $. So, applying these automorphisms to
the orthogonal projection $P_{\Psi }$ on the vacuum state $\Psi \doteq
\left( 1,0,\ldots \right) $, we deduce that the function 
\begin{equation}
t\mapsto f\left( t\right) \doteq \mathrm{tr}\left( P_{\Psi }\iota _{t}\left(
P_{\Psi }\right) \right) =\left\langle \Psi ,\iota _{t}\left( P_{\Psi
}\right) \Psi \right\rangle _{\mathfrak{h}}  \label{def}
\end{equation}%
is continuous on $\mathbb{R}$, with $\mathrm{tr}(\cdot )$ being the usual
trace for operators, while using Equation (\ref{sdsddsddsdsd2}) and (\ref%
{sdsddsddsdsd3}), we get that, for any $t\in \left[ -\delta ,\delta \right] $%
, 
\begin{equation*}
f\left( t\right) =\left\langle \Psi ,\mathbb{U}_{t}\Psi \right\rangle _{%
\mathfrak{h}}\left\langle \Psi ,\mathbb{U}_{t}^{\ast }\Psi \right\rangle _{%
\mathfrak{h}}=\det \left( 1+K_{t}^{\ast }K_{t}\right) ^{-1/2}=\mathrm{e}%
^{-\left( 1/2\right) \mathrm{tr}\ln \left( 1+K_{t}^{\ast }K_{t}\right) }
\end{equation*}%
and the function 
\begin{equation}
t\mapsto g\left( t\right) \doteq \mathrm{tr}\ln \left( 1+K_{t}^{\ast
}K_{t}\right) =-2\ln f\left( t\right)  \label{deg}
\end{equation}%
is thus continuous on $\left[ -\delta ,\delta \right] $. In particular,
since $K_{0}=0$, 
\begin{equation}
\lim_{t\rightarrow 0}\mathrm{tr}\ln \left( 1+K_{t}^{\ast }K_{t}\right) =0\ .
\label{dh}
\end{equation}%
Using the spectral theorem, observe that, for any time $t\in \left[ -\delta
,\delta \right] $, the spectrum of $K_{t}^{\ast }K_{t}$ is an ordered
discrete set $\left\{ \lambda _{k,t}\right\} _{k=1}^{\infty }\subseteq 
\mathbb{R}_{0}^{+}$ of positive eigenvalues satisfying $\lambda _{k,t}\leq
\lambda _{q,t}$ when $k\geq q$ as well as 
\begin{equation}
g\left( t\right) =\sum_{k=1}^{\infty }\ln \left( 1+\lambda _{k,t}\right) \ .
\label{dh2}
\end{equation}%
It is worth recalling that $K_{t}^{\ast }K_{t}$ is here trace-class, i.e., $%
K_{t}\in \mathcal{L}^{2}(\mathfrak{h})$. Additionally, we conclude from (\ref%
{dh}) and $\lambda _{k,t}\geq 0$ that%
\begin{equation*}
\lim_{t\rightarrow 0}\sup_{k\in \mathbb{N}}\lambda _{k,t}=0
\end{equation*}%
as well as%
\begin{equation}
\lim_{t\rightarrow 0}\left\Vert K_{t}\right\Vert _{2}^{2}=\lim_{t\rightarrow
0}\sum_{k=1}^{\infty }\lambda _{k,t}\leq 2\lim_{t\rightarrow
0}\sum_{k=1}^{\infty }\ln \left( 1+\lambda _{k,t}\right) =0\ ,  \label{ddd1}
\end{equation}%
using additionally the inequality $x\leq 2\ln \left( 1+x\right) $ for any $%
x\in \left[ 0,1\right] $. Since $\mathrm{e}^{ith}$ is a Bogoliubov
transformation for all times one has (\ref{sssssssss}). In particular $%
\left\Vert \mathbf{u}_{t}\right\Vert _{\mathrm{op}}\leq 1$ and one can now
perform the limit $t\rightarrow 0$ in (\ref{ddd20}) while using (\ref{ddd1})
to deduce that%
\begin{equation}
\lim_{t\rightarrow 0}\left\Vert \mathbf{v}_{t}\right\Vert _{2}=0\ .
\label{eqdfdf20}
\end{equation}%
Since $\mathrm{tr}(X^{\top })=\mathrm{tr}(X)$ (use, e.g., Remark \ref{remark
idiote copy(1)}, in particular $X^{\top }\doteq \mathcal{C}X^{\ast }\mathcal{%
C}$), we also deduce from (\ref{eqdfdf}) and (\ref{eqdfdf20}) that 
\begin{equation}
\lim_{t\rightarrow 0}\left\Vert \mathbf{1}-\mathbf{u}_{t}^{\ast }\mathbf{u}%
_{t}\right\Vert _{1}=0\ .  \label{eqdfdf3}
\end{equation}%
Assertion (i) is therefore proven and we now show Assertion (ii) of the
proposition. \medskip

\noindent \underline{Step 4:} Observe from (\ref{sdsddsddsdsd3})--(\ref{def}%
) that, for any $t\in {\mathbb{R}}$,%
\begin{equation*}
f\left( t\right) =\mathrm{tr}\left( P_{\Psi }\mathrm{e}^{it\mathrm{H}%
/2}P_{\Psi }\mathrm{e}^{-it\mathrm{H}/2}\right) =\left\langle \Psi ,\mathrm{e%
}^{it\mathrm{H}/2}\Psi \right\rangle _{\mathfrak{h}}\left\langle \Psi ,%
\mathrm{e}^{-it\mathrm{H}/2}\Psi \right\rangle _{\mathfrak{h}}\ ,
\end{equation*}%
from which we deduce that 
\begin{equation*}
\partial _{t}f\left( t\right) =\frac{i}{2}\left( \left\langle \Psi ,\mathrm{e%
}^{it\mathrm{H}/2}\mathrm{H}\Psi \right\rangle _{\mathfrak{h}}\left\langle
\Psi ,\mathrm{e}^{-it\mathrm{H}/2}\Psi \right\rangle _{\mathfrak{h}%
}-\left\langle \Psi ,\mathrm{e}^{it\mathrm{H}/2}\Psi \right\rangle _{%
\mathfrak{h}}\left\langle \Psi ,\mathrm{e}^{-it\mathrm{H}/2}\mathrm{H}\Psi
\right\rangle _{\mathfrak{h}}\right)
\end{equation*}%
as well as 
\begin{equation*}
\partial _{t}^{2}f\left( t\right) =\frac{1}{2}\left( \left\langle \Psi ,%
\mathrm{e}^{it\mathrm{H}/2}\mathrm{H}\Psi \right\rangle _{\mathfrak{h}%
}\left\langle \Psi ,\mathrm{e}^{-it\mathrm{H}/2}\mathrm{H}\Psi \right\rangle
_{\mathfrak{h}}-\mathfrak{Re}\left( \left\langle \mathrm{H}\Psi ,\mathrm{e}%
^{it\mathrm{H}/2}\mathrm{H}\Psi \right\rangle _{\mathfrak{h}}\left\langle
\Psi ,\mathrm{e}^{-it\mathrm{H}/2}\Psi \right\rangle _{\mathfrak{h}}\right)
\right) \ ,
\end{equation*}%
as soon as $\Psi \in \mathcal{D}\left( \mathrm{H}\right) $. In particular, $%
\partial _{t}f\left( 0\right) =0$ and 
\begin{equation*}
\partial _{t}^{2}f\left( 0\right) =\frac{1}{2}\left( \left\langle \Psi ,%
\mathrm{H}\Psi \right\rangle _{\mathfrak{h}}^{2}-\left\langle \mathrm{H}\Psi
,\mathrm{H}\Psi \right\rangle _{\mathfrak{h}}\right) \leq 0\ .
\end{equation*}%
($-\partial _{t}^{2}f\left( 0\right) $ is in fact the quantum fluctuation of
the energy observable $\mathrm{H}$ in the state $\left\langle \Psi ,\left(
\cdot \right) \Psi \right\rangle _{\mathfrak{h}}$, which is positive by the
Cauchy-Schwarz inequality.) As $f\left( t\right) >0$ at least for all $t\in %
\left[ -\delta ,\delta \right] $, it follows from (\ref{deg}) that, for any $%
t\in \left[ -\delta ,\delta \right] $, 
\begin{equation*}
\partial _{t}g\left( t\right) =-2\frac{\partial _{t}f\left( t\right) }{%
f\left( t\right) }\qquad \text{and}\qquad \partial _{t}^{2}g\left( t\right)
=2\frac{\left( \partial _{t}f\left( t\right) \right) ^{2}}{f\left( t\right)
^{2}}-2\frac{\partial _{t}^{2}f\left( t\right) }{f\left( t\right) }\ .
\end{equation*}%
In particular, since $f\left( 0\right) =1$ and $\partial _{t}f\left(
0\right) =0$, 
\begin{equation*}
g\left( 0\right) =0\ ,\quad \partial _{t}g\left( 0\right) =0\quad \text{and}%
\quad \partial _{t}^{2}g\left( 0\right) =-2\partial _{t}^{2}f\left( 0\right)
\ .
\end{equation*}%
Using Taylor's theorem, we conclude that 
\begin{equation*}
\lim_{t\rightarrow 0}\left\vert t^{-2}g\left( t\right) \right\vert =\partial
_{t}^{2}g\left( 0\right) =\left\langle \mathrm{H}\Psi ,\mathrm{H}\Psi
\right\rangle _{\mathfrak{h}}-\left\langle \Psi ,\mathrm{H}\Psi
\right\rangle _{\mathfrak{h}}^{2}\in \mathbb{R}_{0}^{+}
\end{equation*}%
when $\Psi \in \mathcal{D}\left( \mathrm{H}\right) $. By (\ref{dh2}), it
implies in this case that%
\begin{equation*}
\lim_{t\rightarrow 0}\left\vert t^{-2}g\left( t\right) \right\vert
=\lim_{t\rightarrow 0}\sum_{k=1}^{\infty }t^{-2}\ln \left( 1+\lambda
_{k,t}\right) \in \mathbb{R}_{0}^{+}\ .
\end{equation*}%
In the same way we get (\ref{ddd1}), we deduce from this last equality that 
\begin{equation*}
\limsup_{t\rightarrow 0}t^{-1}\left\Vert K_{t}\right\Vert _{2}\in \mathbb{R}%
_{0}^{+}\ .
\end{equation*}%
Combined with (\ref{sssssssss}) and (\ref{ddd20}), this last inequality
yields Assertion (ii).
\end{proof}

\begin{corollary}[Hilbert-Schmidt property of twisted off-diagonal elements]

\label{coro Bach-tech9}\mbox{}\newline
Assume all conditions of Proposition \ref{prop Bach-tech8}.

\begin{enumerate}
\item[\emph{(i)}] If $D_{0}$ can be extended by continuity to a bounded
operator, again denoted $D_{0}\in \mathcal{B}(\mathfrak{h})$, then there is $%
\delta \in \mathbb{R}^{+}$ such that $(\varkappa _{t}(D_{0}))_{t\in \left[
-\delta ,\delta \right] }\subseteq \mathcal{L}^{2}(\mathfrak{h})$ and this
operator family is continuous in $\mathcal{L}^{2}(\mathfrak{h})$ at $t=0$,
i.e., 
\begin{equation*}
\lim_{t\rightarrow 0}\left\Vert \varkappa _{t}\left( D_{0}\right)
\right\Vert _{2}=0\ .
\end{equation*}

\item[\emph{(ii)}] If for some $c\in \mathbb{R}^{+}\cup \{\infty \}$,%
\begin{equation*}
\mathrm{e}^{it\Upsilon _{0}}\mathcal{Y}\subseteq \mathcal{Y}\ ,\qquad t\in
\left( -c,c\right) \ ,
\end{equation*}%
with the operator family $(\overline{D}_{0}\mathrm{e}^{it\Upsilon
_{0}})_{t\in (-c,c)}$ being strongly continuous on $\mathcal{Y}$ and $\Psi
\in \mathcal{D}\left( \mathrm{H}\right) $, then $D_{0}$ can be extended by
continuity to a Hilbert-Schmidt operator, again denoted $D_{0}\in \mathcal{L}%
^{2}(\mathfrak{h})$.
\end{enumerate}
\end{corollary}

\begin{proof}
Combine Proposition \ref{prop Bach-tech8} with Lemma \ref{Lenma Bach-tech4
copy(1)} as well as Corollaries \ref{coro Bach-tech5} and \ref{coro
Bach-tech6}.
\end{proof}

\bigskip

\noindent \textit{Acknowledgments:} We thank Volker Bach, Walter de Siquiera
Pedra, Sascha Lill and anonymous reviewers for their valuable comments. This
work is supported by the Basque Government through the grant IT1615-22 and
the BERC 2022-2025 program as well as as the following grants:

\begin{itemize}
\item Grant PID2020-112948GB-I00 funded by MCIN/AEI/10.13039/501100011033
and by \textquotedblleft ERDF A way of making Europe\textquotedblright .

\item Grant PID2024-156184NB-I00 funded by MICIU/AEI/10.13039/501100011033
and cofunded by the European Union.
\end{itemize}

\end{document}